\documentclass[aps,prd,twocolumn,nofootinbib,superscriptaddress,noshowpacs,noshowkeys,preprintnumbers,amsmath,amssymb,floatfix]{revtex4}
\usepackage{graphicx}
\usepackage{color}
\usepackage[rgb]{xcolor}

\DeclareMathOperator{\im}{Im}
\DeclareMathOperator{\re}{Re}
\DeclareMathOperator{\Abs}{Abs}
\DeclareMathOperator{\Disp}{Disp}
\DeclareMathOperator{\sodot}{{\scriptscriptstyle\odot}}
\def\mpp{M_{\pi^\pm}}
\def\mpn{M_{\pi^0}}

\makeatletter
\newcommand{\doublewidetilde}[1]{{%
		\mathpalette\double@widetilde{#1}%
}}
\newcommand{\double@widetilde}[2]{%
	\sbox\z@{$\m@th#1\widetilde{#2}$}%
	\ht\z@=.9\ht\z@
	\widetilde{\box\z@}%
}
\makeatother

\def\bracketif<#1|#2>{\ensuremath{\left\langle #1 \vert_{f} #2\right\rangle_{i}}}

\def\MLoo{{\cal M}_{00}^L}

\def\MLpo{{\cal M}_{+0}^L}
\def\MLmp{{\cal M}_{x}^L} 

\def\MSpo{{\cal M}_{+0}^S}

\def\MSmp{{\cal M}_{x}^S}

\def\Mpoo{{\cal M}_{x}}
\def\Mpop{{\cal M}_{0+}}
\def\Mppm{{\cal M}_{+-}}
\def\Mppp{{\cal M}_{++}}

\def\Metaoo{{\cal M}_{00}^\eta}

\def\Metapo{{\cal M}_{+0}^\eta}
\def\Metamp{{\cal M}_{x}^\eta} 

\def\Aoo{{A}_{00}}
\def\Apo{{A}_{0+}}
\def\Ax{{A}_{x}}
\def\Apm{{A}_{+-}}
\def\App{{A}_{++}}

\DeclareMathOperator{\Li}{Li_2}
\DeclareMathOperator{\Poly}{\mathcal{J}}

\def\onlytwocolumn#1{} 

\renewcommand{\theequation}{\Roman{section}.\arabic{equation}}
\allowdisplaybreaks

\begin{document}

\title{Dispersive construction of two-loop $P \to \pi\pi\pi$ $(P=K,\eta)$ amplitudes}
\author{K.\ Kampf}
 \email{kampf@ipnp.mff.cuni.cz}
 \affiliation{Institute of Particle and Nuclear Physics, Faculty of Mathematics and Physics, 
Charles University, V Hole{\v{s}}ovi{\v{c}}k{\'a}ch 2, CZ-180 00 Prague 8, Czech Republic}

\author{M.\ Knecht}
 \email{Marc.Knecht@cpt.univ-mrs.fr}
 \affiliation{Centre de Physique Th\'{e}orique, CNRS/Aix-Marseille Univ./Univ. du Sud Toulon-Var (UMR 7332)\\
CNRS-Luminy Case 907, 13288 Marseille Cedex 9, France}

\author{J.\ Novotn{\'y}}
 \email{novotny@ipnp.mff.cuni.cz}
 \affiliation{Institute of Particle and Nuclear Physics, Faculty of Mathematics and Physics, 
Charles University, V Hole{\v{s}}ovi{\v{c}}k{\'a}ch 2, CZ-180 00 Prague 8, Czech Republic}

\author{M.\ Zdr{\'a}hal}
 \email{zdrahal@ipnp.mff.cuni.cz}
 \affiliation{Institute of Particle and Nuclear Physics, Faculty of Mathematics and Physics, 
Charles University, V Hole{\v{s}}ovi{\v{c}}k{\'a}ch 2, CZ-180 00 Prague 8, Czech Republic}

\begin{abstract}
We present and develop a general dispersive framework allowing us to construct
representations of the amplitudes for the processes $P\pi\to\pi\pi$, $P=K,\eta$, 
valid at the two-loop level in the low-energy expansion. The construction proceeds through a
two-step iteration, starting from the tree-level amplitudes and their $S$ and $P$ 
partial-wave projections. The one-loop amplitudes are obtained for all possible
configurations of pion masses. The second iteration is presented in detail in the cases where either all
masses of charged and neutral pions are equal, or for the decay into three
neutral pions. Issues related to analyticity properties of the amplitudes and of
their lowest partial-wave projections are given particular attention.
This study is introduced by a brief survey of the situation, for both experimental
and theoretical aspects, of the decay modes  into three pions of charged and 
neutral kaons and of the eta meson.
\end{abstract}


\maketitle
\allowdisplaybreaks

\section{Introduction}\label{sec:intro}

Our experimental knowledge of the Dalitz-plot structures of the amplitudes for the processes 
$P\to \pi\pi\pi$ has substantially improved during the last decades,
 for $P=K^\pm$
\cite{Batley:2005ax,Batley:2006mu,Batley:2006tt,Batley:2007md,Batley:2007aa,%
Batley:2000zz,Batley:2010fj},
$P=K_L$ \cite{Abouzaid:2008aa}, 
or $P=\eta$ \cite{Bashkanov:2007aa,Ambrosino:2008ht,Adolph:2008vn,Prakhov:2008ff,Unverzagt:2008ny,%
Ambrosinod:2010mj,Adlarson:2014aks,Anastasi:2016cdz,Ablikim:2015cmz,Prakhov:2018tou}.
This situation is likely to improve further still in the future
\cite{Li:2009jd,Gan:2015nyc}. The sizes of the collected data samples in the case of the charged kaons,
for instance, outgrow by orders of magnitude those that were available before. 
This general increase in statistics has prompted various theoretical studies 
\cite{Bijnens:2002vr,Bijnens:2004ku,Bijnens:2004vz,Bijnens:2004ai,Gamiz:2003pi,%
Cabibbo:2004gq,Cabibbo:2005ez,Gamiz:2006km,Colangelo:2006va,Bijnens:2007pr,Bissegger:2007yq,%
Bissegger:2008zz,Bissegger:2008ff,Gullstrom:2008sy,Schneider:2010hs,Kampf:2011wr,Gasser:2011ju,Descotes-Genon:2014tla,Kolesar:2016jwe,Colangelo:2016jmc,Albaladejo:2017hhj,Kolesar:2017xrl,Colangelo:2018jxw}
of these decay modes, often with an emphasis
on isospin-breaking contributions. Indeed, from the theoretical point of view,
these processes are interesting because they provide access to fundamental quantities. For instance,
the rates for the decays $\eta\to \pi\pi\pi$, which are forbidden in the isospin limit,
offer a good possibility to determine the value of the quark 
mass ratio 
\begin{equation}\label{R mass ratio}
R=\frac{m_s-\widehat{m}}{m_d-m_u}\,, 
\end{equation}
where $m_u$, $m_d$, and
$m_s$ denote the masses of the three lightest quark flavors, while ${\widehat m} = (m_u + m_d)/2$.
Furthermore, the processes with two neutral pions 
in the final state exhibit the so-called cusp effect, which contains information on the $\pi\pi$  
scattering lengths in the $S$-wave. Concerning this last aspect, in particular, 
the decay modes $K^\pm\rightarrow\pi^\pm\pi^0\pi^0$ 
and $K_L \to 3\pi^0$ have already been studied from this point of view by the NA48 
\cite{Batley:2005ax,Batley:2000zz} and KTeV \cite{Abouzaid:2008aa} collaborations,     
respectively. The first attempts to measure the same effects
in the decays of the $\eta$ meson into three neutral pions have also been reported \cite{Prakhov:2008ff}.\footnote{Depending on the process,
the cusp can be more or less pronounced, and therefore more or less easy to measure.
A criterion allowing one to estimate the ``visibility'' of the cusp in the different processes mentioned above
has been proposed and discussed in Ref.~\cite{Kampf:2011wr}.}

Traditionally, the processes $P \to \pi\pi\pi$ are most of the time being 
analysed with a polynomial parameterization 
[in terms of slopes and curvatures in appropriately chosen
Dalitz-plot variables] of the amplitude, 
and theoretical expressions have often been given in this form as well. 
It is clear that the study of non-analytic features
of the amplitude, like a cusp, cannot be done within such a simple framework.
The aim of the work presented in this article
is, therefore, the construction of a model-independent form of two-loop amplitudes 
of the processes mentioned above that is valid up to
two loops in the low-energy expansion, and that exhibits the correct unitarity parts coming
from the $\pi\pi$ intermediate states. These are the only states that, up to
that order, give rise to non-analytic structures in the 
corresponding decay amplitudes. Other two-meson intermediate states correspond to more 
remote thresholds for the $P\pi\to\pi\pi$ scattering amplitudes that,
inside the decay region, can be quite appropriately approximated and described by a polynomial. 
Intermediate states with more than two mesons occur only at higher orders in the chiral expansion, 
and will, therefore, not be considered here. Our aim is, however, to include isospin-breaking 
effects induced by the mass difference between charged and neutral pions,
without which no cusp would be seen in the decay distribution. 
Note that the effects of electromagnetic interactions other than those leading to isospin
breaking in the meson masses, i.e.\ exchanges of virtual photons
between charged states and photon emission, are not included in the construction.
A more complete discussion of this aspect is given in Sec.~2.1 of Ref.~\cite{DescotesGenon:2012gv}.

Our construction is adapted from the ``reconstruction theorem,'' 
first established in Ref.~\cite{Stern:1993rg}
for the case of the $\pi\pi$ amplitude in the isospin limit. 
The authors of Ref.~\cite{Stern:1993rg} have shown 
that, up to two loops in the chiral expansion, the analytical form 
of the amplitude is completely fixed by general properties like relativistic 
invariance, unitarity, analyticity, and crossing, supplemented
by chiral counting for the partial waves. Isospin symmetry was also invoked, 
but its role merely served to reduce the scattering amplitudes in the various 
channels to a single amplitude. The construction of the analytical expression 
of the $\pi\pi$ amplitude in this framework was then implemented explicitly in Ref.~\cite{Knecht:1995tr}.   
The method is, however, more general and allows for several extensions. First, it 
also applies to the scattering amplitudes involving other pseudo-scalar mesons \cite{Zdrahal:2008bd}. 
Next, it generalizes in a straightforward manner to 
the situation without isospin symmetry, the main difference being that several 
independent amplitudes will be involved in order to describe the different channels 
\cite{DescotesGenon:2012gv}. Finally, it also applies to other observables, like form factors
\cite{DescotesGenon:2012gv,Bernard:2013faa,Colangelo:2015kha}.
The present work extends the general method of the ``reconstruction theorem'' 
of Ref.~\cite{Stern:1993rg} to the amplitudes of the scattering processes $P\pi\rightarrow\pi\pi$,
$P=K^\pm, K_L, K_S,\eta$, in the threshold region. The amplitudes for the decay processes $P\rightarrow 3\pi$ are then 
obtained by analytic continuation below the threshold and inside the physical decay region
in the Mandelstam plane.

This is the first article devoted to the presentation of the details of the construction 
of such analytical expressions for the decay amplitudes $P \to \pi\pi\pi$ up to two loops in the low-energy expansion.  
In the present article, we give the full isospin-breaking result for all the amplitudes only at the one-loop level, 
while at the first stage, the expressions of the two-loop amplitudes are only worked 
out in the limit where the masses of the  neutral and charged pions are equal. 
This allows describing some general features of our construction
in a simpler framework without having to deal, in addition, with several kinematic
complications that arise only when the intermediate- and final-state pions have unequal masses. 
Incidentally, this is the framework that we have used in Ref.~\cite{Kampf:2011wr}, devoted to the
analysis of the decay of the $\eta$ meson into three pions. We also plan to update
the latter analysis, taking more recent data \cite{Adlarson:2014aks,Anastasi:2016cdz,Prakhov:2018tou}
into account, but this will be left for a separate work \cite{plan1}.
A rather simple extension of this framework, though, allows dealing with isospin breaking in the 
pion masses in the case of the decay channels into three neutral pions, which we will treat in a 
second stage in the present article. 
The results where all isospin-breaking effects due to the mass difference 
between neutral and charged pions are included up to two loops
when the final state contains also charged pions will be discussed in a forthcoming paper \cite{plan2}. 
We also do not address the possible violation of {\it CP} invariance,  for instance, in the $K\to\pi\pi\pi$
processes, although it could be straightforwardly incorporated into our framework if necessary.

The outline of this paper is then as follows. 
In the next section, we briefly summarize all the processes 
in question and list the existing studies of the last few years.
Section III recalls the main aspects and content of the reconstruction theorem and
introduces our notation. In Sec.~IV, we write the results of the first iteration 
of the reconstruction theorem for the $\pi\pi$ scattering and
the $P\pi \to \pi\pi$ amplitudes. Section V then gives the result of the second iteration 
in the limit where the charged and neutral pion masses are identical. 
These results are extended, for the decay modes into three neutral
pions, to the situation where the difference in the pion masses is taken into
account in Sec.~VI. The final section is devoted to a summary and conclusions.
In order not to overload the main text with too many technical issues and
lengthy expressions, some of them have been gathered in four appendices. 

Some aspects of this work have also been discussed in earlier preliminary 
reports \cite{Kampf:2008ts,Zdrahal:2009ns,Zdrahal:2009cp}. A comprehensive
account with more details on some of the technical aspects can also
be found in Ref.~\cite{Zdrahal}.

\section{Processes in question}\label{sec:processes}
\setcounter{equation}{0}

Our analysis covers the following list of processes:
\begin{subequations}\label{procesy}
\begin{align}
K^\pm&\rightarrow\pi^0\pi^0\pi^\pm, \\
K^\pm&\rightarrow\pi^\pm\pi^\pm\pi^\mp, \\
K_L&\rightarrow\pi^0\pi^0\pi^0, \\ 
K_L&\rightarrow\pi^+\pi^-\pi^0, \\
K_S&\rightarrow\pi^+\pi^-\pi^0, \\
\eta&\rightarrow\pi^0\pi^0\pi^0, \\
\eta&\rightarrow\pi^+\pi^-\pi^0. 
\end{align}
\end{subequations}
The natural framework for describing the amplitudes of these processes 
is three-flavor chiral perturbation theory \cite{GL1}, extended, in the case of kaon decays, 
to also include the weak non-leptonic decays \cite{KMW1}.
At the lowest order, the corresponding amplitudes are simply first-order polynomials 
in the Mandelstam variables. Here, as already mentioned, we will ignore {\it CP}-violating contributions, 
so that the coefficients of these lowest-order polynomials are real.  
The computation of higher orders meets technical complications,
as well as the necessity to consider an increasing number of low-energy constants,
which then need to be determined independently.  

In the rest of this section, we are going to describe the present 
situation for the three types of processes, the decay
of a charged or of a neutral {\it CP}-odd kaon into three pions,
the similar decay modes of the $\eta$ meson, and finally the decay
into three pions of a neutral {\it CP}-even kaon.

\subsection*{$K^\pm$ and $K_L^0$ decays into three pions}

These decay amplitudes have been computed using the framework of chiral perturbation theory up to next-to-leading order (NLO) already in 1991 \cite{KMW2}, at that time ignoring all the isospin-breaking effects. Explicit expressions for the one-loop amplitudes in  the
isospin limit have been given in Ref.~\cite{Bijnens:2002vr}. The most comprehensive
work also including isospin breaking, but still stopping at NLO only, 
is contained in the series of papers \cite{Bijnens:2004ku,Bijnens:2004vz,Bijnens:2004ai}.

The interest for these processes has risen after the observation of a
unitarity cusp in the event distribution with respect to the invariant
mass of two neutral pions by the NA48 collaboration
\cite{Batley:2005ax}. In Ref.~\cite{Cabibbo:2004gq}, this cusp has been proposed   
as a potentially clean method allowing for the determination of the 
combination $a_0-a_2$ of $\pi\pi$ scattering lengths. 
Elaborations on this idea appeared in Ref.~\cite{Cabibbo:2005ez}.
Assuming a simplified analytic structure of the amplitude and using unitarity of the 
scattering matrix allows expressing the decay amplitude in the vicinity of the cusp
as an expansion in the scattering lengths $a_i$.
In Ref.~\cite{Gamiz:2006km}, the same assumptions are made, but in addition, the 
isospin symmetric NLO result of chiral perturbation theory is used as an input 
for the real parts of the amplitudes.

It was pointed out by the authors of Refs.~\cite{Colangelo:2006va,Bissegger:2007yq,Gasser:2011ju} that the correct 
analytic structure of the amplitudes should be more complicated than described 
in Refs.~\cite{Cabibbo:2005ez,Gamiz:2006km}, and they have constructed a representation
of the amplitude within the framework of
non-relativistic effective field theory, based on a combined expansion in 
both the scattering lengths and a formal non-relativistic parameter $\varepsilon$.
This expansion is considered to remain valid over the whole decay region,
although the pions emitted at the edge of the decay region are already relativistic.
Finally, taking advantage of working within a Lagrangian formulation, the
effects of real and virtual photons could also be accounted for \cite{Bissegger:2008ff}
within this non-relativistic framework.

\subsection*{$\eta$ decays into three pions}

The decay modes $\eta \to \pi^+\pi^-\pi^0$ and $\eta \to \pi^0\pi^0\pi^0$ are $\Delta I =1$
transitions, and thus require isospin breaking. The latter is provided by two sources,
electromagnetic interactions on the one hand, and the quark mass difference $m_d - m_u$
on the other hand. 
As far as the former is concerned, contributions of the order ${\cal O}(e^2 E^0)$
vanish \cite{Sutherland:1966zz,Bell:1996mi}, and corrections of the order
${\cal O}(e^2 m_q)$, $q=u,d,s$ to the decay rate were found to be quite small \cite{Baur:1995gc,Ditsche:2008cq}.
Thus, to a very good approximation, the amplitudes for the decay modes of the $\eta$ meson
into three pions are proportional to the isospin-breaking quark mass difference $m_d - m_u$,
e.g., $A^{\eta\to \pi^+\pi^-\pi^0} (s,t,u) = (\sqrt{3}/4 R) f (s,t,u)$, 
with the quark-mass ratio $R$ already defined in (\ref{R mass ratio}).
Measuring the corresponding decay rates thus directly gives   
information on $m_d - m_u$,
provided one knows $f (s,t,u)$ sufficiently well.

The amplitude $f (s,t,u)$ has been computed in the chiral expansion, at
orders ${\cal O} (E^2)$ \cite{Cronin:1967jq,Bell:1996mi}, ${\cal O} (E^4)$ \cite{Gasser:1984pr}, 
and ${\cal O} (E^6)$ \cite{Bijnens:2007pr}. The convergence is, however, slow,
due to strong $\pi\pi$ rescattering effects. Furthermore, the two-loop expression
involves many unknown ${\cal O} (E^6)$ low-energy constants (LEC). The very long complete analytical expression of the ${\cal O} (E^6)$ amplitude has not been published, but is 
available as a Fortran code from the authors of Ref.~\cite{Bijnens:2007pr}.
Other approaches have, therefore, been considered in order to improve the situation. For instance,
a more compact explicit representation of $f (s,t,u)$ at NNLO can also be worked out \cite{Schneider:2010hs}
within the non-relativistic framework mentioned above. Notice, however, that in the center 
of the Dalitz plot, the momenta of the outgoing pions in the rest frame of the decaying particle,
which count as order ${\cal O}(\varepsilon)$, can already represent 90\% of their rest energy, which is counted as order ${\cal O}(1)$.

Alternatively, the iterative resummation of the $\pi\pi$ rescattering effects can be handled numerically 
in a dispersive framework \cite{Kambor:1995yc,Anisovich:1996tx}.
In this second approach, one writes unitarity relations with $\pi\pi$ intermediate states and constructs 
dispersion relations of the Khuri-Treiman type \cite{Khuri:1960zz,Kacser,Bonnevay:1963,Bronzan:1964zz,Aitchison:1965,Neveu:1970tn,Anisovich:1993kn}. The amplitude is then obtained
by finding the numerically fixed-point solution of these relations. The determination 
of the subtraction constants arises from matching \cite{Kambor:1995yc,Anisovich:1996tx} with 
the NLO results of chiral perturbation theory.
A more recent analysis \cite{Colangelo:2016jmc} within this framework uses
instead information from the chiral perturbation theory calculation at NNLO 
of Ref.~\cite{Bijnens:2007pr}.

All these studies, but the one of Ref.~\cite{Schneider:2010hs}, address the description of the 
amplitude $f (s,t,u)$ in the isospin limit, thus leaving out the possibility to discuss the 
effect of the cusp in the $\pi^0\pi^0$ invariant mass of the decay of the $\eta$ into three 
neutral pions. Isospin-breaking effects in $f (s,t,u)$ can also be naturally included in the relativistic approach 
we will develop in the present work.

\subsection*{$K_S$ decay into three pions}

Our procedure can easily be applied as well to the {\it CP}-conserving part of the decay
amplitude for the decay $K_S\to\pi^+\pi^-\pi^0$. Nevertheless, the corresponding branching  
ratio is very small, which makes it difficult to measure the energy dependence of the decay 
distribution, although some measurements exist \cite{Angelopoulos:1998aw,Angelopoulos:2003hm}.
Therefore, the only parameter connected with this process that has been measured 
recently \cite{Batley:2005zp} is the amplitude of the {\it CP}-conserving component of 
the decay $K_S\to\pi^+\pi^-\pi^0$ relative to $K_L \to\pi^+\pi^-\pi^0$.
From the theoretical point of view, this process is also covered by the 
chiral perturbation theory computations up to NLO presented in 
Refs.~\cite{Bijnens:2002vr,Bijnens:2004ku,Bijnens:2004vz,Bijnens:2004ai}.

\section{General structure of the two-loop amplitudes}\label{sec:general}
\setcounter{equation}{0}

As mentioned above, we will obtain the amplitudes ${\cal M}(s_1,s_2,s_3)$ 
describing the decay processes 
\begin{equation}\label{rozpad}
P(k)\to \pi(p_1)\, \pi(p_2)\, \pi(p_3)
\end{equation}
from an analytic continuation into the subthreshold region of the
amplitude ${\cal M}(s,t,u)$ of a corresponding scattering process
of the type
\begin{equation}\label{rozptyl}
P(k)\,\pi(p_1)\to \pi(p_2)\, \pi(p_3),
\end{equation}
where $k^2 = M_P^2$, $p_i^2 = M_i^2$.
In the first case, the variables are defined as usual,
$s_i = (k-p_i)^2$, whereas in the scattering region we take
$s=(k+p_1)^2$, $t=(k-p_2)^2$, $u=(k-p_3)^2$. These variables
satisfy $s_1+s_2+s_3 = 3s_0$ and $s+t+u =  3s_0$, respectively, where
\begin{equation}\label{3s0}
3 s_0 = M_P^2 + M_1^2 + M_2^2 + M_3^2.
\end{equation}
Notice that,
depending on the phase convention used for the various states, 
this analytic continuation does not necessarily boil down
to a mere substitution of the variables $(s,t,u)$ by $(s_1,s_2,s_3)$,
but can also generate an additional phase for the amplitude. We shall
specify our convention in that matter later in this section. For the time
being, we use a rather generic notation, in order to keep the discussion as 
general as possible. Later on, when considering specific processes,
we shall refine the notation according to our needs [see in
particular Tables \ref{table_calM} and \ref{pipiTable} below,
which also specify the phase conventions and the notation we will use for the various meson masses].

From a practical point of view, and in order not to obscure the line of reasoning with analyticity 
issues connected with the fact that $M_P > 3 M_\pi$,
it is useful, at a first stage, to replace $M_P$ by a fictitious
mass ${\overline M}_P$, somewhat lower than $3 M_{\pi^0}$, or than 
$3 M_\pi$ if isospin-breaking effects are neglected.\footnote{The fact that
the kaons are also unstable through their decay into two pions is irrelevant here
since this feature would only start to play a role beyond the orders in the
low-energy expansion we are considering. The $\eta$ meson is stable with respect 
to decay into two pions if {\it CP} is conserved, which we have already assumed to be the case.}
Then, the amplitudes possess the usual analyticity properties. In particular, 
they are real analytic \cite{Eden:1966dnq}. After the construction of the two-loop 
amplitude for this fictitious process is completed, one can perform an
analytic continuation in ${\overline M}_P^2$ toward its physical value, provided
it is endowed with a small positive imaginary part $\delta$, ${\overline M}_P^2 \to M_P^2 + {\rm i}\delta$.
The issues related to this analytic continuation will be discussed
in Section~\ref{Sec:Second iteration isospin} and Appendix~\ref{app:integrals}.
Notice simply, at this stage, that in the course of this process, real 
analyticity is lost, and the real and imaginary parts of the amplitude along its cuts 
actually become complex dispersive and absorptive parts, respectively, i.e., schematically,
\begin{equation}\label{Def:disp_abs}
\begin{split}
&\re {\cal M}(s)\rightarrow \Disp{\cal M}(s)=\frac{1}{2}[{\cal M}(s+{\rm i}0)+{\cal M}(s-{\rm i}0)],\\
&\im {\cal M}(s)\rightarrow \Abs{\cal M}(s)=\frac{1}{2{\rm i}}[{\cal M}(s+{\rm i}0)-{\cal M}(s-{\rm i}0)],
\end{split}
\end{equation}
where $0$ indicates an infinitesimal positive number.
In order to keep the notation simple, we will continue denoting the mass of the meson $P$ by $M_P$,
even when it is smaller than $3 M_\pi$. This slight abuse of notation should not cause any confusion.
Let us also point out that, even for values of $M_P$ lower than $3 M_\pi$,
the analytic properties of the partial-wave projections
are more involved than, say, in the $\pi\pi$ case, where,
besides the $s$-channel unitarity cut, there is only a left-hand cut
coming from the unitarity cut in the $u$ channel. In the case of the $P\pi\to\pi\pi$
partial-wave projections, additional circular cuts can be present.
For a more complete discussion, we refer the reader to Refs.~\cite{Kennedy1962} and \cite{Kacser}.

In order to proceed with the construction of the scattering
amplitudes ${\cal M}(s,t,u)$, we write, keeping the preceding discussion
in mind and following Ref.~\cite{Stern:1993rg}, 
thrice subtracted, fixed-$t$, dispersion relations\footnote{Due to the Froissart
bound and Regge phenomenology, two subtractions are enough to ensure convergence of the dispersion
integrals \cite{Kambor:1995yc,Anisovich:1996tx}. However, as shown in \cite{Stern:1993rg}, in order to obtain a
representation that correctly accounts for all two-loop contributions,
it is necessary to start from over-subtracted dispersion relations.
Two subtractions would be enough, though, for the purpose of constructing 
a representation of the amplitudes valid at one loop only. We will make
implicit use of this last remark in Sec.~\ref{Sect:first_iteration}. \label{footnote1}}
\begin{equation}\label{disprel}
\begin{split}
&{\cal M}(s,t,u)=a(t) + (s-u)b(t) + (s-u)^2 c(t)\\
&\qquad{}+\frac{s^3}{\pi}\int_{s_\mathrm{thr}}^\infty \frac{dx}{x^3} 
\frac{\Abs {\cal M}(x,t,3s_0-x-t)}{x-s}\\
&\qquad{}+\frac{u^3}{\pi}\int_{u_\mathrm{thr}}^\infty \frac{dx}{x^3} 
\frac{\epsilon\Abs{\cal M}_u(x,t,3s_0 -x-t)}{x-u}\,,
\end{split}
\end{equation}
where $a(t)$, $b(t)$, and $c(t)$ are arbitrary subtraction functions,
whereas $s_\mathrm{thr}$ and $u_\mathrm{thr}$ denote the thresholds in the 
corresponding channels. In the presence of anomalous thresholds,
the corresponding discontinuities should be included. Otherwise,
$s_\mathrm{thr}$ and $u_\mathrm{thr}$ are fixed by unitarity.
In this expression, $\epsilon$ denotes a crossing phase, which
depends on the phase convention adopted for the various 
amplitudes, see Table \ref{table_calM}. We have chosen a phase
convention that reduces to the one of Condon and Shortley in the isospin limit. 
In practice, this amounts to having a minus sign
for the crossing of a charged pion, whereas the crossing of 
a neutral pion generates no phase. Furthermore, 
${{\cal M}_u}(s,t,u)$ denotes the amplitude
in the crossed $u$-channel. The latter obeys a similar dispersion relation,
with subtraction functions $a_u(t)$, $b_u(t)$, 
and $c_u(t)$, and with ${\cal M}$ replacing ${{\cal M}_u}$
in the $u$-channel integral. 

Notice that for the applications we have in mind,
we need only to assume that the above dispersion relations 
are valid in the region of the kinematic variables where the chiral expansion makes sense.
Then, they merely embody the analyticity
properties of the corresponding amplitudes constructed by the usual
Feynman diagram method in the field-theoretic framework of chiral perturbation theory.
Thus, from this point of view, their existence is actually not an issue in this context.
This link with Feynman diagrams will also provide the basis for the analytic continuation
in $M_P^2$, which has been discussed previously.

The next step consists in writing, for each amplitude ${\cal M}(s,t,u)$ in Table \ref{table_calM},
a decomposition of the form
\begin{equation}
{\cal M}(s,t,u) = 16\pi\left[
t_0(s) + 3 t_1(s)\cos\tilde\theta \right] + 
{\cal M}_{\ell\ge 2}(s,t,u),
\label{PWdecomp}
\end{equation}
where $t_0(s)$ and $t_1(s)$ denote the partial waves for
angular momentum $\ell$ equal to zero and one, respectively. The contributions 
from higher partial waves are contained in ${\cal M}_{\ell\ge 2}(s,t,u)$.
The scattering angle is given by
\begin{equation}
\cos\tilde\theta = \frac{1}{2K_{P1;23}(s)}\left[2t + s - 3s_0 + \frac{\Delta_{P1}\Delta_{23}}{s}\right]
\label{theta_vs_t}
\end{equation}
with $\Delta_{P1}\equiv M_P^2 - M_1^2$ and $\Delta_{23} \equiv M_2^2 - M_3^2$.
The function $K_{P1;23}(s)$ is defined by
\begin{equation}
K_{P1;23}^2(s) = \frac{1}{4s^2}\,\lambda_{P1}(s) \lambda_{23}(s),
\label{Kacser_def}
\end{equation}
where $\lambda_{P1}(s) = \lambda(s,M_P^2,M_1^2)$, $\lambda_{23}(s) = \lambda(s,M_2^2,M_3^2)$
are expressed in terms of the K\"allen or triangle function $\lambda(x,y,z)\equiv x^2 + y^2 + z^2 -2xy - 2xz - 2yz$.
We will refer to $K_{P1;23}(s)$ as the Kacser function since
it generalizes the function introduced in Ref.~\cite{Kacser} to the
case where the mass difference between neutral and charged pions is not neglected. 
The Kacser function $K_{P1;23}(s)$ as an analytic function of $s$ is unambiguously defined by its cuts, 
which are conventionally placed at the real axis\footnote{For the case $M_P>M_1+M_2+M_3$, the cuts  
correspond to the intervals  $((M_2-M_3)^2,(M_2+M_3)^2)$  and $((M_P-M_1)^2,(M_P+M_1)^2)$} and by the values of (\ref{Kacser_def}) as long 
as  $s$ corresponds to the scattering region, $s\ge (M_P+M_1)^2$. 
Eventually, we need 
to continue the amplitude analytically to lower values of $s$. This will require a 
careful study of the analytic properties of $K_{P1;23}(s)$ in the complex $s$-plane
for the different configurations of pion masses in Sections~\ref{Sec:Second iteration isospin} and \ref{Sec:second iteration neutral}.
The same also applies to the partial-wave projections $t_0(s)$ and $t_1(s)$. For
$s\ge (M_P+M_1)^2$, they are defined as usual by the integrals
\begin{equation}
t_\ell(s) = \frac{1}{32\pi}\int_{-1}^{+1}d\cos\tilde\theta\,(\cos\tilde\theta)^\ell{\cal M}(s,t,u),\ \ell=0,1 .   
\label{t_ell_def}     
\end{equation} 

\indent

\begin{table}[t]
\begin{center}
\begin{tabular}{|l||l|l|c|l|}
\hline
\multicolumn{1}{|c||}{process} & \multicolumn{1}{c|}{${\cal M}$} & \multicolumn{1}{c|}{${{\cal M}_u}$} & $\epsilon$  & \multicolumn{1}{c|}{$3 s_0$}
\\
\hline
$K^\pm\pi^\pm \to \pi^\pm\pi^\pm$ & $\Mppp$ & $\Mppm$ & $+1$ & $M_K^2 + 3 M_\pi^2 $
\\
\hline
$K^\pm\pi^\mp \to \pi^\pm\pi^\mp$ & $\Mppm$ & $\Mppp$ & $+1$ & $M_K^2 + 3 M_\pi^2 $
\\
\hline\hline
$K^\pm\pi^\mp \to \pi^0\pi^0$ & $\Mpoo$ & $\Mpop$ & $-1$ & $M_K^2 + M_\pi^2 + 2 M_{\pi^0}^2$
\\
\hline
$K^\pm\pi^0 \to \pi^0\pi^\pm$ & $\Mpop$ & $\Mpoo$ & $-1$ & $M_K^2 + M_\pi^2 + 2 M_{\pi^0}^2$
\\
\hline\hline
$K_L\pi^0 \to \pi^\pm\pi^\mp$ & $\MLmp$ & $\MLpo$ & $-1$ & $M_{K_L}^2 + 2 M_\pi^2 + M_{\pi^0}^2 $
\\
\hline
$K_L\pi^\pm \to \pi^\pm\pi^0$ & $\MLpo$ & $\MLmp$ & $-1$ & $M_{K_L}^2 + 2 M_\pi^2 + M_{\pi^0}^2 $
\\
\hline\hline
$K_L\pi^0 \to \pi^0\pi^0$ & $\MLoo$ & $\MLoo$ & $+1$ & $M_{K_L}^2 + 3 M_{\pi^0}^2 $
\\
\hline\hline
$K_S\pi^0 \to \pi^+\pi^-$ & $\MSmp$ & $\MSpo$ & $-1$ & $M_{K_S}^2 + 2 M_\pi^2 + M_{\pi^0}^2 $
\\
\hline
$K_S\pi^+ \to \pi^+\pi^0$ & $\MSpo$ & $\MSmp$ & $-1$ & $M_{K_S}^2 + 2 M_\pi^2 + M_{\pi^0}^2 $
\\
\hline\hline
$\eta\,\pi^0 \to \pi^\pm\pi^\mp$ & $\Metamp$ & $\Metapo$ & $-1$ & $M_\eta^2 + 2 M_\pi^2 + M_{\pi^0}^2 $
\\
\hline
$\eta\,\pi^\pm \to \pi^\pm\pi^0$ & $\Metapo$ & $\Metamp$ & $-1$ & $M_{\eta}^2 + 2 M_\pi^2 + M_{\pi^0}^2 $
\\
\hline\hline
$\eta\,\pi^0 \to \pi^0\pi^0$ & $\Metaoo$ & $\Metaoo$ & $+1$ & $M_\eta^2 + 3 M_{\pi^0}^2 $
\\
\hline
\end{tabular}
\caption{For each process under consideration, we show the amplitude ${\cal M}(s,t,u)$,
and the corresponding amplitude ${{\cal M}_u}(s,t,u)$ in the crossed $u$-channel
that appears in the dispersion relation (\ref{disprel}). The penultimate column
gives the appropriate crossing phase $\epsilon$, and the last column displays,
for each process, the quantity $3 s_0$. Here, $M_\pi$ and $M_K$ denote
the charged pion and kaon masses, respectively. {\it CP} violation is ignored
so that the same amplitude describes the {\it CP}-conjugate processes. On the other hand,
we keep the distinction between the masses of the neutral kaons $K_L$ and $K_S$.
 }\label{table_calM}
\end{center}
\end{table}
\noindent
In order to construct the analytic continuations of the partial waves
outside of the region of definition consisting of the portion of the
real axis $s\ge (M_P+M_1)^2$, the contour of integration in Eq.~(\ref{t_ell_def})
needs to be appropriately deformed in the complex plane. This issue, which is rather
well documented in the literature \cite{Bronzan,Kacser,Anisovich}
in the case where all the pions have equal masses, 
requires a careful discussion, and we address it when considering the second iteration of the construction process
in Section \ref{Sec:Second iteration isospin} [see also Appendix \ref{app:anomalous}]. 
For the moment, however, we proceed with our general outline. To that effect, 
it is necessary to use our knowledge about the dominant chiral behavior of the various 
quantities appearing in Eq.~(\ref{PWdecomp}). For the dispersive and 
absorptive parts of the lowest partial-wave projections, we have
\begin{equation}
\begin{split}
\Disp t_{\ell=0,1}(s) ={\cal O}(E^2), \ \    \Disp {\cal M}_{\ell\geq 2}={\cal O}(E^4),
\\
\Abs t_{\ell=0,1}(s) = {\cal O}(E^4),\  \ \  \Abs {\cal M}_{\ell\geq2}={\cal O}(E^8).
\end{split}\label{chibeh}
\end{equation}
Therefore, up to order ${\cal O}(E^6)$, only the $S$ and $P$ waves contribute
to the absorptive parts of the dispersion relations. Imposing the crossing relations, the subtraction function reduces to a polynomial ${\cal P}(s,t,u)$ 
of at most third order in the Mandelstam variables \cite{Stern:1993rg}. 
The amplitudes eventually take the form
\begin{equation}
{\cal M}(s,t,u)={\cal P}(s,t,u)+{\cal U}(s,t,u)+{\cal O}(E^8),
\label{M_for_P_to_3pi}
\end{equation}
where ${\cal U}(s,t,u)$ denotes the non-analytic unitarity part,
\begin{equation} 
\begin{split}
{\cal U}(s,t,u) &= 16\pi\big[{\cal W}_0(s)+3(t-u){\cal W}_1(s)
\\
&+
{\cal W}_0^t(t) +3(u-s){\cal W}_1^t(t)
\\
&+
{\cal W}_0^u(u)+3(t-s){\cal W}_1^u(u)
\big]
.
\end{split}
\label{U_for_P_to_3pi}                
\end{equation}
The functions ${\cal W}_\ell(s)$, $\ell=0,1$ are analytic in the
complex $s$-plane, except for a right-hand cut, with discontinuities
provided by the absorptive parts along this same cut of the partial-wave projections,
for instance,
\begin{align}
\Abs {\cal W}_0(s) &= \left[ \Abs t_0(s) + 3\, 
\frac{\Delta_{P1}\Delta_{23}}{s}\,
\frac{\Abs t_1(s)}{2K_{P1;23}(s)} \right]
\nonumber\\
& {}\times \theta (s-s_\mathrm{thr}) ,
\nonumber\\
\Abs {\cal W}_1(s) &=  
\frac{\Abs t_1(s)}{2K_{P1;23}(s)}\, \theta (s-s_\mathrm{thr})
.
\label{discW}
\end{align}
Similar properties hold for the other functions ${\cal W}_\ell^{t,u}(s)$, but
their absorptive parts are now given in terms of the partial waves of the
amplitudes in the crossed channels. The absorptive parts (\ref{discW}) do not
completely fix the functions ${\cal W}_{0,1}(s)$. Without loss of generality,
we shall require that they satisfy the asymptotic conditions \cite{Knecht:1995tr}
\begin{equation}\label{asymptotic conditions}
\lim_{\vert s\vert\to\infty}\,{\cal W}_0(s)/s^4 =0,
\quad \lim_{\vert s\vert\to\infty}\,{\cal W}_1(s)/s^3 =0
\end{equation}
up to arbitrary polynomials in $s$, of at most third order
for ${\cal W}_0$ or at most second order for ${\cal W}_1$, which are absorbed into ${\cal P}(s,t,u)$. 
Note that, even after the implementation of these conditions, the ambiguity in the single-variable polynomials 
is not entirely fixed, however, the form of ${\cal P}(s,t,u)$ is unique (up to inconsequential ambiguities
stemming from the condition $s+t+u=3s_0$). These expressions form the content of
the reconstruction theorem in the present context. Notice that, depending
on the symmetries of the amplitude ${\cal M}(s,t,u)$, these functions are
not all independent. Actually, making use of the arbitrariness of the various functions 
${\cal W}_{0,1}(s)$, one can achieve that ${\cal U}(s,t,u)$ and ${\cal P}(s,t,u)$
separately have the same $s$, $t$, $u$ symmetries as the full amplitude
${\cal M}(s,t,u)$. Finally, as far as their structure is concerned,
the amplitudes of the processes involving the $\eta$ meson can be obtained
from those involving the $K_L$ meson, upon changing the corresponding labels.
We thus need to consider only thirteen distinct functions, which we choose
as indicated in Table~\ref{W_for_P-pi_Table}. Note that a given process involves at most only three distinct functions.
\begin{table}[t]
\begin{center}
\begin{tabular}{|c||c|c|c|c|c|c|}
\hline
 ${\cal M}$ & ${\cal W}_0$ & ${\cal W}_1$ & ${\cal W}_0^t$ & ${\cal W}_1^t$  & ${\cal W}_0^u$  &  ${\cal W}_1^u$
\\
\hline
 ${\cal M}_{++}$  & ${\cal W}_{++}$ & $-$ & ${\cal W}_{+-}^{(0)}$ & ${\cal W}_{+-}^{(1)}$ & ${\cal W}_{+-}^{(0)}$  & ${\cal W}_{+-}^{(1)}$
\\
\hline
 ${\cal M}_{+-}$  & ${\cal W}_{+-}^{(0)}$ & ${\cal W}_{+-}^{(1)}$ & ${\cal W}_{+-}^{(0)}$ & $-{\cal W}_{+-}^{(1)}$  %
  & ${\cal W}_{++}$ & $-$ 
\\
\hline
 ${\cal M}_x$  & ${\cal W}_{x}$  & $-$ & $-{\cal W}_{0+}^{(0)}$ & $-{\cal W}_{0+}^{(1)}$ & $-{\cal W}_{0+}^{(0)}$ & $-{\cal W}_{0+}^{(1)}$
\\
\hline
 ${\cal M}_{0+}$   & ${\cal W}_{0+}^{(0)}$     & ${\cal W}_{0+}^{(1)}$    & ${\cal W}_{0+}^{(0)}$ & $-{\cal W}_{0+}^{(1)}$  %
  &  $- {\cal W}_{x}$ & $-$
 \\
 \hline
 ${\cal M}_x^L$  & ${\cal W}_{L;x}$  & $-$ & $-{\cal W}_{L;+0}^{(0)}$ & $-{\cal W}_{L;+0}^{(1)}$   %
  & $-{\cal W}_{L;+0}^{(0)}$ & $-{\cal W}_{L;+0}^{(1)}$
\\
\hline
 ${\cal M}^L_{+0}$   & ${\cal W}_{L;+0}^{(0)}$     & ${\cal W}_{L;+0}^{(1)}$    & ${\cal W}_{L;+0}^{(0)}$ & $-{\cal W}_{L;+0}^{(1)}$  %
  &  $-{\cal W}_{L;x}$ & $-$ 
\\
\hline
 ${\cal M}_{00}^L$  & ${\cal W}_{L;00}$  & $-$ & ${\cal W}_{L;00}$ & $-$  & ${\cal W}_{L;00}$ & $-$ 
 \\
 \hline
 ${\cal M}_x^S$  & $-$   & ${\cal W}_{S;x}^{(1)}$ & ${\cal W}_{S;+0}^{(0)}$ & ${\cal W}_{S;+0}^{(1)}$ & $-{\cal W}_{S;+0}^{(0)}$ & $-{\cal W}_{S;+0}^{(1)}$
\\
\hline
 ${\cal M}^S_{+0}$   & ${\cal W}_{S;+0}^{(0)}$ & ${\cal W}_{S;+0}^{(1)}$  & $-{\cal W}_{S;+0}^{(0)}$ & ${\cal W}_{S;+0}^{(1)}$  %
  &  $-$ & $-{\cal W}_{S;x}^{(1)}$ 
\\
\hline
\end{tabular}
\caption{Expressions, for each $P\pi\to\pi\pi$ scattering amplitude, of the functions
occurring in the representation of Eq.~(\ref{U_for_P_to_3pi}) in terms of a set 
of thirteen independent functions. The amplitudes involving the $\eta$ meson
have the same structure as those involving the $K_L$ meson, and follow upon 
replacing the label $L$ by $\eta$.}\label{W_for_P-pi_Table}
\end{center}
\end{table}

It remains to transform these dispersive representations into a tool
that will lead to an explicit construction of the two-loop amplitudes.
For this, it is necessary to specify the input for the absorptive parts of the partial
waves that appear in these expressions. This input will be provided by unitarity.
Up to and including two-loop order, these absorptive parts result only from
intermediate states composed of two pseudo-scalar mesons, $\pi\pi$, $K\pi$, $\eta\pi$, etc. 
Except for the $\pi\pi$ case, the singularities induced
by these intermediate states are far from the central region of the
Dalitz plot that describes the $P\to \pi\pi\pi$ decay processes. 
For a description of the latter, and of the corresponding unitarity
cusps due to the $\pi\pi$ intermediate states, it is, therefore, not
necessary to explicitly retain the intermediate states corresponding to these
higher thresholds. Their contributions can be expanded in powers
of the Mandelstam variables divided by the square of a scale, which is at least equal
to the kaon (or eta) mass, so that they will appear only in the polynomial contributions
to the amplitudes. Of course, this approximation would not be suitable if
we intended to describe the scattering processes $P\,\pi\to\pi\,\pi$ themselves.
If necessary, the formalism to be described below could actually be extended
to the full set of possible two-meson states.
Therefore, we have [recall that only the values $\ell = 0,1$ need to be considered] 
\begin{align}
\Abs \, t_\ell^{i\to f}(s) &= \sum_k \frac{1}{S_k}
\frac{\lambda^{1/2}_k(s)}{s}\, t_\ell^{i\to k}(s)\left[ f_\ell^{f\to k}(s)\right]^* 
\nonumber\\
&\qquad\qquad
{}\times\theta(s-s^{\mathrm{thr}}_k)
.
\label{Im_t}
\end{align}  
The sum goes over all the possible two-pions intermediate states $k$, $f_\ell^{f\to k}(s)$
denotes the partial-wave projection of the corresponding $\pi\pi$ scattering
amplitude $f\to k$, while $S_k$ is the symmetry factor, $S_k=2$ for identically charged pions, 
and $S_k=1$ otherwise. Note that for $\ell=1$, one always has $S_k=1$.
Furthermore, $s^{\mathrm{thr}}_k$ denotes the threshold
at which the channel $k$ opens, and $\lambda_k(s)$ is the triangle function
that describes the corresponding phase space. There are only three possibilities, 
$s^{\mathrm{thr}}_k = 4M_\pi^2,(M_\pi+M_{\pi^0})^2,4M_{\pi^0}^2$,
depending on the channel under consideration [$M_\pi$ ($M_{\pi^0}$) denotes
the mass of the charged (neutral) pion].

Thus, as was to be expected, in order to complete our program, we need to consider
at the same time the amplitudes ${A}(s,t,u)$ for the scattering processes
$\pi(p_1)\,\pi(p_2) \to \pi(p_3)\,\pi(p_4)$ in 
the various channels, but only at lowest
and at next-to-leading orders. It can be achieved following the same
path as for the $P \pi\to\pi\,\pi$ amplitudes.
The construction of the amplitudes ${A}(s,t,u)$
in the presence of isospin breaking has already been
described in Ref.~\cite{DescotesGenon:2012gv},
so we can remain brief, and refer the reader to
this reference for details. The starting point is again provided
by thrice subtracted, fixed-$t$, dispersion relations satisfied by the amplitudes 
${A}(s,t,u)$,
\begin{equation}\label{disprelA}
\begin{split}
&{A}(s,t,u)=a(t) + (s-u)b(t) + (s-u)^2 c(t)\\
&\qquad{}+\frac{s^3}{\pi}\int_{s_\mathrm{thr}}^\infty \frac{dx}{x^3} 
\frac{\im{A}(x,t,\Sigma-x-t)}{x-s}\\
&\qquad{}+\frac{u^3}{\pi}\int_{u_\mathrm{thr}}^\infty \frac{dx}{x^3} 
\frac{\epsilon\im{{A}_u}(x,t,\Sigma -x-t)}{x-u}\,,
\end{split}
\end{equation}
where $a(t)$, $b(t)$, and $c(t)$ are arbitrary subtraction functions,
$\Sigma$ stands for the sum of the squared masses of the pions appearing in the process,
and $s_\mathrm{thr}$ and $u_\mathrm{thr}$ denote the thresholds in the 
corresponding channels. 
Although, in order not to overburden the notation at this stage,
we have used the same symbols, these last quantities can, of course, be different from
the ones appearing in Eq.~(\ref{disprel}). Table \ref{pipiTable} shows the notation that will
be used in the sequel when specific channels are considered.

We now decompose the various amplitudes ${A}(s,t,u)$ such as to single out
the lowest, $S$ and $P$, partial waves,
\begin{equation}
{A}(s,t,u) = 16\pi\left[
f_0(s) + 3 f_1(s)\cos\theta \right] + 
{\cal A}_{\ell\ge 2}(s,t,u) ,
\label{pipiPWexp}
\end{equation}
where the scattering angle is given by a formula analogous to (\ref{theta_vs_t}),
\begin{equation}
\cos\theta = \frac{s(t-u) + \Delta_{12}\Delta_{34}}{\lambda_{12}^{1/2}(s)\lambda_{34}^{1/2}(s)}
\label{scatt_angle}
\end{equation}
with $\Delta_{ij}\equiv M_i^2 - M_j^2$, $M_i$ and $M_j$ being pion masses,
so that now the only possibilities are $\Delta_{ij}=0,\pm \Delta_\pi$, $\Delta_\pi \equiv M_\pi^2 - M_{\pi^0}^2$. 
Likewise, $\lambda_{ij}(s)$ is the K\"allen function involving the pion masses
$M_i$ and $M_j$. The dominant chiral behaviour of these various pieces
is given as in Eq.~(\ref{chibeh}), with now $f_{0,1}(s)$ and ${A}_{\ell\geq 2}$
replacing $t_{0,1}(s)$ and ${\cal M}_{\ell\geq 2}$, respectively,
so that again, only the $S$ and $P$ waves contribute
to the absorptive parts of the dispersion relations up to order ${\cal O}(E^6)$. After imposing 
crossing, the amplitudes ${A}(s,t,u)$ eventually take the form
\begin{equation}
{A}(s,t,u)={P}(s,t,u)+{U}(s,t,u)+{\cal O}(E^8),
\end{equation}
where ${U}(s,t,u)$ is the non-analytic unitarity part,
\begin{equation} 
\begin{split}
{U}(s,t,u) &= 16\pi\big[{W}_0(s)+3(t-u){W}_1(s)
\\
&{}+
{W}_0^t(t) +3(u-s){W}_1^t(t)
\\
&{}+
{W}_0^u(u)+3(t-s){W}_1^u(u)
\big]
,
\end{split}
\label{U_for_pi-pi}
\end{equation}
and ${P}(s,t,u)$ is a polynomial of at most third order in the Mandelstam variables 
with the same $s$, $t$, $u$ symmetries as the amplitude ${A}(s,t,u)$. 
Because of crossing relations among subsets of the $\pi\pi$
amplitudes, it is possible to express all the functions that
appear in the representations of the type (\ref{U_for_pi-pi})
in terms of only seven distinct functions, which we denote as $W_{00} (s)$, $W_x (s)$, 
$W_{++} (s)$, $W_{+-}^{(0)} (s)$, $W_{+-}^{(1)} (s)$, $W_{+0}^{(0)} (s)$, $W_{+0}^{(1)} (s)$.
How these functions contribute to the various amplitudes
can be gathered from Table \ref{W_for_pi-pi_Table}.%
\begin{table}[t]
\begin{center}
\begin{tabular}{|l||l|l|c|l|}
\hline
\multicolumn{1}{|c||}{process} & \multicolumn{1}{c|}{${A}$} & \multicolumn{1}{c|}{${{A}_u}$} & $\epsilon$  & \multicolumn{1}{c|}{$s+t+u$}
\\
\hline
$\pi^\pm \pi^\mp \to \pi^\pm\pi^\mp$ & $\Apm$ & $\App$ & $+1$ & $4 M_\pi^2 $
\\
\hline
$\pi^\pm \pi^\pm \to \pi^\pm\pi^\pm$ & $\App$ & $\Apm$ & $+1$ & $4 M_\pi^2 $
\\
\hline\hline
$\pi^\pm \pi^0 \to \pi^\pm \pi^0$ & $A_{+0}$ & $A_{+0}$ & $+1$ & $2 \Sigma_\pi $
\\
\hline
$\pi^\pm \pi^0 \to \pi^0 \pi^\pm$ & $\Apo$ & $\Ax$ & $-1$ & $2 \Sigma_\pi $
\\
\hline
$\pi^\pm \pi^\mp \to \pi^0 \pi^0$ & $\Ax$ & $\Apo$ & $-1$ & $2 \Sigma_\pi $
\\
\hline\hline
$\pi^0 \pi^0 \to \pi^0 \pi^0$ & $\Aoo$ & $\Aoo$ & $+1$ & $4 M_{\pi^0}^2 $
\\
\hline
\end{tabular}
\caption{For each $\pi\pi$ scattering process, the table shows the amplitude ${A}(s,t,u)$,
and the corresponding amplitude ${{A}_u(s,t,u)}$ in the crossed $u$-channel,
which appears in the dispersion relation (\ref{disprelA}). The penultimate column
gives the appropriate crossing phase $\epsilon$, and the last column displays,
for each process, the quantity $s+t+u$. Here, $M_\pi$ denotes
the charged pion mass, and $\Sigma_\pi \equiv M_\pi^2 + M_{\pi^0}^2$.  }\label{pipiTable}
\end{center}
\end{table}
\begin{table}[t]
\begin{center}
\begin{tabular}{|c||c|c|c|c|c|c|}
\hline
 ${A}$ & $W_0$ & $W_1$ & $W_0^t$ & $W_1^t$  & $W_0^u$  &  $W_1^u$
\\
\hline
 $\Apm$  &  $W_{+-}^{(0)}$ & $W_{+-}^{(1)}$ & $W_{+-}^{(0)}$ & $- W_{+-}^{(1)}$ & $W_{++}$  &  $-$
\\
\hline
 $\App$  &  $W_{++}$ & $-$ & $W_{+-}^{(0)}$ & $W_{+-}^{(1)}$  & $W_{+-}^{(0)}$ & $W_{+-}^{(1)}$ 
\\
\hline
 $A_{+0}$  & $W_{+0}^{(0)}$  & $W_{+0}^{(1)}$ & $W_x$ & $-$ & $W_{+0}^{(0)}$ & $W_{+0}^{(1)}$
\\
\hline
 $\Apo$  & $W_{+0}^{(0)}$  & $- W_{+0}^{(1)}$ & $W_{+0}^{(0)}$ & $W_{+0}^{(1)}$ & $W_x$ & $-$
\\
\hline
 $\Ax$   & $- W_x$     & $-$    & $- W_{+0}^{(0)}$ & $W_{+0}^{(1)}$  &  $- W_{+0}^{(0)}$ & $W_{+0}^{(1)}$ 
\\
\hline
 $\Aoo$  & $W_{00}$  & $-$ & $W_{00}$ & $-$  & $W_{00}$ & $-$
\\
\hline
\end{tabular}
\caption{Expressions, for each $\pi\pi$ scattering amplitude, of the functions
occurring in the representation of Eq.~(\ref{U_for_pi-pi}) in terms of the set 
of seven distinct functions.}\label{W_for_pi-pi_Table}
\end{center}
\end{table}
The functions $W_\ell(s)$, $\ell=0,1$, are analytic in the
complex $s$-plane, except for a right-hand cut, with discontinuities
given by the appropriate partial waves, e.g.%
\footnote{Although real analyticity
is preserved for the $\pi\pi$ scattering amplitudes, even in the presence of isospin 
breaking, in order to uniformize the notation, we also call the real and imaginary parts
dispersive and absorptive parts, respectively.}
\begin{align}
\Abs {W}_0(s) &=\left[ \Abs  f_0(s) + 3 \,
\frac{\Delta_{12}\Delta_{34}}{\lambda_{12}^{1/2}(s)\lambda_{34}^{1/2}(s)}\,
\Abs  f_1(s) \right] 
\nonumber\\
& \qquad\qquad {}\times\theta (s - s_\mathrm{thr}),
\label{Im_W0_W1} 
\\
\Abs {W}_1(s) &=  
\frac{s}{\lambda_{12}^{1/2}(s)\lambda_{34}^{1/2}(s)} \, \Abs f_1(s)
 \theta (s - s_\mathrm{thr})
.
\nonumber    
\end{align}
These discontinuities are again provided by unitarity,
\begin{align}
\Abs f_\ell^{i\to f}(s) &=\sum_k \frac{1}{S_k}
\frac{\lambda^{1/2}_k(s)}{s} f_\ell^{i\to k}(s) \left[f_\ell^{f\to k}(s)\right]^\star 
\nonumber\\
&\qquad\qquad
{}\times\theta(s-s^{\mathrm{thr}}_k)
.
\label{Im_f}
\end{align}
For reasons already explained above, only the contributions of two-pion intermediate states
need to be retained.

\indent

\begin{figure*}[thb]
\begin{center}
\setlength{\unitlength}{1mm}%
\begin{picture}(149,69)(0,-54)
\thicklines\large
\put(0,-13){\framebox(20,11){\parbox{20mm}{\small\centering ${\cal O}(E^2)$ $P\pi\to\pi\pi$\\ amplitude}}}
\put(20.5,-8.5){\vector( 2,-1){8}}
%
\put(0,-28){\framebox(20,11){\parbox{20mm}{\small\centering ${\cal O}(E^2)$ $\pi\pi\to\pi\pi$\\amplitude}}}
\put(20.5,-22.5){\vector( 2,1){8}}
\put(20.5,-25.5){\vector( 2,-1){8}}
\put(28.5,-12.5){\line(0,-1){6}}
\put(28.5,-15.5){\vector(1,0){8}}
\put(36.5,-20){\framebox(20,9){\parbox{20mm}{\small\centering Abs $t_{0,1}(s)$\\ at ${\cal O}(E^4)$}}}
\put(56.5,-14.5){\vector(1,0){8}}
\put(56.5,-16.5){\vector(1,0){8}}
\put(64.5,-20){\framebox(20,11){\parbox{20mm}{\small\centering ${\cal O}(E^4)$ $P\pi\to\pi\pi$\\ amplitude}}}
\put(64.5,-4){\makebox(20,15){\parbox{20mm}{\small\centering ${\cal O}(E^4)$ polynomial\\ in $s,t,u$}}}
\put(74.5,4.1){\oval(21,15)}
\put(74.5,-3.2){\vector( 0,-1){6}}
\put(85,-15.5){\vector( 2,-1){8}}
\put(28.5,-29.45){\vector(1,0){8}}
\put(36.5,-34){\framebox(20,9){\parbox{20mm}{\small\centering Abs $f_{0,1}(s)$ \\ at ${\cal O}(E^4)$}}}
\put(56.5,-28.5){\vector(1,0){8}}
\put(56.5,-30.5){\vector(1,0){8}}
\put(64.5,-34){\framebox(20,11){\parbox{20mm}{\small\centering ${\cal O}(E^4)$ $\pi\pi\to\pi\pi$\\amplitude}}}
\put(64.5,-51){\makebox(20,8){\parbox{20mm}{\small\centering ${\cal O}(E^4)$ polynomial\\ in $s,t,u$}}}
\put(74.5,-47.5){\oval(21,15)}
\put(74.5,-40){\vector( 0,1){6}}
\put(85,-29.5){\vector( 2,1){8}}
\put(93,-19.5){\line(0,-1){6}}
\put(93,-22.5){\vector(1,0){8}}
\put(101,-27){\framebox(20,9){\parbox{20mm}{\small\centering Abs $t_{0,1}(s)$\\ at ${\cal O}(E^6)$}}}
\put(121,-21.5){\vector(1,0){8}}
\put(121,-23.5){\vector(1,0){8}}
\put(129,-28.5){\framebox(20,11.5){\parbox{20mm}{\small\centering ${\cal O}(E^6)$ $P\pi\to\pi\pi$\\amplitude}}}
\put(129,-11.5){\makebox(20,15){\parbox{24mm}{\small\centering ${\cal O}(E^6)$ polynomial\\ in $s,t,u$}}}
\put(139,-3){\oval(21,15)}
\put(139,-11){\vector( 0,-1){6}}
\end{picture}
\end{center}
  \caption{Schematic representation of the iterative two-steps reconstruction procedure
for the $P\pi\to\pi\pi$ amplitudes. The absorptive part (of partial waves) denoted by $\Abs$ is defined in Eq.~(\ref{Def:disp_abs}).}\label{scheme}
\end{figure*}
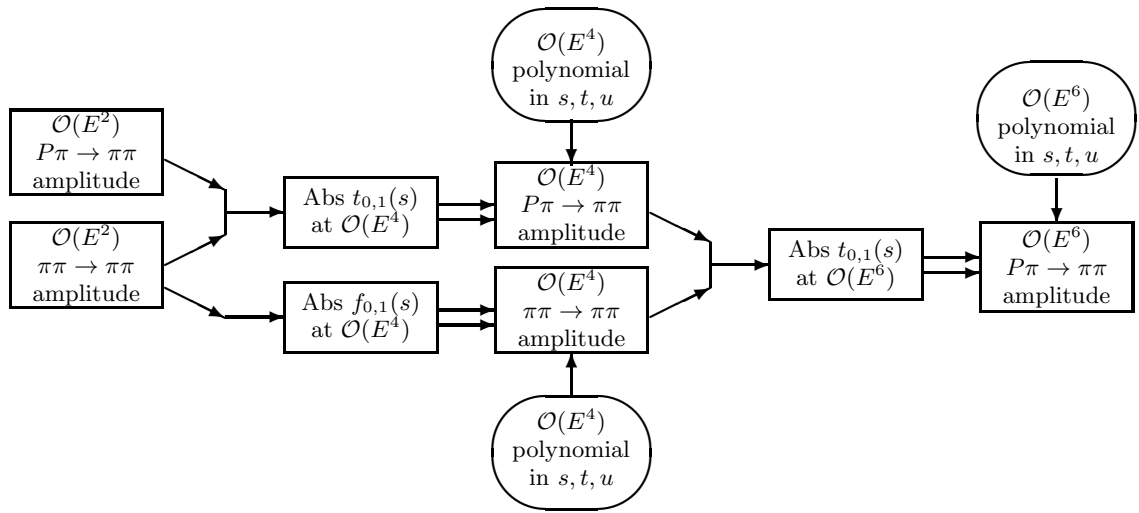

Within the framework we have just presented, the amplitudes for the
processes $P\pi\to\pi\pi$ and $\pi\pi\to\pi\pi$ form a closed set.
This framework also allows for a
two-step iterative construction of the corresponding amplitudes to two loops,
along the lines described in detail in Ref.~\cite{Knecht:1995tr}.
The essential information is provided by the chiral counting of the partial waves, Eq.~(\ref{chibeh})
for the amplitudes ${\cal M} (s,t,u)$, and similar relations 
\cite{Stern:1993rg,Knecht:1995tr,DescotesGenon:2012gv} for the $\pi\pi$ scattering amplitudes.
Indeed, at lowest order, these amplitudes, being ${\cal O}(E^2)$, are real
first-order polynomials in the Mandelstam variables, entirely saturated
by the $S$ and $P$ partial waves, 
\begin{alignat}{2}
f_{\ell}(s)&=\Disp f_{\ell}(s)+ {\cal O}(E^4)& &= \varphi_{\ell}(s) + {\cal O}(E^4),
\nonumber\\
t_{\ell}(s)&=\Disp t_{\ell}(s)+ {\cal O}(E^4)& &= {\widetilde\varphi}_{\ell}(s) + {\cal O}(E^4),
\end{alignat}
for $\ell = 0,1$. Subsequently, they provide the discontinuities of
the $\ell = 0,1$ partial waves along the unitarity cut
at order ${\cal O}(E^4)$,
\begin{equation}
\begin{split}
\Abs f_\ell^{i\to f}(s) &= \sum_k \frac{1}{S_k}
\frac{\lambda^{1/2}_k(s)}{s} \varphi_\ell^{i\to k}(s) \varphi_\ell^{k\to f}(s)
\\
&\qquad\qquad\ {}\times\theta(s-s^{\mathrm{thr}}_k)
 + {\cal O}(E^6)
 ,
\\
\Abs t_\ell^{i\to f}(s) &= \sum_k \frac{1}{S_k}
\frac{\lambda^{1/2}_k(s)}{s} {\widetilde\varphi}_\ell^{i\to k}(s) {\varphi}_\ell^{k\to f}(s)
\\
&\qquad\qquad\ 
{}\times\theta(s-s^{\mathrm{thr}}_k)
 + {\cal O}(E^6)
,
\end{split}
\label{Im_t_Im_f_p4}
\end{equation}
from which the one-loop amplitudes can
be constructed, up to an ambiguity which reduces to a polynomial
of at most second order in the Mandelstam variables. From these one-loop
expressions, one may now compute the order ${\cal O}(E^4)$ dispersive
parts  of the lowest partial waves, 
\begin{equation}
\begin{split}
\Disp f_{\ell}(s) &= \varphi_{\ell}(s) + \psi_{\ell}(s) + {\cal O}(E^6)
,\\
\Disp t_{\ell}(s) &= {\widetilde\varphi}_{\ell}(s) + {\widetilde\psi}_{\ell}(s) + {\cal O}(E^6)
,
\end{split}
\label{Disp_f_and_Disp_t}
\end{equation}
for $s \ge s_{\mathrm{thr}}$, which in turn will provide 
the corresponding absorptive parts at order ${\cal O}(E^6)$, 
\begin{align}
\Abs f_\ell^{i\to f}(s) &= \sum_k \frac{1}{S_k}
\frac{\lambda^{1/2}_k(s)}{s} \varphi_\ell^{i\to k}(s)
\Big[ \varphi_\ell^{k\to f}(s) 
\nonumber\\
&
\qquad{}+ 2 \psi_\ell^{k\to f}(s) 
\Big]
\times\theta(s-s^{\mathrm{thr}}_k)
 + {\cal O}(E^8)
 ,
\nonumber\\
\Abs t_\ell^{i\to f}(s) &= \sum_k \frac{1}{S_k}
\frac{\lambda^{1/2}_k(s)}{s} \bigg\{ {\widetilde\varphi}_\ell^{i\to k}(s)
\Big[
{\varphi}_\ell^{k\to f}(s) 
\nonumber\\
&
\qquad{}+ \psi_\ell^{k\to f}(s) \Big]
+ {\widetilde\psi}_\ell^{i\to k}(s) \varphi_\ell^{k\to f}(s) 
 \bigg\}
\nonumber\\
&
\qquad {}\times\theta(s-s^{\mathrm{thr}}_k) + {\cal O} (E^8)
.
\label{Im_t_Im_f_p6}
\end{align}
\noindent
These equations follow from Eq.~(\ref{Im_t}), from the chiral counting
of the partial waves, combined with {\it T} invariance and the fact that
real analyticity holds for the $\pi\pi$ scattering amplitudes,
so that the quantities $\varphi_{\ell}(s)$ and $\psi_{\ell}(s)$ 
in Eq.~(\ref{Disp_f_and_Disp_t}) are real. Note that the contributions of the absorptive 
parts of the amplitudes on the right-hand sides of these equations cancel, and so there 
appear effectively just the dispersive parts of the partial waves.
From (\ref{Im_t_Im_f_p6}), one can then obtain the full two-loop 
amplitudes. This construction is unique up to a polynomial contribution
of at most third order in the Mandelstam variables, with coefficients that
remain finite in the limit of vanishing pion masses. The main point of this
discussion is that in order to obtain the full two-loop expressions of the
amplitudes, only the dispersive parts of the one-loop $S$ and $P$ partial waves 
need to be computed directly.
Extracting these partial-wave projections from the corresponding one-loop
amplitudes, however, constitutes also the most
demanding difficulty in this reconstruction procedure, the main steps of which 
are summarized in Fig.~\ref{scheme}.
\begin{figure}[b]
\begin{center}
  \includegraphics[width=0.40\textwidth]{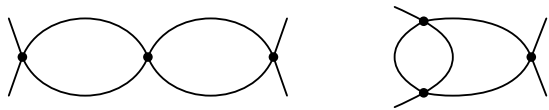}\\
  \caption{The two-loop topologies that contribute to the
  absorptive parts.}\label{topol}
\end{center}
\end{figure}

Let us close this general overview by briefly specifying the link with
the usual framework of the three-flavor chiral perturbation theory \cite{GL1}.
The latter considers an expansion in the Mandelstam variables $s$, $t$, $u$
and in the light-quark masses, all counting as order ${\cal O}(E^2)$. In the
approach presented here, the expansion is in the Mandelstam variables
and in the meson masses, which also count as order ${\cal O}(E^2)$.
In the chiral perturbation framework, amplitudes are worked out upon
computing all the relevant Feynman diagrams generated, at a given order, by the
effective Lagrangian, including tree-level and tadpole diagrams.
In the dispersive approach followed here, the absorptive parts
of the Feynman diagram are accounted for by the unitarity part ${\cal U}(s,t,u)$
[or $ U(s,t,u)$],
whereas their dispersive parts contribute both to ${\cal U}(s,t,u)$ and ${\cal P}(s,t,u)$
[or $U(s,t,u)$ and $P(s,t,u)$].
At two loops, these diagrams correspond to the two topologies illustrated
in Fig.~\ref{topol}, and describe direct rescattering (diagram on the left) and rescattering
in the crossed channels (diagram on the right).
Contributions from tree or tadpole diagrams are accounted for solely by the subtraction 
polynomial ${\cal P}(s,t,u)$ [or $ P(s,t,u)$].

\section{First iteration}\label{Sect:first_iteration}
\setcounter{equation}{0}

In this section, we construct the one-loop expressions
of the amplitudes for $\pi\pi$ scattering and for the $P\pi\to\pi\pi$ decay processes,
thus fulfilling the first step in our program. As a starting point, we need the
lowest-order expressions of the corresponding amplitudes, which are, according to chiral counting, 
of order ${\cal O}(E^2)$. It means that they are first-order polynomials in the Mandelstam variables, 
with coefficients that remain finite in the chiral limit. In the case of the $\pi\pi$ amplitudes, 
for instance, this leads to 
\begin{equation}
A(s,t,u)= 16\pi \left[ {\rm a} + {\rm b} \, \frac{s-\mu_+}{F_\pi^2} + {\rm c} \, \frac{t-u-\mu_-}{F_\pi^2} \right] 
+ {\cal O} (E^4)
,
\label{a_b_c}
\end{equation}
where a, b and c are constants, $F_\pi$ denotes the decay
constant of the charged pion, with $F_\pi =f_\pi/\sqrt{2}= 92.28(10)\,\textrm{MeV}$ from \cite{Tanabashi2019}, 
and $\mu_\pm$ specify some reference point in the Mandelstam plane. 
In the chiral limit, a and $\mu_\pm$ vanish, while b and c
remain non zero and finite, so that the corresponding amplitudes
satisfy the current-algebra consistency conditions \cite{Adler:1964um,Adler:1965ga}. 
Besides, from a practical point of view, 
the subtraction points $\mu_\pm$ should lie in the region of the Mandelstam plane where it
makes sense to apply the chiral expansion, but are otherwise arbitrary. 
Their choice will depend on the applications one has in mind, 
and will also determine the interpretation of the parameters a, b, and c. 

For instance, in Ref.~\cite{Knecht:1995tr}, the reference point
was chosen to be the center of the Dalitz plot [$\mu_+=4M_\pi^2/3$, $\mu_-=0$
in the isospin limit], which led to the interpretation of these
parameters as subthreshold coefficients. It might be an appropriate choice
if one wants to discuss quantities for which the expansion in light quark masses
converges rapidly. Table \ref{alpha_beta_Table} shows the corresponding
parameters for the $\pi\pi$ scattering amplitudes in the various channels. 
For later use, we have defined values for $\mu_+$ ($\mu_- = 0$ in
all cases)
even in the case where ${\rm b}$ vanishes.
Notice also that in the cases where there are two
identical pions in the initial and/or final state, Bose symmetry implies
that the coefficient ${\rm c}$ vanishes.
All these parameters are not independent since all the amplitudes 
involving a given total number of, say, charged pions
are related by crossing. In terms of the parameters of Table \ref{alpha_beta_Table},
these relations read
\begin{align}
\alpha_{+-} &= \alpha_{++} + {\cal O} (E^2),
&
\alpha_{+0} &= \alpha_{x} + {\cal O} (E^2),
\nonumber\\
\beta_{+-} &= \beta_{++} + {\cal O} (E^2),
&
\beta_{+0} &= \beta_{x} + {\cal O} (E^2),
\\
\gamma_{+-} &= \beta_{++} + {\cal O} (E^2),
&
\gamma_{+0} &= \beta_{x} + {\cal O} (E^2)
.
\nonumber
\end{align}
It is, however, more convenient to treat these quantities
as independent in a first stage, and to make the replacement,
including the corrections from higher orders indicated above,
only at the end of the calculation. It will also
help to visualize the origin of various contributions in the higher-order expressions of the amplitudes.

\begin{table*}[t]
\begin{center}
\begin{tabular}{|c||c|c|c|c||c|c|c|}
\hline
 $A$ & $16\pi$a$F_\pi^2/M_{\pi^0}^2$ & $16\pi$b & $16\pi$c & $\mu_+$  & $\lambda_s$ & $\lambda_t$ & $\lambda_u$ 
\\
\hline
 $\Apm$  &  $2\alpha_{+-} /3$ & $\beta_{+-}/2$ & $\gamma_{+-}/2$ & $4 M_\pi^2 /3$ %
 & $ \lambda^{(1)}_{+-} + \lambda^{(2)}_{+-}$ & $ \lambda^{(1)}_{+-} + \lambda^{(2)}_{+-}$ & $ 2 \lambda^{(2)}_{+-}$
\\
\hline
 $\App$  &  $2\alpha_{++} /3$ & $-\beta_{++}$ & $0$ & $4 M_\pi^2/3 $ %
 & $ 2 \lambda^{(2)}_{+-}$ & $ \lambda^{(1)}_{+-} + \lambda^{(2)}_{+-}$ & $ \lambda^{(1)}_{+-} + \lambda^{(2)}_{+-}$
\\
\hline
 $A_{+0}$  & $\alpha_{+0} /3$  & $-\beta_{+0}/2$ & $+\gamma_{+0}/2$ & $2 \Sigma_\pi/3$ %
 & $+ \lambda^{(2)}_x$ & $+ \lambda^{(1)}_x$ & $+ \lambda^{(2)}_x$ 
\\
\hline
 $\Apo$  & $\alpha_{+0} /3$  & $-\beta_{+0}/2$ & $-\gamma_{+0}/2$ & $2 \Sigma_\pi/3$ %
 & $+ \lambda^{(2)}_x$ & $+ \lambda^{(2)}_x$ & $+ \lambda^{(1)}_x$ 
\\
\hline
 $\Ax$   & $-\alpha_x /3$     & $- \beta_x$    &  $0$ & $ 2 \Sigma_\pi /3 $ %
 & $- \lambda^{(1)}_x$ & $- \lambda^{(2)}_x$ &$- \lambda^{(2)}_x$ 
\\
\hline
 $\Aoo$  & $\alpha_{00} $  & $0$ & $0$ & $4 M_{\pi^0}^2/3$ %
 & $3 \lambda_{00}^{(1)}$ & $3 \lambda_{00}^{(1)}$ & $3 \lambda_{00}^{(1)}$ 
\\
\hline
\end{tabular}
\caption{Definition of the parameters a, b and c of Eq.~(\ref{a_b_c})
for each $\pi\pi$ scattering process, in the case where one opts for a 
description in terms of subthreshold parameters. In these expressions,
we have defined
$\Sigma_\pi \equiv M_\pi^2 + M_{\pi^0}^2$, and $\mu_- = 0$
in all cases. 
Also shown are the expressions, in terms of the independent
quantities $\lambda_{+-}^{(1)}$, $\lambda_{+-}^{(2)}$, 
$\lambda_x^{(1)}$, $\lambda_x^{(2)}$, and $\lambda_{00}^{(1)}$,
of the constants $\lambda_{s,t,u}$ appearing in the subtraction polynomials
$P(s,t,u)$ at one-loop order defined in Eq.~(\ref{P_one-loop}).}\label{alpha_beta_Table}
\end{center}
\end{table*}

\begin{table}[b]
\begin{center}
\begin{tabular}{|c||c|c|c|c|c|}
\hline
 $A$ & a & b & c & $\mu_+$  & $\mu_-$
\\
\hline
 $\Apm$  &  ${a}_{+-}$ & ${\rm b}_{+-}$ & ${\rm c}_{+-}$ & $4 M_\pi^2 $ & $ 0 $
\\
\hline
 $\App$  &  ${a}_{++}$ & ${\rm b}_{++}$ & $0$ & $4 M_\pi^2 $  & $0$
\\
\hline
 $A_{+0}$  & ${a}_{+0}$  & ${\rm b}_{+0}$ & $+{\rm c}_{+0}$ & $(M_\pi + M_{\pi^0})^2$ & $-(M_\pi - M_{\pi^0})^2$
\\
\hline
 $\Apo$  & ${a}_{+0}$  & ${\rm b}_{+0}$ & $-{\rm c}_{+0}$ & $(M_\pi + M_{\pi^0})^2$ & $+(M_\pi - M_{\pi^0})^2$
\\
\hline
 $\Ax$   & ${a}_x$     & ${\rm b}_x$    & $ 0 $ & $4 M_\pi^2$  &  $0$
\\
\hline
 $\Aoo$  & ${a}_{00}$  & $0$ & $0$ & $4 M_{\pi^0}^2$ & $0$
\\
\hline
\end{tabular}
\caption{Definition of the parameters a, b and c of Eq.~(\ref{a_b_c})
for each $\pi\pi$ scattering process, in the case where one opts for a 
description in terms of scattering lengths. }\label{a_b_Table}
\end{center}
\end{table}

On the other hand, subthreshold parameters are not observables
that are particularly suited for the discussion of experimental data on $\pi\pi$ scattering.
Furthermore, since the aim of the present study is to provide a parameterization 
of the $P\to\pi^0\pi^0\pi$ amplitudes which displays the dependence of the cusp  
on the $\pi\pi$ scattering lengths, the latter appears,
in this context, as a better choice of parameters.
In this case, $\mu_\pm$ are to be chosen such as to correspond to the threshold values of $s$, $t$, and $u$.
In order that the parameters a in (\ref{a_b_c}) retain their meaning as scattering lengths up to the two-loop level,
the subtraction polynomials have to be adjusted appropriately. How this can be done has been
described in Appendix F of Ref.~\cite{DescotesGenon:2012gv}. The various parameters
that enter the $\pi\pi$ amplitudes at the lowest order are shown in Table \ref{a_b_Table}.
Again, there are crossing relations among subsets of them. These relations, with next-to-leading
corrections included, are also displayed in Appendix F of Ref.~\cite{DescotesGenon:2012gv}.
Notice that the normalization
of the scattering lengths differs by a factor of two from the one
usually adopted. The definition used here is in agreement with the normalization
of the partial-wave projections in Eq.~(\ref{pipiPWexp}). However, when
expressing these scattering lengths in terms of the two $S$-wave scattering
lengths $a_0^0$ and $a_0^2$ in the isospin limit, we shall use, for the latter,
the familiar normalization. According to Eqs. (2.18), (2.21), and (2.22) of Ref.~\cite{KnechtUrech98},
at order ${\cal O}(E^2)$, these relations read\footnote{Our quantities $a_0^0$ and $a_0^2$ 
correspond to those denoted as $(a_0^0)_{\mbox{\scriptsize{str}}}$ 
and $(a_0^2)_{\mbox{\scriptsize{str}}}$, respectively, in Ref.~\cite{KnechtUrech98}.}
\begin{align}
{a}_{+-}&=\frac{2}{3}\,a^0_0+\frac{1}{3}\,a^2_0-2 a^2_0\,\frac{\Delta_\pi}{M_{\pi}^2}\,,\notag\\
{a}_{++} &= 2 a^2_0-2 a^2_0\,\frac{\Delta_\pi}{M_{\pi}^2}\,,\notag\\
a_{+0}&=a^2_0-a^2_0\,\frac{\Delta_\pi}{M_{\pi}^2}\,,
\\
a_x &= - \frac{2}{3}\,a^0_0+\frac{2}{3}\,a^2_0+ a^2_0\,\frac{\Delta_\pi}{M_{\pi}^2}\,,\notag\\
a_{00}&=\frac{2}{3}\,a^0_0+\frac{4}{3}\,a^2_0-\frac{2}{3}\left(a^0_0 + 2 a^2_0\right)\frac{\Delta_\pi}{M_{\pi}^2}\, ,\notag
\end{align}
where $\Delta_\pi$ was defined after Eq.~(\ref{scatt_angle}).

In the case of the $P\to\pi\pi\pi$ amplitudes, we shall adopt a parameterization more
akin to the usual description of their Dalitz-plot structure in terms of, say, slopes and
curvatures with respect to appropriately chosen kinematic variables. Details will be provided in Section~\ref{one loop P-pi}.

\subsection{$\pi\pi$ scattering}

Since the construction of the $\pi\pi$ scattering amplitudes with $M_\pi \neq M_{\pi^0}$
has already been discussed at some length in Refs.~\cite{DescotesGenon:2012gv,Bernard:2013faa},
our account will remain brief. 
First, we need to notice that at one loop, 
the discontinuities of these functions are given by Eqs.~(\ref{Im_W0_W1})
and (\ref{Im_t_Im_f_p4}). It is a straightforward matter to extract the corresponding
order ${\cal O}(E^2)$ $S$ and $P$ partial waves $\varphi_\ell(s)$, $\ell=0,1$, from
the general representation (\ref{a_b_c}) of the lowest-order
$\pi\pi$ scattering amplitudes:
\begin{align}
\varphi_0 (s) & = 
{\rm a} + {\rm b} \, \frac{s - \mu_+ }{F_\pi^2} - \frac{\rm c}{F_\pi^2} \left( \mu_- + \frac{ \Delta_{12} \Delta_{34} }{s} \right)
,
\nonumber\\
\varphi_1 (s) & = 
\frac{\rm c}{3 F_\pi^2} \, \frac{\lambda_{12}^{1/2}(s) \lambda_{34}^{1/2}(s)}{s}\,
.
\end{align}
\noindent
In view of the form of $\varphi_1(s)$, it is useful to
remember, as evidenced by Eqs.~(\ref{Im_W0_W1}) and (\ref{Im_t_Im_f_p4}), 
that the discontinuities only involve the combination [in hopefully
self-explanatory notation]
\begin{equation}
\frac{\varphi_1^{i\to k}(s) \varphi_1^{k\to f} (s)}{\lambda_i^{1/2} (s) \lambda_f^{1/2} (s)}
=
\frac{{\rm c}^{i\to k} {\rm c}^{f\to k}}{9 F_\pi^4} \, \frac{\lambda_k (s)}{s}
\,.
\label{varphi1_remark}
\end{equation}
One then obtains the following expressions:
\begin{align}
W_{00}(s)&=
16\pi\bigg\{
\frac{1}{2}
\left[\varphi^{00}_0(s) \right]^2\!{\bar J}_0 (s)
+
\left[\varphi^{x}_0(s) \right]^2\!{\bar J} (s)
\bigg\}
\nonumber\\
&\qquad
{} + {\cal O}(E^6)
,\\
\begin{split}
W_{x}(s) &= -16\pi \varphi^{x}_0(s)\left[
 \frac{1}{2}
\varphi^{00}_0(s)  \,{\bar J}_0 (s)
 +
\varphi^{+-}_0(s)\,{\bar J} (s)
\right]
\\
&\qquad
{}+{\cal O}(E^6)
,
\end{split}\\
{W}_{+0}^{(0)}(s) &=
16\pi\bigg\{
\frac{{\rm c}_{+0}^2}{3 F_\pi^4} \Delta_\pi^2
\left[
1 - 2 \frac{\Sigma_\pi}{s} - 3\frac{\Delta_\pi^2}{s^2}
\right]
\nonumber\\
&
+\left[ \varphi^{+0}_0(s) \right] ^2\bigg\}{\bar J}_{\pm 0} (s)
+
16\pi\,\frac{4 {\rm c}_{+0}^2}{3 F_\pi^4}\frac{\Delta_\pi^4}{s^2}\,{\bar{{\bar J}}}_{\pm 0} (s)
\nonumber\\
&\qquad
{}+{\cal O}(E^6)
,\\
{W}_{+0}^{(1)}(s) &=
16\pi \frac{{\rm c}_{+0}^2}{9 F_\pi^4} \frac{\lambda_{\pm 0}(s)}{s}\,{\bar J}_{\pm 0} (s)
+{\cal O}(E^6)
,\\
W_{++}(s) &=
16\pi \,\frac{1}{2}\, 
\left[ \varphi^{++}_0(s) \right] ^2 {\bar J} (s)+ {\cal O}(E^6) 
,\\
\begin{split}
W_{+-}^{(0)}(s) &= 16\pi \bigg\{ \! 
\left[  \varphi^{+-}_0(s) \right] ^2
{\bar J}(s) 
+
\frac{1}{2}\,
\left[ \varphi^{x}_0(s) \right] ^2 {\bar J}_0 (s)
\bigg\}
\\
&
{}+ {\cal O}(E^6)
,
\end{split}\raisetag{\baselineskip}
\\
W_{+-}^{(1)}(s) &=  16\pi \, \frac{{\rm c}_{+-}^2}{9 F_\pi^4}
\,\left( s  - {4}M_{\pi}^2 \right){\bar J} (s)\,+\, {\cal O}(E^6)
.
\end{align}
Here, ${\bar J}_0(s)$, ${\bar J} (s)$, and ${\bar J}_{\pm 0} (s)$ denote the dispersive integrals\footnote{
These functions are dispersive representations of the 
two-point one-loop integrals subtracted at $s=0$,
\begin{align}
J(p^2;m_1 , m_2) &= \frac{1}{i} \int \frac{d^4 k}{(2\pi)^4}
\frac{1}{k^2-m_1^2} \frac{1}{(k-p)^2-m_2^2}\,
\nonumber
,\\
{\bar J}(s;m_1 , m_2) &= {J}(s;m_1 , m_2)-{J}(0;m_1 , m_2)
\nonumber
\end{align}
with ${\bar J} (s) = {\bar J}(s; M_\pi , M_\pi)$,
${\bar J}_0 (s) = {\bar J}(s; M_{\pi^0} , M_{\pi^0})$,
${\bar J}_{\pm 0} (s) = {\bar J}(s; M_\pi , M_{\pi^0})$.
Their closed forms are given, e.g., in Ref.~\cite{Gasser:1984pr}. We also use
$$
\bar{J}(0)=0, \quad \bar{J}_0(4M_{\pi^0}^2)=\frac{1}{8\pi^2}=\bar{J}_{\pm 0}\left (\mu_+\right )+\bar{J}_{\pm 0}\left (\mu_-\right ).
$$
}
\begin{align}
{\bar J}_0 (s) &= \frac{s}{16\pi^2}\int_{4M_{\pi^0}^2}^{\infty}\frac{dx}{x}\,\frac{1}{x-s-{\rm i}0}\,\sigma_0 (x)
,
\nonumber\\
{\bar J} (s) &= \frac{s}{16\pi^2}\int_{4M_{\pi}^2}^{\infty}\frac{dx}{x}\,\frac{1}{x-s-{\rm i}0}\,\sigma (x)
,
\label{Jbar_functions}
\\
{\bar J}_{\pm 0} (s) &=
 \frac{s}{16\pi^2}\int_{(M_{\pi^\pm}  + M_{\pi^0})^2}^{\infty}
 \frac{dx}{x}\,\frac{1}{x-s-{\rm i}0}\,\frac{\lambda_{\pm 0}^{1/2} (x)}{x}\,
,
\nonumber
\end{align}
with
\begin{equation}
\sigma_{0}(s)=\sqrt{1 - \frac{4M_{\pi^0}^2}{s}},\quad \sigma(s)=\sqrt{1 - \frac{4M_{\pi}^2}{s}}
,
\label{sigma_def}
\end{equation}
and $\lambda_{\pm 0}(s) = \lambda(s,M_\pi^2,M_{\pi^0}^2)$.
Finally, 
\begin{equation}
{\bar{{\bar J}}}(s)\equiv {\bar J}(s) - s {\bar J}^\prime(0)
,
\end{equation}
and
\begin{equation}
\Sigma_\pi \equiv M_\pi^2 + M_{\pi^0}^2
.
\label{Sigma_pi}
\end{equation}

The subtraction polynomials have the following form at the one-loop order
[cf.\ also footnote \ref{footnote1}]:
\begin{equation}
\begin{split}
P (s , t , u) &=
16\pi \left[ {\rm a} + {\rm b} \, \frac{s-\mu_+}{F_\pi^2} + {\rm c} \, \frac{t-u-\mu_-}{F_\pi^2} \right] 
- w 
\\
&
+ \frac{\lambda_s}{F_\pi^4}  \left( s - \frac{3}{2} \mu_+ \right)^2
+ \frac{\lambda_t}{F_\pi^4} \left( t - \frac{3}{2} \mu_+ \right)^2
\\
&
+ \frac{\lambda_u}{F_\pi^4}  \left( u - \frac{3}{2} \mu_+ \right)^2
+ {\cal O} (E^6)
.
\end{split}\raisetag{\baselineskip}
\label{P_one-loop}
\end{equation}
Crossing and Bose symmetry (when applicable) restrict to five the
number of constants $\lambda_{s,t,u}$ that are independent. Their
expressions in terms of the independent parameters 
introduced in Ref.~\cite{DescotesGenon:2012gv}
are shown in Table \ref{alpha_beta_Table}. These expressions
also apply to the case where one opts for the parameterisation
in terms of scattering lengths, and the constraints
coming from the crossing properties have already been
accounted for.
The contribution denoted by $w$ is not a new parameter, 
but is expressed in terms of the ones already present.
In the case one chooses to retain the parameterisation
in terms of the subthreshold parameters, $w$ is
fixed so as to reproduce the expressions for the polynomials
given in Ref.~\cite{DescotesGenon:2012gv}, which were
themselves constructed so that they reproduce the polynomial part of the
$\pi\pi$ amplitude in the isospin limit as given in Ref.~\cite{Knecht:1995tr}. 
Explicitly, this means 
\begin{equation}
w_{+0} = - w_x = \frac{\Delta_\pi^2}{F_\pi^4} \times \lambda_x^{(1)}
+ {\cal O} (E^6)
,
\end{equation}
and $w=0$ in the remaining cases.
For the parameterisation in terms of scattering lengths,
the value of $w$ is simply
\begin{align}
w &= \frac{\lambda_s}{4} \frac{\mu_+^2}{F_\pi^4}
+
\frac{\lambda_t}{F_\pi^4}  \left( t_{\rm thr} - \frac{3}{2} \mu_+ \right)^2
+ 
\frac{\lambda_u}{F_\pi^4}  \left( u_{\rm thr} - \frac{3}{2} \mu_+ \right)^2
\nonumber\\
&+
\re U (\mu_+ , t_{\rm thr}  , u_{\rm thr} )
+ {\cal O} (E^6)
,
\end{align}
with $t_{\rm thr} - u_{\rm thr} = \mu_-$ and the values for $\mu_+$ and $\mu_-$ as given in Table \ref{a_b_Table}.
This choice of $w$, therefore, ensures that the parameter ${\rm a}$ in Eq.~(\ref{a_b_c})
indeed corresponds to the scattering length, i.e.,
\begin{equation}
\re A (\mu_+ , t_{\rm thr}  , u_{\rm thr}) \equiv 16 \pi {\rm a} + {\cal O} (E^6)
.
\end{equation}

At next-to-leading order, the constraints arising from the 
crossing property read
\begin{equation} 
{\rm c}_{+-}= {\rm b}_{+-}, \qquad {\rm b}_{++}+2 {\rm b}_{+-}=0 
,
\end{equation}
\begin{multline}
{\rm a}_{+-} - {\rm a}_{++} + \frac{4M_{\pi^\pm}^2}{F_\pi^2}\,{\rm b}_{++} = 
\frac{2}{\pi} ({\rm a}_{+-})^2
 - \frac{({\rm a}_{++})^2}{\pi}\qquad\\ \qquad+\frac{2}{\pi}\frac{\lambda_{+-}^{(2)}-\lambda_{+-}^{(1)}}{F_\pi^4}\,M_\pi^4 
+16\pi\,\frac{({\rm a}_x)^2}{2}\, \re{\bar J}_0(4M_{\pi}^2) 
,
\end{multline}
\begin{equation} 
{\rm c}_{+0}= -{\rm b}_{+0}
,
\end{equation}
\begin{equation} 
{\rm b}_{x}-2 {\rm b}_{+0}=\frac{3}{16\pi} \frac{\lambda_{x}^{(2)}-\lambda_{x}^{(1)}}{F_\pi^2}(M_\pi-M_{\pi^0})(3M_\pi+M_{\pi^0})
,
\end{equation}
and
\begin{multline}
{\rm a}_{x} + {\rm a}_{+ 0} - 4\,\frac{M_{\pi^\pm}^2}{F_\pi^2}\,{\rm b}_{x} = \frac{2}{\pi}\, {\rm a}_x {\rm a}_{+-}+\frac{2}{\pi}\, {\rm a}_{+0}^2\\
\begin{aligned}
&+16\pi\,\frac{{\rm a}_x {\rm a}_{00}}{2}\, \re {\bar J}_0(4M_\pi^2)
-16\pi\,\frac{4}{3}\frac{{\rm b}_x^2}{F_\pi^4}\,M_\pi^2\Delta_\pi^2 {\bar J}'(0)
\\
&-32\pi\left({\rm a}_{x}^2-4M_\pi^2{\rm a}_{x}\,\frac{{\rm b}_{x}}{F_\pi^2} +\frac{8}{3}\frac{{\rm b}_{x}^2}{F_\pi^4}\,M_\pi^4\right){\bar J}_{\pm 0}(-\Delta_\pi)\\
&-\frac{M_\pi^2}{\pi} \frac{\lambda_{x}^{(2)}(3M_\pi^2-M_{\pi^0}^2)-2\lambda_{x}^{(1)}M_{\pi^0}^2}{F_\pi^4}\,.
\end{aligned}
\raisetag{\baselineskip}
\end{multline}

\subsection{$P\pi\rightarrow\pi\pi$ scattering}\label{one loop P-pi}

The implementation of the first iteration for the amplitudes of the processes 
$P\pi\rightarrow\pi\pi$ follows rather closely the previous case of $\pi\pi$
scattering, so that the main use of this section is to establish the notation in a more precise way. 
The starting point is now given by the order ${\cal O}(E^2)$ expressions for the amplitudes
in Table \ref{table_calM} that are (anti-)symmetric under exchange of $t$ and $u$,
which we now parameterize in the following form,
\begin{align}
\Mppp (s,t) &= A_{++} + B_{++}\, \frac{s-s_0}{F_\pi^2}\,,
\nonumber\\
\Mpoo (s,t) &= A_x + B_{x}\, \frac{s-s_0}{F_\pi^2}\,,
\nonumber\\
\MLmp (s,t) &= A^L_x + B^L_x\, \frac{s-s_0}{F_\pi^2}\,,
\label{LO_P3piAmp1}\\
\MLoo (s,t) &= A^L_{00},
\nonumber\\
\MSmp (s,t) &= B^S_x \,\frac{t-u}{F_\pi^2}\, .\nonumber
\end{align}
The crossing property then furnishes the remaining amplitudes 
\begin{align}
\Mppm (s,t) &= A_{++} - \frac{B_{++}}{2 F_\pi^2}\left[ (s-s_0) + (t-u)\right],
\nonumber\\
\Mpop (s,t) &= - A_{x} + \frac{B_{x}}{2 F_\pi^2}\left[ (s-s_0) + (t-u)\right],
\nonumber\\
\MLpo (s,t) &= - A_{x}^L + \frac{B_{x}^L}{2 F_\pi^2}\left[ (s-s_0) + (t-u)\right],
\nonumber\\
\MSpo (s,t) &= \frac{B_{x}^S}{2 F_\pi^2}\left[ 3(s-s_0) - (t-u)\right]
.
\label{LO_P3piAmp2}
\end{align}
The value of $s_0$ depends on the process under consideration, see (\ref{3s0}).
We have not written the amplitudes describing the two processes involving
the $\eta$ meson. As far as their structure is concerned, they can be obtained
in what follows from the amplitudes involving the $K_L$ meson, upon replacing
the mass of the latter by the mass of the former, and upon changing the notation
for the coefficients appearing in the polynomial part [e.g., $A^\eta_x$ instead
of $A^L_x$, and so on]. The computation of the ${\cal O}(E^2)$ $S$ and $P$
partial waves ${\widetilde\varphi}_\ell(s)$, $\ell = 0,1$, from the amplitudes 
(\ref{LO_P3piAmp1}) and (\ref{LO_P3piAmp2}), presents no particular difficulty, 
and we merely display the resulting expressions,\footnote{The remark made in
Eq.~(\ref{varphi1_remark}) above also applies to the product 
${\widetilde\varphi}_1^{i \to k}(s) \varphi_1^{k\to f} (s)$.\\
In addition, when referring to the Kacser function for a definite process, we denote the pions by their charges, 
e.g., $K_{K_L \pm;\pm 0}(s)$ or $K_{K\mp;00}(s)$, and so on. For a generic case, 
we write just $K(s)$.}
\begin{align}
{\widetilde{\varphi}}_{0}^{++}(s) &= \frac{1}{16\pi}
\left[A_{++} + \frac{B_{++}}{F_\pi^2}
\left(s - \frac{M_K^2}{3}-M_\pi^2\right)\right]
,
\nonumber\\
{\widetilde{\varphi}}_{0}^{+-}(s) &= \frac{1}{32\pi}
\left[ 2 A_{++} - \frac{B_{++}}{F_\pi^2}
\left(s - \frac{M_K^2}{3}-M_\pi^2\right) \right]
,
\nonumber\\
{\widetilde{\varphi}}_{1}^{+-}(s) &= 
- \frac{B_{++}}{48\pi} \, \frac{K_{K\mp;\pm\mp}(s)}{F_\pi^2}
\,,
\end{align}
\begin{align}
{\widetilde{\varphi}}_{0}^{x}(s) &= \frac{1}{16\pi}\left[A_{x}  +  
\frac{B_{x}}{3F_\pi^2} (3s - M_K^2 - M_\pi^2 - 2 M_{\pi^0}^2) \right]
,
\nonumber\\
{\widetilde{\varphi}}_{0}^{0+}(s) &= -\frac{A_{x}}{16\pi}
-\frac{B_{x}}{96\pi F_\pi^2} \bigg[M_K^2 + M_\pi^2 + 2 M_{\pi^0}^2  
\nonumber\\
&
\qquad 
-\,3 s - \frac{3}{s}(M_{K}^2 - M_{\pi^0}^2)\Delta_\pi \bigg]
,
\nonumber\\
{\widetilde{\varphi}}_{1}^{0+}(s) &= 
\frac{B_{x}}{48\pi}\,\frac{K_{K 0; 0\pm} (s)}{F_\pi^2}\,
,
\end{align}
\begin{align}
{\widetilde{\varphi}}_{0}^{L;x}(s) &=  \frac{1}{16\pi}
\left[  A_{x}^L + \frac{B_{x}^L}{3F_\pi^2} (3s - M_{K_L}^2 - 2 M_\pi^2 - M_{\pi^0}^2)  \right]\!
,
\nonumber\\
\begin{split}
{\widetilde{\varphi}}_{0}^{L;+0}(s) &=  -\frac{A_{x}^L}{16\pi}
-\frac{B_{x}^L}{96\pi F_\pi^2}  \bigg[M_{K_L}^2 + 2 M_\pi^2 + M_{\pi^0}^2 
\\
&\qquad 
- 3 s + \frac{3}{s}(M_{K_L}^2 - M_{\pi}^2)\Delta_\pi\bigg]
,
\end{split}
\nonumber\\
{\widetilde{\varphi}}_{1}^{L;+0}(s) &=  
\frac{B_{x}^L}{48\pi}\,\frac{K_{K_L \pm; \pm 0} (s)}{F_\pi^2}
\,,
\end{align}
\begin{align}
{\widetilde{\varphi}}_{0}^{L;00}(s) &= \frac{1}{16\pi}\,A_{00}^L 
,  
\end{align}
\begin{align}
{\widetilde{\varphi}}_{1}^{S;x}(s)  &= 
\frac{B_{x}^S}{24\pi}\,\frac{K_{K_S 0; + -} (s)}{F_\pi^2}\,
,
\nonumber\\
{\widetilde{\varphi}}_{0}^{S;+0}(s) &=\frac{B_x^S}{32\pi F_\pi^2}\bigg[ 3s - M_{K_S}^2 - 2 M_\pi^2 - M_{\pi^0}^2 
\nonumber\\
&
\qquad 
+\frac{1}{s}(M_{K_S}^2 - M_{\pi}^2)\Delta_\pi\bigg]
,
\\
{\widetilde{\varphi}}_{1}^{S;+0}(s) &= 
- \frac{B_{x}^S}{48\pi}\,\frac{K_{K_S \pm; \pm 0} (s)}{F_\pi^2}\,.
\nonumber
\end{align}

In the case $P=K^\pm$, there are two independent channels to
consider, one involving charged pions only and the other involving
two neutral pions. The structure of the corresponding two-loop 
amplitudes, as inferred from the reconstruction theorem, is given by 
\begin{align}
&\Mppp(s,t,u) =  {\cal P}_{++}(s,t,u)  \,
+16\pi {\cal W}_{++}(s)\hspace{6em}
\nonumber\\
&\  
+16\pi \left[{\cal W}^{(0)}_{+-}(t) + 3 (u-s){\cal W}^{(1)}_{+-}(t)\right]
\nonumber\\
&\  
+16\pi \left[{\cal W}^{(0)}_{+-}(u) + 3 (t-s){\cal W}^{(1)}_{+-}(u)\right]
+ {\cal O}(E^8) 
,
\end{align}
and
\begin{align}
&\Mpoo(s,t,u) =  {\cal P}_{x}(s,t,u)  \,
+16\pi {\cal W}_{x}(s)\hspace{6em}
\nonumber\\
&\  
+16\pi \left[{\cal W}^{(0)}_{0+}(t) + 3 (u-s){\cal W}^{(1)}_{0+}(t)\right]
\nonumber\\
&\  
+16\pi \left[{\cal W}^{(0)}_{0+}(u) + 3 (t-s){\cal W}^{(1)}_{0+}(u)\right]
+ {\cal O}(E^8) 
.
\end{align}
The various functions that appear in these expressions
have discontinuities that are given by unitarity. At the one-loop order,
and when restricted to two-pions intermediate states, these read
[cf. Eqs.~(\ref{discW}) and (\ref{Im_t_Im_f_p4})]
\begin{align}
\begin{split}
\Abs {\cal W}_{++}(s) &=
\frac{1}{2}\,\sigma (s)\, 
{\widetilde\varphi}^{++}_0(s) \varphi^{++}_0(s)
\theta (s-4M_{\pi}^2)\\
&
+{\cal O}(E^6),
\end{split}
\nonumber\\
\Abs{\cal W}_{+-}^{(0)}(s) &=
\bigg\{ \sigma (s)\,
{\widetilde\varphi}^{+-}_0(s)\varphi^{+-}_0(s) 
\,\theta (s-4M_{\pi}^2)
\nonumber\\   
&\quad+
\frac{1}{2}\,\sigma_0 (s)\,
{\widetilde\varphi}^{x}_0(s) \varphi^{x}_0(s) 
\,\theta (s-4M_{\pi^0}^2)
\bigg\}
\nonumber\\
&+ {\cal O}(E^6),
\nonumber\\
\Abs{\cal W}_{+-}^{(1)}(s) &= \sigma (s)
\frac{{\widetilde\varphi}^{+-}_1(s)\varphi^{+-}_1(s) }{2 K_{K\mp;\pm\mp}(s)}
\,\theta (s-4M_{\pi}^2)
\nonumber\\
&+  {\cal O}(E^6),
\nonumber\\
\Abs {\cal W}_{x}(s) &=
\bigg\{ \frac{1}{2}\,\sigma_0 (s)\, 
{\widetilde\varphi}^{x}_0(s) \varphi^{00}_0(s)  
\,\theta (s-4M_{\pi^0}^2)
\nonumber\\
&\quad
+
\sigma (s)\,
{\widetilde\varphi}^{+-}_0(s) \varphi^{x}_0(s) 
\,\theta (s-4M_{\pi}^2)
\bigg\}
\nonumber\\
&+ {\cal O}(E^6),
\nonumber\\
\Abs {\cal W}_{0+}^{(0)}(s) &=
\frac{\lambda_{\pm 0}^{1/2}(s)}{s}
\bigg\{
\varphi^{+0}_0(s) {\widetilde{\varphi}}_{0}^{0+}(s)
\nonumber\\
&\quad-
3\,\frac{\Delta_\pi  (M_K^2 - M_{\pi^0}^2)}{s} \frac{\varphi^{+0}_1(s) 
{\widetilde{\varphi}}_{1}^{0+}(s)}{2 K_{K 0; 0\pm} (s)}\bigg\}
\nonumber\\
&\qquad
\times
\theta \left(s-(M_{\pi} + M_{\pi^0})^2\right) + {\cal O}(E^6),
\nonumber\\
\Abs {\cal W}_{0+}^{(1)}(s) &=
\frac{\lambda_{\pm 0}^{1/2}(s)}{s} \,
\frac{\varphi^{+0}_1(s)\, {\widetilde{\varphi}}_{1}^{0+}(s)}{2 K_{K 0; 0\pm} (s)}
\nonumber\\
&\times
\theta \left(s-(M_{\pi} + M_{\pi^0})^2\right) 
+ {\cal O}(E^6)
.
\end{align}
Up to a polynomial ambiguity, this fixes then these functions to read
\begin{align}
{\cal W}_{++}(s) &= 
16\pi \,\frac{1}{2}\, 
{\widetilde\varphi}^{++}_0(s) \varphi^{++}_0(s)  {\bar J} (s)
+ {\cal O}(E^6)
,
\nonumber\\
{\cal W}_{+-}^{(0)}(s) &= 16\pi \bigg\{ 
{\widetilde\varphi}^{+-}_0(s)\varphi^{+-}_0(s)  
{\bar J}(s) 
\nonumber\\
&\qquad
+
\frac{1}{2}\,
{\widetilde\varphi}^{x}_0(s) \varphi^{x}_0(s)   {\bar J}_0 (s)
\bigg\}
+ {\cal O}(E^6)
,\\
{\cal W}_{+-}^{(1)}(s) &=   - \frac{B_{++} {\rm c}_{+-}}{18 F_{\pi}^4}
\, (s  - {4}M_{\pi}^2) {\bar J} (s) 
+ {\cal O}(E^6) 
,\nonumber
\end{align}
and
\begin{align}
{\cal  W}_{x}(s) &= 16\pi\bigg[
\frac{1}{2} {\widetilde\varphi}^{x}_0(s) \varphi^{00}_0(s)  \,{\bar J}_0 (s)
\nonumber\\
&
\qquad +
{\widetilde\varphi}^{+-}_0(s) \varphi^{x}_0(s)\,{\bar J} (s)
\bigg]
+ {\cal O}(E^6)
 ,
\nonumber\\
{\cal  W}^{(0)}_{0+}(s) &= 
\bigg\{\!
- \frac{B_x {\rm c}_{+0}}{6F_\pi^4}  
\left[
1 - 2 \frac{M_\pi^2 + M_{\pi^0}^2}{s}- 3\,\frac{\Delta_\pi^2}{s^2}
\right]
\nonumber\\
&\hspace{-1em}
\times(M_K^2 - M_{\pi^0}^2)\Delta_\pi
+16\pi \varphi^{+0}_0(s) {\widetilde{\varphi}}_{0}^{0+}(s)\bigg\}{\bar J}_{\pm 0} (s)
\nonumber\\
&\hspace{-1em}
-\frac{2 B_x {\rm c}_{+0}}{3 F_\pi^4}\frac{(M_K^2 - M_{\pi^0}^2)\Delta_\pi^3}{s^2}
\,{\bar{{\bar J}}}_{\pm 0} (s)
+ {\cal O}(E^6)
,
\nonumber\\
{\cal W}_{0+}^{(1)}(s) &=
\frac{B_x {\rm c}_{+0}}{18 F_\pi^4} \frac{\lambda_{\pm 0}(s)}{s}\,{\bar J}_{\pm 0} (s)
+{\cal O}(E^6)
.
\end{align}

We may proceed similarly in the case of the reactions
with $P=K_L$ (or $\eta$). The amplitude for $K_L\pi^0\to\pi^\mp\pi^\pm$ reads
\begin{align}
&\MLmp(s,t,u) = {\cal P}_{L;x}(s,t,u) +16\pi {\cal W}_{L;x}(s)\hspace{7em} 
\nonumber\\
&\!
-16\pi\!\left[{\cal W}^{(0)}_{L;+0}(t) + 3 (u-s){\cal W}^{(1)}_{L;+0}(t)\right]
\nonumber\\
&\!-16\pi\!\left[{\cal W}^{(0)}_{L;+0}(u) + 3 (t-s){\cal W}^{(1)}_{L;+0}(u)\right]
+ {\cal O}(E^8) ,\raisetag{\baselineskip}
\label{WLx}
\end{align}
whereas for $K_L\pi^0\to\pi^0\pi^0$, we obtain
\begin{align}
\MLoo(s,t,u) &=  
16\pi\left[ {\cal W}_{L;00}(s)
+ {\cal W}_{L;00}(t)
+ {\cal W}_{L;00}(u) \right]
\nonumber\\
&
+\, {\cal P}_{L;00}(s,t,u)  
+{\cal O}(E^8).
\end{align}
At the one-loop order, the various functions that
appear in these expressions are given as
\begin{align}
{\cal  W}_{L;x}(s) &= 16\pi\bigg[
\frac{1}{2}\, {\widetilde\varphi}^{L;00}_0(s) \varphi^{x}_0(s)  \,{\bar J}_0 (s)
\nonumber\\
&
\qquad + {\widetilde\varphi}^{L;x}_0(s) \varphi^{+-}_0(s)\,{\bar J} (s)
\bigg]
+{\cal O}(E^6)
,
\nonumber\\
{\cal  W}^{(0)}_{L;+0}(s) &= 
\bigg\{
\frac{B_x^L {\rm c}_{+0}}{6F_\pi^4} 
\left[
1 - 2 \frac{M_\pi^2 + M_{\pi^0}^2}{s} - 3 \frac{\Delta_\pi^2}{s^2}
\right]
\nonumber\\
&\hspace{-2em}
\times(M_{K_L}^2 - M_{\pi}^2)\Delta_\pi
+16\pi \varphi^{+0}_0(s) {\widetilde{\varphi}}_{0}^{L;+0}(s)\bigg\}{\bar J}_{\pm 0} (s)
\nonumber\\
&\hspace{-2em}
+\frac{2B_x^L {\rm c}_{+0}}{3 F_\pi^4}\frac{(M_{K_L}^2 - M_{\pi}^2)\Delta_\pi^3}{s^2}
\,{\bar{{\bar J}}}_{\pm 0} (s)
\,+\,{\cal O}(E^6)
,
\nonumber\\
{\cal W}_{L;+0}^{(1)}(s) &=
\frac{B_x^L {\rm c}_{+0}}{18 F_\pi^4} \frac{\lambda_{\pm 0}(s)}{s}\,{\bar J}_{\pm 0} (s)
+{\cal O}(E^6)
,
\label{WLx_1loop}
\end{align}
and
\begin{equation}
\begin{split}
{\cal  W}_{L;00}(s) &= 16\pi\bigg[
\frac{1}{2} {\widetilde\varphi}^{L;00}_0(s) \varphi^{00}_0(s)  \,{\bar J}_0 (s)
\\
&
\qquad+
{\widetilde\varphi}^{L;x}_0(s) \varphi^{x}_0(s)\,{\bar J} (s)
\bigg]
+{\cal O}(E^6)
.
\label{calW_00}
\end{split}
\end{equation}

At the same one-loop accuracy, the polynomial contributions
to the four amplitudes that we have just discussed, namely
$\Mppp$, $\Mpoo$, $\MLmp$, $\MLoo$, can be written as
[cf. footnote \ref{footnote1}]
\begin{multline}
{\cal P} (s,t,u) = A  +  B\, \frac{s-s_0}{F_\pi^2} + 
C\, \frac{(s-s_0)^2}{F_\pi^4}\\
+ \frac{D}{F_\pi^4}  \left[(t-s_0)^2 + (u-s_0)^2 \right]
+ {\cal O}(E^6) 
,
 \label{calP}
\end{multline}
putting appropriate labels on the coefficients, e.g., $A_{++}$, $A_x$,
and so on. In the case of ${\cal P}^L_{00} (s,t,u)$, the additional
restrictions $B^L_{00}=0$ and $C^L_{00}=D^L_{00}$, due to Bose symmetry, apply.
These coefficients are in one-to-one correspondence with the Dalitz-plot
parameters of the $K\to\pi\pi\pi$ amplitudes.

\indent

Finally, it remains to discuss the case $P=K_S$ briefly. The corresponding two-loop amplitude
for the process $K_S\pi^0\to\pi^+\pi^-$ has the form
\begin{multline}
{\cal M}_x^S (s,t,u) = {\cal P}_{S,x} (s,t,u)
+ 48 \pi (t-u) {\cal W}^{(1)}_{S;x} (s)
\\
\begin{aligned}
&+ 16\pi \left[ {\cal W}_{S;+0}^{(0)} (t) + 3 (u-s) {\cal W}_{S;+0}^{(1)} (t) \right]
\\
&- 16\pi \left[ {\cal W}_{S;+0}^{(0)} (u) + 3 (t-s) {\cal W}_{S;+0}^{(1)} (u) \right]
+ {\cal O}(E^8).
\end{aligned}
\end{multline}
At the one-loop order, the functions involved in this expression read
\begin{equation}
{\cal W}^{(1)}_{S;x} (s) = \frac{B_x^S {\rm c}_{+0}}{9 F_\pi^4} (s - 4 M_\pi^2) {\bar J} (s)
+{\cal O}(E^6)
,
\end{equation}
and
\begin{align}
&{\cal W}_{S;+0}^{(0)} (s) =
\bigg\{-
\frac{B_x^S {\rm c}_{+0}}{6F_\pi^4}  
\left[
1 - 2 \frac{M_\pi^2 + M_{\pi^0}^2}{s} - 3 \frac{\Delta_\pi^2}{s^2}
\right]
\nonumber\\
&\qquad
\times(M_{K_S}^2 - M_{\pi}^2)\Delta_\pi
+16\pi \varphi^{+0}_0(s) {\widetilde{\varphi}}_{0}^{S;+0}(s)\bigg\}{\bar J}_{\pm 0} (s)
\nonumber\\
&\qquad
-\frac{2B_x^S {\rm c}_{+0}}{3 F_\pi^4}\frac{(M_{K_S}^2 - M_{\pi}^2)\Delta_\pi^3}{s^2}
\,{\bar{{\bar J}}}_{\pm 0} (s)
+{\cal O}(E^6)
,
\nonumber\\
&{\cal W}_{S;+0}^{(1)}(s) =-
\frac{B_x^S {\rm c}_{+0}}{18 F_\pi^4} \frac{\lambda_{\pm 0}(s)}{s}\,{\bar J}_{\pm 0} (s)
+{\cal O}(E^6)
.
\end{align}
The one-loop subtraction polynomial, in this case, is given by
\begin{equation}
{\cal P}^S_x (s,t,u) = \frac{t-u}{F_\pi^2} \bigg[ B^S_x + D^S_x\, \frac{s_0-s}{F_\pi^2} \bigg] +{\cal O}(E^6)
.
\label{calP^S}
\end{equation}
Again, $B^S_x$ and $D^S_x$ are related to Dalitz-plot parameters of the $K_S\to\pi^0\pi^+\pi^-$ amplitude.

\section{Second iteration: equal-mass pions}\label{Sec:Second iteration isospin}  
\setcounter{equation}{0}
The construction of the scattering amplitudes for the $\pi\pi\to\pi\pi$
and $P\pi\to\pi\pi$ processes could be performed in a rather straightforward manner
at next-to-leading order:
from the expressions of their discontinuities at one loop, the various functions
$W(s)$ and ${\cal W}(s)$ have been obtained up to an ambiguity consisting
in polynomials of at most second order in the Mandelstam variables. The resulting
amplitudes at one loop display the correct analytic properties expected at this
order, and satisfy all required crossing relations. In order to proceed toward
obtaining two-loop expressions for all the $P\pi\to\pi\pi$ amplitudes, it is necessary
to supplement the subtraction polynomials with ${\cal O}(E^6)$ terms. For 
$\Mppp$, $\Mpoo$, $\MLmp$, $\MLoo$, we take the generic form\footnote{Note that in Ref.~\cite{Kampf:2011wr} for the case of $P=\eta$, we have used the same form of the polynomial with a different normalization of the coefficients $A^\eta_x,\dots, F^\eta_x$ and $A^\eta_{00},\dots, E^\eta_{00}$.}
\begin{multline}
{\cal P} (s,t,u) = A  +  B\, \frac{s-s_0}{F_\pi^2} + 
C\, \frac{(s-s_0)^2}{F_\pi^4}\\
+ \frac{D}{F_\pi^4}  \left[(t-s_0)^2 + (u-s_0)^2 \right]
+ E\, \frac{(s-s_0)^3}{F_\pi^6}\\
+ \frac{F}{F_\pi^6}  \left[(t-s_0)^3 + (u-s_0)^3 \right]
+ {\cal O}(E^8)\label{Op6 polynomial}
\end{multline}
with appropriate labels. Due to the Bose symmetry, in the case of ${\cal P}^L_{00} (s,t,u)$, the additional constraint $F^L_{00}=E^L_{00}$ holds (in addition to the restrictions discussed after (\ref{calP})).
For $\MSmp$ the polynomial is
\begin{multline}
{\cal P}^S_x (s,t,u) = \frac{t-u}{F_\pi^2} \bigg[ B^S_x + D^S_x\, \frac{s_0-s}{F_\pi^2} + E^S_x\, \frac{(s_0-s)^2}{F_\pi^4}\\
+ \frac{F^S_x}{F_\pi^4}\, \left((s_0-t)^2+(s_0-u)^2\right)\bigg]
+{\cal O}(E^8).
\end{multline}

Then, we start from the discontinuities of the functions ${\cal W}(s)$ that hold at
this order, as given by Eqs.~(\ref{discW}), (\ref{Im_W0_W1}), and (\ref{Im_t_Im_f_p6}).
For instance, for the amplitude ${\cal  W}_{L;00}(s)$ it gives
\begin{equation}
\begin{split}
&\Abs {\cal  W}_{L;00}(s)  =  
\frac{\sigma_0(s)}{2} \bigg[ {\widetilde\varphi}^{L;00}_0(s) \varphi^{00}_0(s) 
+ {\widetilde\varphi}^{L;00}_0(s) \psi^{00}_0(s)
\\
&\qquad\quad
+ {\widetilde\psi}^{L;00}_0(s) \varphi^{00}_0(s) 
\bigg] \theta (s-4M_{\pi^0}^2)
\\
&\qquad +
\sigma (s) \bigg[ {\widetilde\varphi}^{L;x}_0(s) \varphi^{x}_0(s)
+ {\widetilde\varphi}^{L;x}_0(s) \psi^{x}_0(s)
\\
&\qquad\quad
+ {\widetilde\psi}^{L;x}_0(s) \varphi^{x}_0(s)
\bigg] \theta (s-4M_{\pi}^2)
+{\cal O}(E^8) .
\end{split}\raisetag{1\baselineskip}
\label{W00_2loop}
\end{equation}
As illustrated by this example, schematically these expressions now involve, in addition to the lowest-order
partial waves represented by the functions $\varphi(s)$ and
${\widetilde\varphi}(s)$, the one-loop dispersive part $\psi(s)$ (${\widetilde\psi}(s)$) of 
the $\pi\pi\to\pi\pi$ ($P\pi\to\pi\pi$) $S$ and $P$ partial-wave projections. 
These are to be obtained from the one-loop expressions of these 
amplitudes that have just been computed in the preceding section. 
As compared to the leading-order case, their expressions at the next-to-leading order
are much more complicated. In the case of the
$\pi\pi$ scattering amplitudes, the difficulty is purely algebraic, and
explicit formulas can be obtained in rather closed forms \cite{DescotesGenon:2012gv,Bernard:2013faa}. 
They essentially generalize the results obtained in the case without isospin breaking 
\cite{Knecht:1995tr} to the situation where the pion mass difference is taken into account.
In the case of the $P\pi\to\pi\pi$ amplitudes, the situation is somewhat 
more involved, due to the existence, in the central region of the
Mandelstam plane, of a bounded region corresponding to the decay process
$P\to\pi\pi\pi$. This feature makes the analytic properties of the
amplitude less simple, and requires a more elaborate analysis. In particular, when performing the
partial-wave projection by the usual integration over the variable $t$,
one has to be careful to find the appropriate prescription for deforming
the path of integration in order to avoid any singularity. The situation has been
well studied in the case that corresponds to the isospin limit: the masses
of the three particles in the final state (i.e., pions for the cases at hand), but also in the intermediate 
states, are identical. As explained before, the appropriate procedure consists in starting from the situation
where $M_P < 3 M_{\pi^0}$, so that the decay region disappears. In this case,
dispersion relations exist under the usual conditions, and one can proceed as outlined in Fig.~\ref{scheme}.
After that, one performs the analytic continuation in the mass, $M_P^2 \to M_P^2 + i\delta$,
to the region where $M_P > 3 M_{\pi^0}$ \cite{Bronzan,Kacser,Anisovich}. In practice,
this analytic continuation is provided by the prescription to deform
the contour of integration in the partial-wave projection. We show in Appendix \ref{app:anomalous}
that this prescription, which was shown to give the correct result in the limit
of equal-mass pions, also provides the appropriate prescription for the processes involving three neutral
external pions, even when there appear pairs of heavier charged pions as
intermediate states. The more
general situation, with also charged pions in the external states, and lighter
neutral pions as intermediate states, involves anomalous thresholds and requires 
a dedicated study. It will be the subject of a subsequent article \cite{plan2}.
Therefore, our analysis in this section is restricted to the case
of equal-mass pions, in both final and intermediate states. The common
pion mass will be taken as $M_\pi$, the mass of the charged pion.
The case $P\pi^0\to\pi^0\pi^0$, $P=K_L,\eta$, with $M_{\pi^0} \neq M_\pi$, will
be considered in the next section.

For equal-mass pions, the discontinuities of the functions ${\cal W}(s)$
at next-to-leading order take the schematic form
[in the case of the functions ${\cal W}^{(1)}(s)$, an additional factor
$1/2K(s)$ is understood]
\begin{equation}
\Abs {\cal W} (s) \sim \sigma(s) \bigg[
{\tilde\varphi}(s) \varphi(s) + {\tilde\varphi}(s) \psi(s)
+ {\tilde\psi}(s) \varphi(s) \bigg]
.
\end{equation}
This leads to the following convenient decomposition of the functions
${\cal W} (s)$ at two loops,
\begin{equation}
{\cal W} (s) = {\cal W}^{\rm 1 loop} (s) + {\cal W}^{\pi\pi} (s) + {\cal W}^{P\pi} (s)
,
\end{equation}
where each term results from the corresponding term in the preceding decomposition
of $\Abs {\cal W} (s)$. 
The preceding section was devoted to the evaluation of the NLO amplitudes ${\cal W}^{\rm 1 loop} (s)$, 
with $\Abs {\cal W}^{\rm 1 loop} (s) \sim \sigma(s) {\tilde\varphi}(s) \varphi(s)$.
The next two terms give the contribution at next-to-next-to-leading order (NNLO), 
and their computation\footnote{It is useful, 
	for what follows, to keep in mind that the functions ${\tilde\varphi}(s)$ and $\varphi(s)$ 
	all simply become first-order polynomials in $s$ when $M_{\pi^0} = M_\pi$.} 
will be addressed in turn in the remainder of the present section. The results for  ${\cal W}^{\pi\pi} (s)$ can be found in Section~\ref{subsec:second pipi} (cf.~(\ref{WpipiC})--(\ref{WpipiS})), whereas those for ${\cal W}^{P\pi} (s)$ are listed in Appendix~\ref{App:w_tilde}. 
Note that for $P=\eta$, we reproduce the expressions that were implicitly used (and the general structure of which was shown) already in Section~IV of Ref.~\cite{Kampf:2011wr}.

\subsection{NLO $\pi\pi$ partial waves and the functions ${\cal W}^{\pi\pi} (s)$}\label{subsec:second pipi}

In the case where the pions have all a common mass $M_\pi$,
the partial-wave projections at one loop $\psi_\ell(s)$, $\ell = 0,1$, for $\pi\pi$
scattering have already been worked out quite some time ago in Ref.~\cite{Knecht:1995tr}. 
We will use the expressions as given in Ref.~\cite{DescotesGenon:2012gv}, with which we 
also share the normalization. The partial-wave projections in the various channels are
expressed as combinations, weighted by the appropriate Clebsh-Gordan
coefficients, of the isospin projections $\psi_I(s)$, $I=0,1,2$, e.g.,
[with the Condon and Shortley phase convention],
\begin{equation}
\left(
\begin{tabular}{c}
$\psi_0^{++}$\\
$\psi_0^{+-}$\\
$\psi_0^x$   \\
$\psi_0^{+0}$\\
$\psi_0^{00}$
\end{tabular}
\right)
=
\left(
\begin{tabular}{cc}
$0$     &   $1$   \\
$\frac{1}{3}$  &  $\frac{1}{6}$   \\
$-\frac{1}{3}~~$  &  $\frac{1}{3}$  \\
$0$     &   $\frac{1}{2}$   \\
$\frac{1}{3}$  &  $\frac{2}{3}$
\end{tabular}
\right)
\left(
\begin{tabular}{c}
$\psi_0$\\
$\psi_2$
\end{tabular}
\right)
,
\label{psi_isospin}
\end{equation}
whereas
$\psi_1^{+-} = \psi_1^{+0} = \psi_1/2$.
The isospin projections $\psi_I(s)$ themselves are expressed in the form
\begin{equation}
\psi_{I}(s) =  2\,\frac{M_\pi^4}{F_\pi^4}\,
\sqrt{\frac{s}{s-4M_\pi^2}}\,
\sum_{i=0}^4 \xi^{(i)}_I(s)
k_i(s) ,
\label{lim_psi_I} 
\end{equation}
where the functions $k_i(s)$ read, for $s\ge 4 M_\pi^2$,
\begin{align}
k_0(s) &= \frac{1}{16\pi}\,\sqrt{\frac{s - 4 M_\pi^2}{s}}\,,
\qquad
k_1(s) = \frac{1}{8\pi}\,L(s),
\nonumber\\
k_2(s) &=  \frac{1}{8\pi}\left(1-\frac{4 M_\pi^2}{s}\right)L(s),
\nonumber\\
k_3(s) &=\frac{3}{16\pi}\,\frac{M_\pi^2}{\sqrt{s(s - 4 M_\pi^2)}}\,L^2(s),
\nonumber\\
k_4(s) &=\frac{1}{16\pi}\frac{M_\pi^2}{\sqrt{s(s - 4 M_\pi^2)}}
\left\{
1+\sqrt{\frac{s}{s - 4 M_\pi^2}}\,L(s)
\right.
\nonumber\\
&
\left.\qquad
+\frac{M_\pi^2}{s - 4 M_\pi^2}\,L^2(s)
\right\}
,
\label{k_n}
\end{align}
with\footnote{Note that for $s\ge 4 M_\pi^2$, one has
	$$
	{\bar J} (s) = \frac{1}{16\pi^2} \bigg[ 2 + \sqrt{1-\frac{4 M_\pi^2}{s}}\, ( L(s) + i \pi) \bigg].
	$$
}
\begin{equation}
L(s) = \ln \frac{1 - \sqrt{\frac{s - 4 M_\pi^2}{s}}}{1 + \sqrt{\frac{s - 4 M_\pi^2}{s}}}
\qquad [s\ge 4 M_\pi^2]
.
\label{L_funct_def}
\end{equation}
The polynomials $\xi^{(i)}_I(s)$ have been worked out in Ref.~\cite{Knecht:1995tr}.
We will use them in the form given in terms of subthreshold parameters in Eq.~(C.8)
of Ref.~\cite{DescotesGenon:2012gv}. Their expressions in terms of the scattering
lengths can be obtained from the formulas (F.8) to (F.11) given in Appendix F of that same reference,
upon taking the limit where the neutral and charged pion masses coincide.
Note that the function $k_4 (s)$ appears only in the $P$-wave projection $\psi_1(s)$, and
the polynomials $\xi^{(4)}_0 (s)$ and $\xi^{(4)}_2 (s)$ vanish identically. Due to the relation
	\begin{equation}
	\left( \frac{s}{M_\pi^2} - 4 \right) k_4(s)
	= k_0 (s) + 2 k_1 (s) + \frac{1}{3} k_3 (s)
	\end{equation}
	there is an ambiguity in the definition of the polynomials
	$\xi_1^{(i)} (s)$. This, however, does not matter in practice and $\psi_1 (s)$ is well defined.

Following Ref.~\cite{Knecht:1995tr}, the expressions of the various functions
$W_0(s)$ and $W_1(s)$ can be written in terms of integrals
\begin{equation}
{\bar K}_i (s) = \frac{s}{\pi} \int_{4 M_\pi^2}^\infty 
\frac{dx}{x} \frac{k_i(x)}{x-s-{\rm i}0}
.
\end{equation}
These integrals can be expressed in terms of the function ${\bar J}(s)$, 
and the corresponding formulas are given in Eq. (3.49) of Ref.~\cite{Knecht:1995tr}.

We may now display, for each function ${\cal W} (s)$ appearing in Table \ref{W_for_P-pi_Table},
its NNLO contribution ${\cal W}^{\pi\pi} (s)$. These expressions read
\begin{align}
{\cal W}_{++}^{\pi\pi}(s) &= 
{\widetilde\varphi}^{++}_0(s) \frac{M_\pi^4}{F_\pi^4} \sum_{i=0}^3 \xi^{(i)}_2 (s) {\bar K}_i (s)
,
\nonumber\\
{\cal W}_{+-}^{(0)\pi\pi}(s) &= 
\frac{1}{3} {\widetilde\varphi}^{+-}_0(s) \frac{M_\pi^4}{F_\pi^4} \sum_{i=0}^3 [2\xi_0^{(i)} (s) + \xi^{(i)}_2 (s)] {\bar K}_i (s)
\nonumber\\
&-
\frac{1}{3} {\widetilde\varphi}^{x}_0(s)  \frac{M_\pi^4}{F_\pi^4} \sum_{i=0}^3 [\xi_0^{(i)} (s) - \xi^{(i)}_2 (s)] {\bar K}_i (s)
,
\nonumber\\
{\cal W}_{+-}^{(1)\pi\pi}(s) &=   - \frac{B_{++}}{96 \pi F_{\pi}^2} 
\frac{M_\pi^4}{F_\pi^4} \sum_{i=0}^4 \xi_1^{(i)} (s) {\bar K}_i (s)
,
\end{align}
\begin{align}
\begin{split}
{\cal  W}_{x}^{\pi\pi}(s) &= 
\frac{1}{3} {\widetilde\varphi}^{x}_0(s)  
\frac{M_\pi^4}{F_\pi^4} \sum_{i=0}^3 [\xi_0^{(i)} (s) + 2 \xi^{(i)}_2 (s)] {\bar K}_i (s)
\\
&- \frac{2}{3}
{\widetilde\varphi}^{+-}_0(s)
 \frac{M_\pi^4}{F_\pi^4} \sum_{i=0}^3 [\xi_0^{(i)} (s) - \xi^{(i)}_2 (s)] {\bar K}_i (s)
 ,
\end{split}
\nonumber\\
{\cal W}_{0+}^{(0)\pi\pi}(s) &= 
{\widetilde\varphi}^{0+}_0(s)  \frac{M_\pi^4}{F_\pi^4} \sum_{i=0}^3 \xi^{(i)}_2 (s) {\bar K}_i (s)
 ,
\nonumber\\
{\cal W}_{0+}^{(1)\pi\pi}(s) &=
\frac{B_x}{96 \pi F_\pi^2} \frac{M_\pi^4}{F_\pi^4} \sum_{i=0}^4 \xi^{(i)}_1 (s) {\bar K}_i (s)
,\label{WpipiC}
\end{align}
\begin{align}
{\cal  W}_{L;x}^{\pi\pi}(s) &= 
\frac{1}{3} {\widetilde\varphi}^{L;x}_0(s)
\frac{M_\pi^4}{F_\pi^4} \sum_{i=0}^3 [2 \xi_0^{(i)} (s) + \xi^{(i)}_2 (s)] {\bar K}_i (s)
\nonumber\\
&- \frac{1}{3} {\widetilde\varphi}^{L;00}_0(s)
\frac{M_\pi^4}{F_\pi^4} \sum_{i=0}^3 [\xi_0^{(i)} (s) - \xi^{(i)}_2 (s)] {\bar K}_i (s)
,
\nonumber\\
{\cal  W}^{(0)\pi\pi}_{L;+0}(s) &= 
{\widetilde{\varphi}}_{0}^{L;+0}(s) 
\frac{M_\pi^4}{F_\pi^4} \sum_{i=0}^3 \xi^{(i)}_2 (s) {\bar K}_i (s)
,
\nonumber\\
{\cal W}_{L;+0}^{(1)\pi\pi}(s) &=
\frac{B_x^L}{96 \pi F_\pi^2}
\frac{M_\pi^4}{F_\pi^4} \sum_{i=0}^4 \xi^{(i)}_1 (s) {\bar K}_i (s)
,
\end{align}
\begin{align}
{\cal  W}^{\pi\pi}_{L;00}(s) &= 
\frac{1}{3} {\widetilde\varphi}^{L;00}_0 (s)  
\frac{M_\pi^4}{F_\pi^4} \sum_{i=0}^3 [\xi_0^{(i)} (s) + 2 \xi^{(i)}_2 (s)] {\bar K}_i (s)
\nonumber\\
&-\frac{2}{3} {\widetilde\varphi}^{L;x}_0(s)  
\frac{M_\pi^4}{F_\pi^4} \sum_{i=0}^3 [\xi_0^{(i)} (s) - \xi^{(i)}_2 (s)] {\bar K}_i (s)
,\raisetag{0.5em}
\end{align}
and finally,
\begin{align}
{\cal W}_{S;x}^{(1)\pi\pi} (s) &=
\frac{B_x^S}{48 \pi F_\pi^2} 
\frac{M_\pi^4}{F_\pi^4} \sum_{i=0}^4 \xi^{(i)}_1 (s) {\bar K}_i (s)
,
\nonumber\\
{\cal  W}^{(0)\pi\pi}_{S;+0}(s) &=
{\widetilde\varphi}_0^{S;+0} (s)\, 
\frac{M_\pi^4}{F_\pi^4} \sum_{i=0}^3 \xi^{(i)}_2 (s) {\bar K}_i (s)
,
\nonumber\\
{\cal  W}^{(1)\pi\pi}_{S;+0}(s) &=
- \frac{B_x^S}{96 \pi F_\pi^2}
\frac{M_\pi^4}{F_\pi^4} \sum_{i=0}^4 \xi^{(i)}_1 (s) {\bar K}_i (s)
.\label{WpipiS}
\end{align}

\subsection{NLO partial waves of the $P\pi\to\pi\pi$ amplitudes and the functions ${\cal W}^{P\pi} (s)$}

Above the physical threshold [or equivalently for $M_P^2<(\sqrt{s}-M_{\pi,\pi^0})^2$], the partial-wave 
projections of the $P\pi\to\pi\pi$ amplitudes in the form given
by Eqs.~(\ref{M_for_P_to_3pi}) and (\ref{U_for_P_to_3pi}) are defined as integrals
over the scattering angle, cf.~Eq.~(\ref{t_ell_def}). This integration can be traded for 
an integration over the Mandelstam variable $t$, upon using relation (\ref{theta_vs_t}).
Starting from the one-loop expression of the amplitude given in Sec.~\ref{one loop P-pi},
and written in the form shown in Eqs.~(\ref{M_for_P_to_3pi}) and (\ref{U_for_P_to_3pi}),
this leads to
\begin{widetext}
\begin{align}\label{t0_t1_proj}
t_0 (s) &= {\cal W}_0 (s) +  \frac{1}{64 \pi K(s)} \int_{t_-}^{t_+} dt
\big[ {\cal P} (s,t,3s_0-t-s) + {\cal P} (s,3s_0-t-s,t) \big]
\nonumber\\
&
+ \frac{1}{2 K(s)} \int_{t_-}^{t_+} dt
\big\{ {\cal W}_0^t (t) + {\cal W}_0^u (t) 
- 3 (2s+t-3s_0) \big[ {\cal W}_1^t (t) + {\cal W}_1^u (t) \big] \big\}
,
\nonumber\\
t_1 (s) &= 2 K(s) {\cal W}_1 (s) + \frac{1}{128 \pi K^2(s)} \int_{t_-}^{t_+} \! dt  (2t + s -  3s_0)
\big[ {\cal P} (s,t,3s_0-t-s) - {\cal P} (s,3s_0-t-s,t) \big]
\nonumber\\
&
+ \frac{1}{4 K^2(s)} \int_{t_-}^{t_+} dt (2t + s -  3s_0)
\big\{ {\cal W}_0^t (t) - {\cal W}_0^u (t) 
- 3 (2s+t-3s_0) \big[ {\cal W}_1^t (t) - {\cal W}_1^u (t)\big] \big\}
\end{align}
\end{widetext}
for the corresponding $S$- and $P$-wave projections.
The limits of integration are given by
\begin{equation}
t_\pm (s) = \frac{3s_0 - s}{2} \pm K(s)
,
\label{limits_int}
\end{equation}
where $K(s)$ is the Kacser function defined in general by the formula (\ref{Kacser_def}).
This $t$-integral representation of the partial-wave projections is well suited for the analytic continuation in $M_P^2$ mentioned above.

In the above formulas, all ${\cal W}$ functions stand for the one-loop expressions
worked out in Sec.~\ref{one loop P-pi}.
Making the connection with the discussion at the end of Sec.~\ref{sec:general},
the contributions from direct rescattering [the first diagram in Fig.~\ref{topol}]
are contained in the functions ${\cal W}_{0,1} (s)$, whereas rescattering in
the crossed channels [the second diagram, with the fish topology, in Fig.~\ref{topol}]
is contained in the integrals involving the functions ${\cal W}_{0,1}^{t,u}(t)$.
The integrals over the polynomial parts present no difficulty. For the polynomial
given in Eq.~(\ref{calP}), one obtains 
\begin{multline}
\frac{1}{4 K(s)} \int_{t_-}^{t_+} dt
\big[ {\cal P} (s,t,3s_0-t-s) + {\cal P} (s,3s_0-t-s,t) \big]
\\
=
A + B \, \frac{s-s_0}{F_\pi^2} 
+ \left( C + \frac{D}{2} \right) \frac{(s-s_0)^2}{F_\pi^4} 
+ \frac{2}{3}D\, \frac{K^2(s)}{F_\pi^4}
\end{multline}
and, for the polynomial $P^S_x(s,t,u)$ given in Eq.~(\ref{calP^S}), 
\begin{multline}
\frac{1}{4 K^2(s)} \int_{t_-}^{t_+} dt (2t + s -  3s_0) 
\big[ {\cal P}^S_x (s,t,3s_0-t-s)
\hspace{2em}
\\
- {\cal P}^S_x (s,3s_0-t-s,t) \big]
 = \frac{4}{3} \bigg[ \frac{B^S_x}{F_\pi^2} - \frac{D^S_x}{F_\pi^4} (s-s_0) \bigg] K(s)
.
\end{multline} 
As for the remaining integrals, for equal-mass
pions, some simplifications occur. For instance, the three functions
${\bar J}(s)$, ${\bar J}_0 (s)$ and ${\bar J}_{\pm 0} (s)$ that appear
in the expressions for the one-loop amplitudes become identical
and equal to ${\bar J}(s)$. Furthermore, contributions involving the twice-subtracted
function ${\bar{\bar J}}_{\pm 0} (s)$ vanish, being proportional to powers of
$\Delta_\pi$. 
From the expressions obtained at NLO, the remaining integrands
can then be written as polynomials in $t$ times the loop function ${\bar J} (t)$,
\begin{multline}
\big\{ \! {\cal W}_0^t (t) \pm {\cal W}_0^u (t) -
3 (2s+t-3s_0) \big[ {\cal W}_1^t (t) \pm {\cal W}_1^u (t) \big] \! \big\}_{{\cal O}(E^4)}
\hspace{1em}
\\
= \left[\sum_{n=0}^2 {\widetilde w}_\pm^{(n)}  \left(\frac{t}{F_\pi^2}\right)^n 
+ \frac{{\cal B}_\pm}{3} \frac{s(t-4M_\pi^2)}{F_\pi^4} \right]{\bar J} (t)
,
\end{multline}
so that one obtains
\begin{widetext}
\begin{multline}
\label{t0_proj}
t_0 (s) = {\cal W}_0 (s) +  \frac{1}{64 \pi K(s)} \int_{t_-}^{t_+} dt
\big[ {\cal P} (s,t,3s_0-t-s)
+ {\cal P} (s,3s_0-t-s,t) \big] 
\\
+ \frac{1}{2 K(s)}  \int_{t_-}^{t_+} dt
\sum_{n=0}^2  \! \bigg[ {\widetilde w}_+^{(n)}
+ (2-n) \frac{{\cal B}_+}{3}  \frac{s}{F_\pi^2}
\left(-2\frac{M_\pi^2}{F_\pi^2}\right)^{1-n}
 \bigg] \left(\frac{t}{F_\pi^2}\right)^n 
{\bar J} (t)
,
\end{multline}
\begin{multline}
t_1 (s) = 2 K(s) {\cal W}_1 (s) + \frac{1}{128 \pi K^2(s)} \int_{t_-}^{t_+} \! dt (2t + s -  3s_0)
\big[ {\cal P} (s,t,3s_0-t-s) - {\cal P} (s,3s_0-t-s,t) \big]
\\
+ \frac{1}{4 K^2(s)} \int_{t_-}^{t_+} dt \sum_{n=0}^2 {\widetilde w}_-^{(n)}  (2t - 3 s_0)
\left(\frac{t}{F_\pi^2}\right)^n  {\bar J} (t)
\\
+ \frac{s}{4 K^2(s)} \int_{t_-}^{t_+} dt \sum_{n=0}^2  \bigg[{\widetilde w}_-^{(n)} 
+ \frac{{\cal B}_-}{3} 
\left(-2\frac{M_\pi^2}{F_\pi^2}\right)^{1-n}
\frac{n(n-3) 2^{3-n} M_\pi^2 - (n-2)(s-3s_0)}{F_\pi^2}
\bigg] 
\left(\frac{t}{F_\pi^2}\right)^n  {\bar J} (t)
.
\label{t1_proj}
\end{multline}
\end{widetext}
The coefficients ${\widetilde w}_\pm^{(n)}$ and ${\cal B}_\pm$ are process dependent,
and their expressions for the various channels are collected in Appendix~\ref{App:w_tilde}.

Following the same path in the case of the partial-wave projections
of the $\pi\pi$ amplitudes, we would obtain the expressions for the functions 
${\psi}_I(s)$ explicitly, namely, in the case of equal-mass pions, we would reproduce the 
formulas (\ref{psi_isospin})-(\ref{L_funct_def})  from the previous subsection. 
The situation in the case of the processes $P\pi\to\pi\pi$ is more
involved, for the reasons already explained previously.   
The main difference from the case of $\pi\pi$ scattering (and the source of the complication) lies in the integration itself.
In the case of $\pi\pi$ scattering, the above integrals are computed
in the usual way, with the limits of integration given by $t_+ (s) =0$
and $t_- (s) = - (s - 4 M_\pi^2)$. 
In the case of the partial-wave projections
of the $P\pi\to\pi\pi$ amplitudes, we have to perform an analytic continuation in  
$M_P^2$ from $M_P=M_\pi$ to $M_P>3M_\pi$ keeping $s$ real and above the two-pion 
unitarity threshold. During the process of this analytic continuation, the physical 
threshold of the $P\pi\rightarrow\pi\pi$ scattering moves above the two-pion unitarity 
threshold, and the unphysical region appears, where the  limits of integration (\ref{limits_int}) 
become complex. Finally, also the decay region emerges. The correct prescription for such analytic 
continuation corresponds to approaching the real values of $M_P^2$ from the upper complex half-plane, 
i.e., taking $M_P^2\rightarrow M_P^2+{\rm i}\delta$. It fixes the position of the endpoints infinitesimally 
below or above the real axis when necessary, as described in detail in \cite{Kacser} for the equal-mass 
pions. As a result, we get in this case
\begin{equation}
t_{\pm }(s) =\frac{3s_{0}-s}{2}+{\cal K}_{\pm}(s),
\end {equation}
where ${\cal K}_{\pm}(s)$ is defined as 
\begin{equation}
{\cal K}_\pm(s)=\left\{
\begin{array}{ll}
\,\mp\vert K(s) \vert,\phantom{\pm \mathrm{i}} &s>M_+^2 \\ 
\pm\mathrm{i}|K(s)|,\phantom{\pm \mathrm{i}\delta} &M_+^{2}>s>M_-^{2} \\ 
 \pm\vert K(s) \vert\pm \mathrm{i}\delta , &M_-^{2}>s>\frac{1}{2}\Delta _{P\pi } \\ 
\pm\vert K(s) \vert +\mathrm{i}\delta , &\frac{1}{2}\Delta _{P\pi}>s>4M_{\pi }^{2}
,
\end{array}
\right.
\label{Kacser_analytic}
\end{equation}
and $K(s)$ denotes the Kacser function for the equal-mass pions. According to the general definition (\ref{Kacser_def}), 
it reads
\begin{equation}
K^2(s)=\frac{1}{4} \left( 1 - \frac{4 M_\pi^2}{s} \right) \lambda_{P\pi} (s).
\end{equation}
In the above formulas, we have introduced the shorthand notation $M_{\pm}=M_P\pm M_\pi$ and $\Delta_{P\pi}=M_P^2- M_\pi^2$.
At the same time, the path of integration itself has to be deformed into the complex-$t$ plane in order that the integration
avoids the two-pion unitarity cut.

As shown in Appendix~\ref{app:integrals}, the required contour integrals of $t^n{\bar J} (t)$,  
with $n=0,1,2,3$, can be easily computed in closed form.
Note that the integrands of (\ref{t0_proj}) and (\ref{t1_proj}) are analytic functions. A detailed analysis of 
the analytic structure of these integrands reveals
[cf.\ Appendix~\ref{app:integrals}] that in the case of equal-mass pions, the above contour 
integrals can be obtained upon taking the difference at the endpoints of the primitives of the functions
occurring in these integrals.
The results can be written compactly as 
[recall that the condition $s \ge 4 M_\pi^2$ holds]
\begin{equation}
\pi \int_{t_- (s)}^{t_+ (s)} dt\, t^n {\bar J} (t)
=  \frac{2K(s)}{\sigma (s)} \sum_{i=0}^3
\kappa^{(n)}_i (s) {\widetilde k}_i (s)
,
\label{Jbar_integrals}
\end{equation}
where the functions ${\widetilde k}_i (s)$ are similar to the
functions $k_i (s)$ introduced in Eq.~(\ref{k_n}) in the case
of the $\pi\pi$ partial-wave projections:
\begin{align}
{\widetilde k}_0(s) &= \frac{1}{16 \pi} \sigma (s) ,
\qquad
{\widetilde k}_1(s) = \frac{1}{16 \pi} L (s) ,
\nonumber\\
{\widetilde k}_2(s) &= \frac{1}{16 \pi} \sigma (s)\, s\, \frac{M (s)}{\lambda_{P\pi}^{1/2} (s)} ,
\nonumber\\
{\widetilde k}_3(s) &= - \frac{1}{16 \pi} M_\pi^2 \frac{M (s)}{\lambda_{P\pi}^{1/2} (s)}\, L (s)
.
\label{tilde_k_functions}   
\end{align}
Actually, $k_0(s)={\widetilde k}_0(s)$
and  $k_1(s)=2{\widetilde k}_1(s)$, but we prefer to keep different notations for them, such as 
to clearly separate the present discussion from the case of the $\pi\pi$ scattering amplitude; of course,
in the limit $M_P\to M_\pi$,  ${\widetilde k}_2 (s) \to k_1(s)/2$
and ${\widetilde k}_3 (s) \to -k_3(s)/3$.
These expressions involve the functions $\sigma (s)$ and $L(s)$,
already defined in Eqs.~(\ref{sigma_def}) and (\ref{L_funct_def}), respectively.
An additional function $M(s)$ appears. For $s>4M_\pi^2$, it is defined as
\begin{align}
M (s) &= - \ln \left[ 1 - \frac{\Delta_{P\pi}}{s} + \frac{\lambda^{1/2}_{P\pi} (s)}{s} \right]
\nonumber\\
&\qquad
- \ln \left[ 1 - \frac{\Delta_{P\pi}}{s} - \frac{\lambda^{1/2}_{P\pi} (s)}{s} \right]^{-1}
\nonumber\\
&= 
- 2 \ln \left[ 1 - \frac{\Delta_{P\pi}}{s} + \frac{\lambda^{1/2}_{P\pi} (s)}{s} \right] + \ln \frac{4 M_\pi^2}{s}\,
.
\end{align}
(Note that in the limit $M_P\to M_\pi$, the function $M(s)$ tends to $L(s)$.) 
Another issue is raised by these expressions, namely, the determination
of the square root of the triangle function $\lambda_{P\pi} (s)$.  Nevertheless, an
inspection of the formula given above for the function $M(s)$
shows that the functions ${\widetilde{k}_i(s)}$ in Eq.~(\ref{Jbar_integrals}) do not depend on the
way one defines $\lambda^{1/2}_{P\pi} (s)$. Given the discussion
preceding Eq.~(\ref{Jbar_integrals}), it seems actually natural to define 
$\lambda^{1/2}_{P\pi} (s)$ as the square root of the function
$\lambda (s , M_P^2 + {\rm i} \delta, M_\pi^2)$, which we will
assume to be the case in the remainder of this section.
The functions ${\widetilde k}_n (s)$ for $n=2,3$ are represented in Fig.~\ref{plot_tilde_k} in Appendix \ref{App:K-tilde}, 
both for $M_P < M_\pi$, where they are real, and 
for $M_P > M_\pi$, where an imaginary part appears.\footnote{Up to trivial changes in the labels and normalization factors, 
${\widetilde k}_n (s)$ are identical with
the functions ${\cal F}_i (s)$ defined in Eqs.~(A.7) to (A.11) of Ref.~\cite{Kampf:2011wr}. For $n=0,1$ they furthermore coincide 
with the imaginary parts of the corresponding functions ${\widetilde K}_n(z)$ shown in Fig.~\ref{plot_tilde_K}.
}

The functions $\kappa_i^{(n)} (s)$, for $i=0,1,2$, are not polynomials in $s$, but have the following general structure,
\begin{equation}
\kappa_i^{(n)} (s) = {\bar\kappa}_i^{(n)} (s) + c_i^{(n)} \frac{\Delta_{P\pi}}{s} + d_i^{(n)} \frac{\Delta_{P\pi}^2}{s^2}
,
\end{equation}
where ${\bar\kappa}_i^{(n)} (s)$ are now  polynomials in $s$, and $c_i^{(n)}$, $d_i^{(n)}$ are numerical coefficients.
The polynomials ${\bar\kappa}^{(n)}_i (s)$ are given by
\begin{align}
{\bar\kappa}_0^{(0)} &= 3 ,
\qquad {\bar\kappa}_0^{(1)} = \frac{1}{4} (-5 s + 5 M_P^2 + 11 M_\pi^2 ) ,
\nonumber\\
{\bar\kappa}_0^{(2)} &= \frac{7}{9} (s - M_P^2)^2 + \frac{9}{2} M_\pi^2 ( M_P^2+M_\pi^2 - s),
\nonumber\\
{\bar\kappa}_0^{(3)} &= \frac{721}{144} s^2 M_\pi^2 - \frac{1442}{144} M_P^2 M_\pi^2 s - \frac{231}{16} M_\pi^4 s + \frac{153}{16} M_\pi^6
\nonumber\\
&+ \frac{883}{144} M_P^4 M_\pi^2 + \frac{195}{16} M_P^2 M_\pi^4 
+ \frac{9}{16} (M_P^2 - s )^3
,
\nonumber\\
{\bar\kappa}_1^{(0)} &= \frac{1}{2} ,
\qquad {\bar\kappa}_1^{(1)} = \frac{1}{4} (M_P^2+M_\pi^2- s),
\nonumber\\
{\bar\kappa}_1^{(2)} &= \frac{1}{6} \left[ (s - M_P^2)^2 - M_\pi^2 
\left( 5 s + 2 M_\pi^2 - 7 M_P^2 \right)  \right]
,
\nonumber\\
{\bar\kappa}_1^{(3)} &= \frac{M_\pi^2}{24} \big[ 25 s^2 - (56 M_P^2 + 53 M_\pi^2) s
+ 43 M_P^4 
\nonumber\\
&\qquad
+ 67 M_P^2 M_\pi^2 - 53 M_\pi^4 \big] + \frac{1}{8} (M_P^2 - s )^3,
\nonumber\\
{\bar\kappa}_2^{(0)} &= \frac{1}{2}\,,
\qquad {\bar\kappa}_2^{(1)} = - \frac{1}{4} \left( s - 2 M_P^2 \right) ,
\nonumber\\
{\bar\kappa}_2^{(2)} &= \frac{1}{6} 
\big[ s^2 - s ( 3 M_P^2 + 4 M_\pi^2)  
\nonumber\\
&\qquad 
+ 3 (M_P^4 + 2 M_P^2 M_\pi^2 - M_\pi^4) \big] ,
\nonumber\\
{\bar\kappa}_2^{(3)} &= \frac{M_\pi^2}{12} \big[ 11 s^2 - 10 s (3 M_P^2 + 2 M_\pi^2) + 30  M_P^4 
\nonumber\\
& \qquad
+ 18  M_P^2 M_\pi^2 - 24  M_\pi^4 \big]
\nonumber\\
&- \frac{1}{8} (s^3 - 4 s^2 M_P^2 + 6 s M_P^4 - 4 M_P^6),
\nonumber\\
{\bar\kappa}_3^{(0)} &= 1 ,
\ {\bar\kappa}_3^{(1)} = M_\pi^2 ,
\ {\bar\kappa}_3^{(2)} = 2 M_\pi^4 ,
\ {\bar\kappa}_3^{(3)} = 5 M_\pi^6
.           
\label{kappa}    
\end{align}
The only non-vanishing coefficients $d_i^{(n)}$ are
\begin{align}
d_1^{(3)} &= \frac{M_\pi^4}{4}\, \Delta_{P\pi},  \qquad d_2^{(2)} = \frac{M_\pi^2}{6}\, \Delta_{P\pi},
\nonumber\\
d_2^{(3)} &= \frac{M_\pi^2}{12} ( 3 M_P^2 + 5 M_\pi^2) \Delta_{P\pi}
,
\end{align}
while the non-vanishing coefficients $c_i^{(n)}$ read
\begin{align}
&
c_0^{(2)} = - \frac{7}{9} M_\pi^2 \Delta_{P\pi},  
\ c_0^{(3)} = - \frac{M_\pi^2}{72}   (81 M_P^2 + 239 M_\pi^2)\Delta_{P\pi},
\nonumber\\
&
c_1^{(1)} = - \frac{1}{2} M_\pi^2 , \qquad c_1^{(2)} = - \frac{1}{2} M_\pi^2 (M_P^2 + M_\pi^2) ,
\nonumber\\
&
c_1^{(3)} = - \frac{1}{4} M_\pi^2 (2 M_P^4 + 7 M_P^2 M_\pi^2 + M_\pi^4) ,
\nonumber\\
&
c_2^{(0)} = - \frac{1}{2}\,, \ c_2^{(1)} = - \frac{\Delta_{P\pi}}{4} , 
\ c_2^{(2)} = - \frac{1}{6} (M_P^2 + 5 M_\pi^2) \Delta_{P\pi} ,
\nonumber\\
&
c_2^{(3)} = - \frac{1}{24} (3 M_P^4 + 34 M_P^2 M_\pi^2 + 59 M_\pi^4) \Delta_{P\pi}
.
\end{align}

\subsubsection{$S$-wave projections}

From expression (\ref{t0_proj}) for the one-loop $S$-wave
projection, we can extract its dispersive part ${\widetilde\psi}_0 (s)$ as
\begin{equation}
{\widetilde\psi}_0 (s)=t_0 ^{\rm 1loop}(s)-{\rm i}\Abs t_0 ^{\rm 1loop}(s).
\end{equation}
In accord with  Eq.~(\ref{discW}), for $s>4M_\pi^2$, we have   $\Abs t_0^{\rm 1loop} (s)=\Abs {\cal W}_0^{\rm 1loop} (s)$, where ${\cal W}_0^{\rm 1loop} (s)$ denotes the one-loop expression
of ${\cal W}_0 (s)$ constructed through the first iteration in Sec.~\ref{one loop P-pi}. Then, using (\ref{t0_proj}), we get finally
\begin{align}
\sigma(s) {\widetilde\psi}_0 (s) &=  \frac{\sigma (s)}{16 \pi}
\bigg\{ 
16\pi \Disp\,{\cal W}_0^{\rm 1loop} (s)
+
\frac{C}{F_\pi^4} (s-s_0)^2 
\nonumber\\
&\quad
+ \frac{D}{F_\pi^4} \bigg[ \frac{2}{3} K^2(s) + \frac{(s-s_0)^2}{2} \bigg]
\bigg\}
\nonumber\\
&+ \frac{1}{\pi} \sum_{i=0}^3 \bigg\{
\sum_{n=0}^2 {\widetilde w}_+^{(n)}   \frac{\kappa_i^{(n)} (s)}{F_\pi^{2n}}
\nonumber\\
&\quad
+ \frac{{\cal B}_+}{3 F_\pi^4} \frac{s}{ F_\pi^2}
 \frac{\kappa_i^{(1)} (s) - 4 M_\pi^2 \kappa_i^{(0)} (s)}{F_\pi^2} \bigg\}
{\widetilde k}_i (s) 
.
\label{S-wave_proj}
\end{align} 

The expression of $\Abs {\cal W}_0 (s)$
at next-to-leading order involves the product 
\begin{equation}
\Abs {\cal W}_0 (s) \sim  \sigma(s) \varphi_0 (s) {\widetilde\psi}_0 (s)
\,\theta(s-4M_\pi^2)
,
\label{calW_0}
\end{equation}
where $\varphi_0 (s)$ is a first-order polynomial in $s$.
The next step consists in constructing a function in the complex-$s$ plane,
which has a cut along the positive real axis, and whose discontinuity is given by
Eq.~(\ref{calW_0}). This is straightforward for the contribution between the first curly brackets 
in Eq.~(\ref{S-wave_proj}) since
\begin{equation}
\sigma (s) \re {\bar J} (s) =  8 \pi \im {\bar J}^2 (s)
.
\end{equation}
For the remaining terms, given by the sum in Eq.~(\ref{S-wave_proj}),
this can be achieved in the following way. 
Let us introduce functions defined through dispersive integrals of the functions ${\widetilde k}_i(s)$,
\begin{align}
\label{widetilde_K}
{\widetilde K}_i (s) &= \frac{s}{\pi} \int_{4 M_\pi^2}^\infty \frac{dx}{x} \frac{{\widetilde k}_i(x)}{x-s-{\rm i}0}
,
\\
\doublewidetilde{K}_i(s) &= \frac{s^2}{\pi} \int_{4 M_\pi^2}^\infty \frac{dx}{x^2} \frac{{\widetilde k}_i(x)}{x-s-{\rm i}0}
= {\widetilde K}_i (s) - s{\widetilde K}_i' (0)
.
\nonumber
\end{align}
Then, the function
\begin{align}
&
[\varphi_0 \sodot {\widetilde\xi}_0](s) =
\nonumber\\
&
=
\frac{1}{16\pi^2 F_\pi^2} [ \varphi_0 (s) - \varphi_0 (0) ]\sum_{n=0}^2
{\widetilde w}_+^{(n)}  \sum_{i=0}^3 \kappa_i^{(n)} (s) 
{\widetilde K}_i (s)
\nonumber\\
&
+
\frac{1}{16\pi^2 F_\pi^2}\, \varphi_0 (0) \sum_{n=0}^2
{\widetilde w}_+^{(n)}  \sum_{i=0}^3 \left[\kappa_i^{(n)} (s) - d_i^{(n)}\frac{\Delta_{P\pi}^2}{s^2} \right]\!
{\widetilde K}_i (s)
\nonumber\\
&
+ \frac{s{\cal B}_+}{16\pi^2 F_\pi^2}\, \varphi_0 (s)
 \sum_{i=0}^3 \big[ \kappa_i^{(1)} (s) - 4 M_\pi^2 \kappa_i^{(0)} (s) \big]
{\widetilde K}_i (s)
\nonumber\\
&
+ \frac{1}{16\pi^2 F_\pi^2}\, \varphi_0 (0) \sum_{n=0}^2 
{\widetilde w}_+^{(n)}  \sum_{i=0}^3 d_i^{(n)}\frac{\Delta_{P\pi}^2}{s^2}\, 
{\doublewidetilde{K}}_i (s)
\end{align}
has a discontinuity along the positive real axis given by the sum
in Eq.~(\ref{S-wave_proj}), multiplied by $\varphi_0 (s)$.
Notice that all the functions ${\widetilde k}_i(x)$ are bounded on the real axis
by $\ln x$ as $x \to + \infty$ since
\begin{multline}
\lim_{x\to +\infty} {\widetilde k}_i(x) =
\bigg\{ \frac{1}{16\pi} \, ; \, \frac{1}{16\pi} \ln \frac{M_\pi^2}{x} \, ; 
\, \frac{1}{16\pi}  \ln \frac{M_\pi^2}{x} \, ; 
\\
-\frac{1}{16\pi} \frac{M_\pi^2}{x} \ln^2 \frac{M_\pi^2}{x} \bigg\}
.
\end{multline}
Therefore, the once-subtracted dispersive integrals in Eq.~(\ref{widetilde_K}) are convergent, and the
functions ${\doublewidetilde{K}}_i(s)$ are defined without ambiguity. Actually, the dispersive integral
for ${\widetilde k}_3(x)$ would already converge without subtraction. Finally, one easily finds expressions 
in terms of ${\bar J} (s)$ \cite{Knecht:1995tr,Kampf:2011wr} for the two first functions, i.e.,
${\widetilde K}_0(s) = {\bar J} (s)$, and
\begin{equation}
{\widetilde K}_1(s) = \frac{1}{2} \, \frac{s}{s - 4 M_\pi^2}
\left[ 16 \pi^2 {\bar J}^{2} (s) - 4 {\bar J} (s) +\frac{1}{4 \pi^2} \right]
.
\label{K0_and_K1}
\end{equation}

\subsubsection{$P$-wave projections}
The dispersive part of the $P$-wave projection can be obtained as
\begin{align}
{\widetilde\psi}_1 (s)&=t_1 ^{\rm 1loop}(s)-{\rm i}\Abs\,t_1 ^{\rm 1loop}(s)\nonumber\\
&=t_1 ^{\rm 1loop}(s)-2{\rm i}K(s)\Abs {\cal W}_1 ^{\rm 1loop}(s),
\end{align}
where we used (\ref{discW}).
Starting from the structure of $\Abs {\cal W}_1 (s)$, i.e.,
\begin{equation}
\Abs {\cal W}_1 (s) \sim  \sigma(s)\, \frac{\varphi_1 (s)}{2 K(s)}\, {\widetilde\psi}_1 (s)
\,\theta(s-4M_\pi^2)
,
\end{equation}
one may notice that ${\widetilde\psi}_1 (s)$ occurs in $\Abs {\cal W}_1 (s)$ 
through the combination [$\varphi_1(s)$ is proportional to $s-4M_\pi^2$]\footnote{In this formula, 
the second  term in the rectangular brackets is present only for $K_S\pi^0\rightarrow\pi^+\pi^-$ scattering, see Appendix~\ref{App:w_tilde}.}
\begin{align}
&\frac{\sigma(s)(s-4M_\pi^2)}{2 K (s)} {\widetilde\psi}_1(s)
=
\nonumber\\
&
=\frac{\sigma(s)(s-4M_\pi^2)}{32\pi}\left[\Disp {\cal W}_1^{\rm 1loop} (s)-\frac{D^S_x}{F_\pi^4}(s-s_0)\right]
\nonumber\\
&\quad
+ \frac{1}{\pi} \sum_{n=0}^3
{\widetilde w}_-^{(n)} \sum_{i=0}^3  
\frac{2 \kappa_i^{(n+1)} (s) - 3 s_0 \kappa_i^{(n)} (s)}{F_\pi^{2n}}
\frac{s {\widetilde k}_i (s)}{\lambda_{P\pi} (s)}
\nonumber\\
&\quad
+ \frac{1}{\pi}\sum_{i=0}^3 \bigg\{
\sum_{n=0}^3
{\widetilde w}_-^{(n)} \, \frac{\kappa_i^{(n)} (s)}{F_\pi^{2n}}
+ \frac{{\cal B}_-}{3 F_\pi^4} \Big[ 2 \kappa_i^{(2)} (s)
\nonumber\\
&\quad\ 
+ (s-3s_0-8M_\pi^2) \kappa_i^{(1)} (s) 
- 4 M_\pi^2 (s-3s_0) \kappa_i^{(0)} (s) \Big] \! \bigg\}
\nonumber\\
&\quad\ 
\times
\frac{s^2{\widetilde k}_i (s)}{\lambda_{P\pi} (s)}\,
,
\label{disc_calW_1}
\end{align}
The main difference from the previous
case of the $S$-wave projection lies in the presence of additional polynomial 
$\lambda_{P\pi}(s)$ in the denominator of the right-hand side. This feature can
be handled by writing
\begin{equation}
\lambda_{P\pi} (s) = (s - M_+^2) (s-M_-^2),\quad M_\pm = M_P \pm M_\pi
,
\end{equation}
and by using the decomposition of a product of fractions.
As a result, we define the additional functions
\begin{align}
{\widetilde K}_i^{(\lambda)} (s) &=
\frac{1}{4}
\left[ \frac{M_\pi^2}{s - M_+^2} \left(
{\widetilde K}_i(s) - \frac{s}{M_+^2} {\widetilde K}_i(M_+^2)
\right)
\right. 
\nonumber\\
&\qquad
\left. 
- \frac{M_\pi^2}{s - M_-^2} \left(
{\widetilde K}_i(s) - \frac{s}{M_-^2} {\widetilde K}_i(M_-^2)
\right)
\right]
\nonumber\\
&\equiv 
\frac{s}{\pi} \int_{4 M_\pi^2}^\infty \frac{dx}{x} 
\, \frac{M_P M_\pi^3}{\lambda_{K\pi} (x)} \frac{{\widetilde k}_i(x)}{x-s-{\rm i}0}
\,,
\label{widetilde_K_lambda0}
\end{align}
which are actually characterized (but not entirely) by the conditions
${\widetilde K}_i^{(\lambda )}(0) = 0$ and their absorptive parts
along the cut on the positive real axis
\begin{equation} 
\Abs {\widetilde K}_i^{(\lambda)}(s) = 
\frac{M_P M_\pi^3}{\lambda_{K\pi} (s)} \, {\widetilde k}_i(s)\,
\theta (s-4M_\pi^2)
.
\end{equation}
Then, the discontinuity along the real positive axis of the function
\begin{equation}
\begin{split}
&
{\widetilde\xi}_1 (s) =
\\
&
=
\frac{1}{\pi} \sum_{n=0}^2
\frac{s {\widetilde w}_-^{(n)}}{M_P M_\pi^3}  
\!\sum_{i=0}^3 \!  {\widetilde K}_i^{(\lambda)}(s)
\frac{2 \kappa_i^{(n+1)} + (s - 3 s_0) \kappa_i^{(n)} (s)}{F_\pi^{2n}}
\\
&
+\frac{1}{\pi}\sum_{i=0}^3 \frac{{\cal B}_-}{3} \Big[ 2 \kappa_i^{(2)} (s)
+ (s-3s_0-8M_\pi^2) \kappa_i^{(1)} (s) 
\\
&\quad
-4 M_\pi^2 (s-3s_0) \kappa_i^{(0)} (s) \Big] 
\frac{s^2}{M_P M_\pi^3 F_\pi^4} \, {\widetilde K}_i^{(\lambda)}(s)
\end{split}
\end{equation}
reproduces the terms displayed in Eq.~(\ref{disc_calW_1}).

\indent

\section{Second iteration: $P\to\pi^0\pi^0\pi^0$ with  $M_\pi \neq M_{\pi^0}$}
\label{Sec:second iteration neutral}
\setcounter{equation}{0}

We next consider the second iteration in the case
where the difference between neutral and charged pion
masses is taken into account, but for the two processes
where all the external pions are neutral, i.e., 
$K_L\to\pi^0\pi^0\pi^0$ and $\eta\to\pi^0\pi^0\pi^0$.
These amplitudes are fully symmetric under exchanges
of the three Mandelstam variables, and are described
by the polynomial of the form (\ref{Op6 polynomial}), 
\begin{multline}
{\cal P} (s,t,u) = A  + \frac{C}{F_\pi^4}  \left[(s-s_0)^2+(t-s_0)^2 + (u-s_0)^2 \right]\\
+ \frac{E}{F_\pi^6}  \left[(s-s_0)^3+(t-s_0)^3 + (u-s_0)^3 \right]
+ {\cal O}(E^8),
\end{multline}
and a single function [see Table~\ref{W_for_P-pi_Table}]
that we call simply ${\cal W}_{L, 00} (s)$. Its expression
at one loop is given in Eq.~(\ref{calW_00}).
According to (\ref{W00_2loop}), at two loops,  the absorptive part of 
${\cal W}_{L, 00} (s)$ is determined in terms of the dispersive 
parts of the $\ell =0$ partial-wave projections of the one-loop amplitudes  
${\cal M}^L_{00}$, ${\cal M}^L_{x}$  and $A_{00}$, $A_{x}$. However, these have now to 
be calculated for $M_\pi \neq M_{\pi^0}$. Regarding the $\pi\pi$
amplitudes $A_{00}$, $A_{x}$, this has already been done in Ref.~\cite{DescotesGenon:2012gv},
where explicit expressions can be found, see Section IV.A therein. As discussed below, 
in the case of the amplitudes ${\cal M}^L_{00}$ and ${\cal M}^L_{x}$, this calculation 
does not raise any difficulties of principle, although the resulting 
formulas become much more complicated than in the case of the equal-mass pions.

\subsection{NLO partial waves of the $P\pi^0\to\pi^0\pi^0$ amplitude}

The additional 
algebraic complexity generated by $M_\pi \neq M_{\pi^0}$ is, in this case,
compensated to some extent by the absence of a $P$ wave.
Discarding the contribution from the polynomial part, which is
trivial to handle, we need to compute two types of integrals, cf.\  (\ref{calW_00}).
The first type is analogous to the one in Eq.~(\ref{Jbar_integrals}),
but involves now the function ${\bar J}_0 (s)$ instead
of ${\bar J}$, while at the same time the limits of
integration become  
\begin{equation}
t^{\pi^0}_\pm (s) = \frac{3 s_0^{\pi^0} - s}{2}+{\cal K}_{\pi^0;\pm} (s) 
,
\end{equation}
with $s_0^{\pi^0} \equiv M_{\pi^0}^2 + M_P^2/3$, and
${\cal K}_{\pi^0;\pm} (s)$ is the same function as ${\cal K}_\pm(s)$, defined in 
Eq.  (\ref{Kacser_analytic}), but
with the charged pion mass replaced by the neutral one.
The result of this integration is given by the formula
(\ref{Jbar_integrals}), provided one replaces the functions
$\kappa_i^{(n)} (s)$ and ${\widetilde k}_i (s)$ by functions 
$\kappa_{\pi^0;i}^{(n)} (s)$ and ${\widetilde k}_i^{\pi^0} (s)$,
respectively, obtained from the former upon performing everywhere
the substitution $M_\pi \to M_{\pi^0}$, e.g.,
\begin{equation}
\pi \int_{t^{\pi^0}_- (s)}^{t^{\pi^0}_+ (s)} \!\!\! dt\, t^n  {\bar J}_0 (t)
=  \frac{2K_{\pi^0}(s)}{\sigma_0 (s)} \sum_{i=0}^3
\kappa^{(n)}_{\pi^0 ;i} (s) {\widetilde k}_i^{\pi^0} (s)
.
\label{Jbar_integrals_pi0}
\end{equation}
The second type of integral that is needed is of a new type.
It involves the loop function ${\bar J} (s)$ for charged pions,
but integrated with the kinematics corresponding to neutral pions.
This second integral can also be done analytically, and the
result is cast into the form
\begin{equation}
\pi\int_{t_-^{\pi^0} (s)}^{t_+^{\pi^0} (s)} \!\!\!\! dt\, t^n {\bar J} (t)
=\frac{2K_{\pi^0}(s)}{\sigma_0 (s)}
\sum_{i=0}^3 \kappa^{(n)}_{\nabla ;i} (s) {\widetilde k}_{\nabla ; i} (s)
.
\label{Jbar_integrals_mixed}
\end{equation}
The set of functions ${\widetilde k}_{\nabla ; i} (s)$ in terms of which
it is expressed differs from the set ${\widetilde k}_i (s)$ given in
Eq.~(\ref{tilde_k_functions}). Explicitly, they read
\begin{align}
\label{tilde_k_nabla_functions}
{\widetilde k}_{\nabla ;0}(s) &= \frac{1}{16 \pi}\, \sigma_0 (s) ,
\nonumber\\
{\widetilde k}_{\nabla ;1}(s) &= \frac{1}{8 \pi} \frac{\sigma_0 (s)}{M_\pi^2}
[ t_+^{\pi^0} (s) \sigma_+ (s) \ln \tau_+ (s) 
\nonumber\\
&\qquad
+ t_-^{\pi^0} (s) \sigma_- (s) \ln \tau_- (s)  ],
\nonumber\\
{\widetilde k}_{\nabla ;2}(s) &= \frac{1}{8 \pi}\, \sigma (s)\, s\, \frac{1}{\lambda_{P\pi}^{1/2} (s)}
[ t_+^{\pi^0} (s) \sigma_+ (s) \ln \tau_+ (s) 
\nonumber\\
&\qquad
- t_-^{\pi^0} (s) \sigma_- (s) \ln \tau_- (s)  ] ,
\\
{\widetilde k}_{\nabla ;3}(s) &= - \frac{1}{16 \pi} \frac{M_\pi^2}{\lambda_{P\pi}^{1/2} (s)} 
[ \ln^2 \tau_+ (s) - \ln^2 \tau_- (s) ]
,
\nonumber
\end{align}
with $\sigma_\pm (s) \equiv \sigma (t_{\pm}^{\pi^0} (s))$ and, likewise, $\tau_\pm (s) \equiv \tau (t_{\pm}^{\pi^0} (s))$,
where
\begin{equation}
 \tau_{\pm} (s) = \frac{\sigma_\pm (s)-1}{\sigma_\pm (s)+1}
 .
\end{equation}
In the limit $M_{\pi^0} \to M_\pi$, one recovers the previous set of functions:
\begin{align}
\label{tilde_k_nabla_functions_limit}
{\widetilde k}_{\nabla ;0}(s) &\to {\widetilde k}_0 (s), \quad 
{\widetilde k}_{\nabla ;3}(s)  \to  {\widetilde k}_3 (s) ,
\nonumber\\
{\widetilde k}_{\nabla ;1}(s) &\to  \left(
\frac{3 s_0 -s}{M_\pi^2} - 4\, \frac{\Delta_{P\pi}}{s} \right) {\widetilde k}_1 (s)
- \frac{\lambda_{P\pi} (s)}{s M_\pi^2} {\widetilde k}_1 (s) ,
\nonumber\\
{\widetilde k}_{\nabla ;2}(s) &\to {\widetilde k}_1 (s) + \left( 1 - \frac{\Delta_{P\pi}}{M_\pi^2} \right) {\widetilde k}_2 (s)
.
\end{align}

As for the functions $\kappa^{(n)}_{\nabla ;i} (s)$, one has 
$\kappa^{(n)}_{\nabla ;3} (s) = \kappa^{(n)}_{3} (s)$, and the remaining ones are given by
\begin{align}
\kappa_{\nabla ;0}^{(0)} &= 3 ,
\quad \kappa_{\nabla ;0}^{(1)} = \frac{1}{4} (-5 s + 5 M_P^2 + 15 M_{\pi^0} - 4 M_\pi^2 ) ,
\nonumber\\
\kappa_{\nabla ;0}^{(2)} &= \frac{7}{9} (s - M_P^2)^2 
- \frac{7}{9} M_{\pi^0}^2 \frac{\Delta_{P\pi^0}^2}{s}
\nonumber\\
&
- \frac{1}{6} \Big[ (s-M_P^2) (28 M_{\pi^0}^2 - M_\pi^2) + 12 M_\pi^4 
\nonumber\\
&\quad
+ 3 M_\pi^2 M_{\pi^0}^3 - 42 M_{\pi^0}^4  \Big] ,
\nonumber\\
\kappa_{\nabla ;0}^{(3)} &= \frac{9}{16} (3 s_0^{\pi^0} - s)^3 - \frac{M_\pi^2}{18} (3 s_0^{\pi^0} - s)^2
\nonumber\\
&
- \frac{5}{12} M_\pi^4 (3 s_0^{\pi^0} - s) - 5 M_\pi^6
\nonumber\\
&
- \frac{9}{8}M_{\pi^0}^2(3 s_0^{\pi^0} - s) \frac{\Delta_{P\pi^0}^2}{s}
+ \frac{1}{18} M_\pi^2 M_{\pi^0}^2 \frac{\Delta_{P\pi^0}^2}{s}\,
,
\nonumber\\
\kappa_{\nabla ;1}^{(0)} &= 0 ,
\quad \kappa_{\nabla ;1}^{(1)} = \frac{M_\pi^2}{8}   ,
\nonumber\\
\kappa_{\nabla ;1}^{(2)} &= \frac{M_\pi^2}{12} ( M_\pi^2 + 2 M_{\pi^0}^2 - s)
,
\nonumber\\
\kappa_{\nabla ;1}^{(3)} &= \frac{M_\pi^2}{16} (s-3 s_0^{\pi^0})^2
+ \frac{M_\pi^4}{24} (s - 3 s_0^{\pi^0})
\nonumber\\
&
- \frac{5}{24} M_\pi^6 - \frac{1}{16} M_\pi^2 M_{\pi^0}^2 \frac{\Delta_{P\pi^0}^2}{s}\,
,
\nonumber\\
\kappa_{\nabla ;2}^{(0)} &= \frac{1}{2}  , \quad
\kappa_{\nabla ;2}^{(1)} = - \frac{1}{8} ( s -  M_P^2 + 4 M_\pi^2 - 3 M_{\pi^0}^2) ,
\nonumber\\
\kappa_{\nabla ;2}^{(2)} &= \frac{1}{12}  (s - 3 s_0^{\pi^0})^2  
-  \frac{1}{6} M_{\pi^0}^2 \frac{\Delta_{P\pi^0}^2}{s}
\nonumber\\
&
+ \frac{M_\pi^2}{12} (s - M_P^2 - 3 M_{\pi^0}^2 - 12 M_\pi^2),
\nonumber\\
\kappa_{\nabla ;2}^{(3)} &= \frac{1}{16} (3 s_0^{\pi^0} - s)^3 - \frac{M_\pi^2}{24} (3 s_0^{\pi^0} - s)^2
\nonumber\\
&
- \frac{5}{24} M_\pi^4 (3 s_0^{\pi^0} - s) - \frac{5}{2} M_\pi^6
\\
& \qquad
- \frac{3}{16}M_{\pi^0}^2(3 s_0^{\pi^0} - s) \frac{\Delta_{P\pi^0}^2}{s}
+ \frac{1}{12} M_\pi^2 M_{\pi^0}^2 \frac{\Delta_{P\pi^0}^2}{s}\, .
\nonumber
\end{align}
We were not able to find simpler expressions for the functions 
${\widetilde k}_{\nabla ;i}(s)$, i.e., comparable to those given in 
Eq.~(\ref{tilde_k_functions}). The origin of the difficulty can, for
instance, be understood upon considering the square of the functions
$\sigma_\pm (s)$,
\begin{multline}
\sigma_\pm^2 (s) = \frac{s}{s-4M_{\pi^0}^2} \, \frac{M_\pi^2}{M_{\pi^0}^2}
\left( \frac{s-4M_{\pi^0}^2 \pm 2{\cal K}_{\pi^0} (s)}{\Delta_{P\pi^0}}  \right)^2\\
- \frac{\Delta_\pi}{M_{\pi^0}^2} \left( 1 - \frac{4 M_{\pi^0}^2}{s} \right),
\end{multline}
and comparing it to the expression for $\sigma_{T_\pm} (s)$ given in Eq.~(\ref{sigma_T}):
as $\Delta_\pi \to 0$, $\sigma_\pm (s) \to 1/\sigma_{T_\pm} (s)$, but no such simple 
expression is available when $\Delta_\pi \neq 0$.

\subsection{NLO $S$-wave projection of the $P\pi^0\to\pi^+\pi^-$  amplitude}

The one-loop representation of the $P\pi^0\to\pi^+\pi^-$ amplitude is given  by Eqs.~(\ref{WLx}) and (\ref{WLx_1loop}). 
In order to obtain the corresponding $S$-wave projection,  we need to compute the contour integrals involving 
the functions ${\bar J}_{\pm0}$ with the endpoints given by 
\begin{equation}
t_{\pm}^x(s)=\frac{3s_0^x-s}{2}+{\cal K}_{x;\pm}(s),
\end{equation}
where now $3s_0^x=M_P^2+M_{\pi^0}^2+2 M_{\pi}^2$ and
\begin{equation}
{\cal K}_{x;\pm}(s)=
\begin{cases}
\mp|K_x(s)|,&s>M_{+x}^2 \\ 
\pm\mathrm{i}|K_x(s)|,&M_{+x}^{2}>s>M_{-x}^{2} \\ 
 \pm|K_x(s)|\pm \mathrm{i}\delta ,&M_{-x}^{2}>s>\frac{M_\pi}{M_\pi+M_{\pi^0}}\Delta _{P\pi^0 } \\ 
\pm|K_x(s)|+\mathrm{i}\delta ,&\frac{M_\pi}{M_\pi+M_{\pi^0}}\Delta _{P\pi^0 }>s>4M_{\pi }^{2}
\end{cases}
\!.
\label{Kacser_analytic_x}
\end{equation}
Here, $M_{\pm x}=M_P\pm M_{\pi^0}$ and $K_x(s)$ is the corresponding  Kacser's function,
given by 
\begin{equation}
K_x^2(s)=\frac{1}{4} \left( 1 - \frac{4 M_\pi^2}{s} \right) \lambda_{P\pi^0} (s).
\end{equation}
Since for $s>4M_\pi^2$  the path corresponding to the movement of $t^x_{\pm}(s)$ in 
the complex $t$-plane does not cross the cut of the function ${\bar J}_{\pm0}(t)$, the 
contour integrals can be again calculated as the differences of the corresponding primitive 
functions at the above endpoints, as described in Appendix~\ref{app:integrals}.  
In analogy with (\ref{Jbar_integrals_mixed}), we express them in terms of elementary 
functions ${\widetilde k}_{+ ; i} (s)$ as
\begin{equation}
\pi\int_{t_-^{x} (s)}^{t_+^{x} (s)} dt\, t^n {\bar J} (t)
=\frac{2K_{x}(s)}{\sigma(s)}
\sum_{i=0}^3 \kappa^{(n)}_{+ ;i} (s)\, {\widetilde k}_{+ ; i} (s)
.
\label{Jbar_integrals_x}
\end{equation}
The explicit form of the latter functions reads
\begin{align}
{\widetilde k}_{+ ;0} (s)&=\frac{1}{16\pi}\,\sigma(s),\quad {\widetilde k}_{+ ;1} (s)=\frac{1}{16\pi}\,L(s),\nonumber\\
{\widetilde k}_{+ ;2} (s)&=\frac{1}{16\pi}\,\sigma(s)\,s\,\frac{M(s)}{\lambda^{1/2}_{P\pi^0}(s)}\,,
\nonumber\\
{\widetilde k}_{+ ;3} (s)&=-\frac{1}{16\pi}\frac{\Sigma_\pi}{2}\frac{M(s)}{\lambda^{1/2}_{P\pi^0}(s)}\,L(s),\nonumber\\
{\widetilde k}_{+ ;4} (s)&=-\frac{1}{16\pi}\,\Delta_\pi\,\frac{{\cal J}(\tau^x_+(s))-{\cal J}(\tau^x_-(s))}{\lambda^{1/2}_{P\pi^0}(s)}\,,
\end{align}
where $\Sigma_\pi \equiv M_\pi^2 + M_{\pi^0}^2$, $\Delta_\pi \equiv M_\pi^2 - M_{\pi^0}^2$,
and the functions $L(s)$ and $\sigma_(s)$ 
are given by (\ref{L_funct_def}) and (\ref{sigma_def}), respectively. 
In the above formulas,
\begin{align}
\tau^x_\pm(s)&=\frac{\sigma_{\pm 0}(t^x_{\pm}(s))-1}{\sigma_{\pm 0}(t^x_{\pm}(s))+1}\,,
\nonumber\\
\\
\sigma_{\pm 0}(t)&=\sqrt{\frac{t-(M_\pi+M_{\pi^0})^2}{t-(M_\pi-M_{\pi^0})^2}}\,.
\nonumber
\end{align}
In the limit of equal-mass pions, we recover the functions ${\widetilde k}_{i}$, namely,
${\widetilde k}_{+ ;i} (s)\to {\widetilde k}_{i}(s)$ for $i=1,\dots,3$. The new function 
${\widetilde k}_{+ ;4} (s)$, which vanishes in the limit $M_{\pi^0} \to M_\pi$, is defined 
in terms of the function ${\cal J}(\tau)$
\begin{align}
{\cal J}(\tau)=\log\frac{M_{\pi^0}}{M_\pi}\,\log\tau &+ {\rm Li}_2\Big(1-\frac{M_{\pi^0}}{M_\pi}\,\tau\Big)
\nonumber\\
&- {\rm Li}_2\Big(1-\frac{M_{\pi}}{M_{\pi^0}}\,\tau\Big).
\end{align}
Here, we assume the standard definition of the dilogarithm, with a cut along the
interval $(1,\infty)$ of the real axis and ${\rm Li}_2(x)={\rm Li}_2(x-{\rm i}0)$ for $x>1$.
Let us denote for short
\begin{equation}
\xi=\frac{M_{\pi }^{2}}{\Delta _{\pi }}\frac{\Delta _{P\pi ^{0}}^{2}}{%
\Delta _{P\pi ^{+}}^{2}}\,.
\end{equation}
Then the  rational functions $ \kappa^{(n)}_{+ ;i} (s)$ are given by 
\begin{align}
\kappa _{+;0}^{\left(- 2\right) } &=-\frac{1}{2\Delta _{P\pi ^{+}}\Delta
_{\pi }}\frac{s}{s-\xi }\,\log \frac{M_{\pi ^{0}}}{M_{\pi }} 
\nonumber\\
&\times \left( \frac{s}{s-\xi }\frac{s-2\Sigma _{\pi }-\Delta _{P\pi ^{0}}}{%
\Delta _{P\pi ^{+}}}+1\right)\!,  
\nonumber\\
\kappa _{+;0}^{\left( -1\right) } &=-\frac{1}{\Delta _{P\pi ^{+}}}\frac{s}{%
s-\xi }\frac{\Sigma _{\pi }}{\Delta _{\pi }}\,\log \frac{M_{\pi ^{0}}}{M_{\pi}
}\,, 
\nonumber\\
\kappa _{+;0}^{\left( 0\right) }  &=2-\frac{\Sigma _{\pi }}{\Delta _{\pi }}%
\,\log \frac{M_{\pi ^{0}}}{M_{\pi }}\,, 
\nonumber\\
\kappa _{+;0}^{\left( 1\right) } &=\frac{1}{4}\left[ 3\left(
M_{P}^{2}-s\right) +4M_{\pi }^{2}+M_{\pi ^{0}}^{2}\right]  
\nonumber\\
&-\frac{1}{2}\frac{\Sigma _{\pi }}{\Delta _{\pi }}\log \frac{M_{\pi ^{0}}}{%
M_{\pi }} 
\left( 8\frac{M_{\pi }^{2}M_{\pi ^{0}}^{2}}{\Sigma _{\pi }}+\Delta
_{P\pi ^{0}}-s\right)\!,  
\nonumber\\
\kappa _{+;0}^{\left( 2\right) } &=\frac{1}{36}\bigg[16s^{2}-s(32\Delta
_{P\pi ^{0}}+61\Sigma _{\pi })+16\Delta _{P\pi ^{0}}^{2}  
\nonumber\\
&\quad+\Delta _{P\pi ^{0}}^{2}(77M_{\pi }^{2}+45M_{\pi ^{0}}^{2}) 
\nonumber\\
&\quad+4(9\Sigma _{\pi }^{2}+4M_{\pi }^{2}M_{\pi ^{0}}^{2})-16M_{\pi }^{2}%
\,\frac{\Delta _{P\pi ^{0}}^{2}}{s}\bigg]  
\nonumber\\
&-\frac{1}{6}\frac{\Sigma _{\pi }}{\Delta _{\pi }}\log \frac{M_{\pi ^{0}}}{%
M_{\pi }}\bigg[ 2s^{2}+2\Delta _{P\pi ^{0}}^{2}  
\nonumber\\
&\quad -s\left( 4\Delta _{P\pi ^{0}}+5\Sigma _{\pi }+12\,\frac{M_{\pi
}^{2}M_{\pi ^{0}}^{2}}{\Sigma _{\pi }}\right)  
\nonumber\\
&\quad+\Delta _{P\pi ^{0}}\left( 3\Sigma _{\pi }+4\,\frac{M_{\pi
}^{2}+4M_{\pi ^{0}}^{2}}{\Sigma _{\pi }}\right)  
\nonumber\\
&\quad+32M_{\pi }^{2}M_{\pi ^{0}}^{2}-3M_{\pi }^{2}\,\frac{\Delta _{P\pi
^{0}}^{2}}{s}\bigg],  
\nonumber\\
\kappa _{+;1}^{\left(- 2\right) } &=\frac{1}{2\Delta _{P\pi ^{+}}\Delta
_{\pi }}\frac{1}{s-\xi }\bigg[ \frac{s}{2}  
-\frac{2\Delta _{P\pi ^{0}}M_{\pi }^{2}-s\Delta_{\pi}}{\Delta_{\pi}}
\nonumber\\
&\quad\times\left( \frac{\Sigma _{\pi }}{\Delta _{\pi }}-\frac{s}{s-\xi }\frac{%
s-2\Sigma _{\pi }-\Delta _{P\pi ^{0}}}{\Delta _{P\pi ^{+}}}\right)
\!\bigg],
\nonumber\\
\kappa _{+;1}^{\left( -1\right) } &=\frac{2\Delta _{P\pi ^{0}}M_{\pi
}^{2}-s\Delta _{\pi }}{2\Delta _{P\pi ^{+}}\Delta _{\pi }\left( s-\xi
\right) }  , \quad
\kappa _{+;1}^{\left( 0\right) } = \frac{1}{2}\, ,
\nonumber\\
\kappa _{+;1}^{\left( 1\right) } &=\frac{1}{4s} \left( s - 2M_{\pi }^{2}\right) 
\left( \Delta _{P\pi^{0}} - s\right),  
\nonumber\\
\kappa _{+;1}^{\left( 2\right) } &=\frac{1}{12}\bigg\{\!\left( s-\Delta
_{P\pi ^{0}}\right) \bigg[ 2s+2\,\frac{\Delta _{P\pi ^{0}}}{s}-6\,\frac{\Delta
_{P\pi ^{+}}}{s}  
\nonumber\\
&\quad -\left( 2M_{P}^{2}+9M_{\pi }^{2}+3M_{\pi ^{0}}^{2}\right)\! 
\bigg]  -2M_{\pi ^{0}}^{2}\frac{\Delta _{P\pi ^{0}}}{s}\bigg\},  
\nonumber\\
\kappa _{+;2}^{\left(- 2\right) } &=\frac{1}{4\Delta _{P\pi ^{+}}\Delta
_{\pi }}\left[ s-\Delta _{P\pi ^{0}}
+\frac{\Delta _{P\pi ^{0}}\Sigma _{\pi }-s\Delta _{\pi }}{%
\Delta _{\pi }}\right.  
\nonumber\\
&\quad\left.\times\left( \frac{\Sigma _{\pi }}{\Delta _{\pi }}-\frac{s}{s-\xi }\frac{%
s-2\Sigma _{\pi }-\Delta _{P\pi ^{0}}}{\Delta _{P\pi ^{+}}}\right) \right], 
\nonumber\\
\kappa _{+;2}^{\left( -1\right) } &=-\frac{\Delta _{P\pi ^{0}}\Sigma _{\pi
}-s\Delta _{\pi }}{2\Delta _{P\pi ^{+}}\Delta _{\pi }\left( s-\xi \right) }\,,
\nonumber\\
\kappa _{+;2}^{\left( 0\right) }  &= \frac{1}{2}\left( 1-\frac{\Delta _{P\pi
^{0}}}{s}\right),  
\nonumber\\
\kappa _{+;2}^{\left( 1\right) } &=-\frac{1}{4}\left( s - 2M_{P}^{2}+%
\frac{\Delta _{P\pi ^{0}}}{s}\right),  
\nonumber\\
\kappa _{+;2}^{\left( 2\right) } &=\frac{1}{12}\bigg\{
6M_{\pi ^{0}}^{2}\Sigma _{\pi }\frac{\Delta _{P\pi ^{0}}^{2}}{s}
+\left( 1-\frac{%
\Delta _{P\pi ^{0}}}{s}\right)  
\nonumber\\
&\quad\times \bigg[ 2s^{2}-s\left( 4M_{P}^{2}+5\Sigma _{\pi }\right)
+2\Delta _{P\pi ^{0}}^{2}  
\nonumber\\
&\quad+\Delta _{P\pi ^{0}}\left( 7M_{\pi }^{2}+3M_{\pi
^{0}}^{2}\right) +6M_{\pi ^{0}}^{2}\Sigma _{\pi }  
\nonumber\\
&\quad+2\left( 3M_{\pi ^{0}}^{2}\Sigma _{\pi }-M_{\pi }^{2}\Delta
_{P\pi ^{0}}\right) \frac{\Delta _{P\pi ^{0}}}{s}\bigg] \bigg\},  
\nonumber\\
\kappa _{+;3}^{\left( -2\right) } &=0 , \quad
\kappa _{+;3}^{\left( -1\right) } = -\frac{1}{\Sigma _{\pi }}\, , \quad
\kappa _{+;3}^{\left( 0\right) } = 1,
\nonumber\\
\kappa _{+;3}^{\left( 1\right) } &=2\,\frac{M_{\pi }^{2}M_{\pi ^{0}}^{2}}{%
\Sigma _{\pi }}\, , \quad
\kappa _{+;3}^{\left( 2\right) } = 2M_{\pi }^{2}M_{\pi ^{0}}^{2},
\nonumber\\
\kappa _{+;4}^{\left( -2\right) } &= 2\,\frac{M_{\pi }^{2}M_{\pi ^{0}}^{2}}{%
\Delta _{\pi }^{4}} , \quad
\kappa _{+;4}^{\left( -1\right) } =\frac{\Sigma _{\pi }}{\Delta _{\pi }^{2}%
}\, , \quad
\kappa _{+;4}^{\left( 0\right) }  = -1,
\nonumber\\
\kappa _{+;4}^{\left( 1\right) } &=
\kappa _{+;4}^{\left( 2\right) } = 0
.
\end{align}
Let us note that the single and the double poles that appear in the coefficients $\kappa_{+;i}^{(n)}$ for $s=\xi$ 
for $n<0$, are in fact spurious artifacts corresponding to the partition of the integrals in Eq.~(\ref{Jbar_integrals_x}) 
into the individual terms. In the full sum on the right-hand side of (\ref{Jbar_integrals_x}), the various pole 
contributions cancel each other.
Note also that the analogous integrals which are necessary for the calculation of the dispersive parts of the  NLO partial-wave 
projections of the pion scattering amplitudes $A_{00}$ and $A_x$, and which are given explicitly in Section IV.B of Ref.~\cite{DescotesGenon:2012gv},  
can be formally obtained from the above formulas in the limit $M_P\to M_{\pi^0}$. This limit requires, however, enlarging the 
definition of the function $L(s)$ to the region $4M_{\pi^0}^2<s<4M_\pi^2$, namely, to set in this region
\begin{equation}
L(s)=\log\left(1-\sigma(s)\right)-\log\left(1+\sigma(s)\right)
,
\end{equation}
where we assume the principal branch of the logarithm with a cut on the interval $(-\infty,0)$ 
along the real axis, and $\log (x)=\log(x+{\rm i}0)$ for $x<0$.

Using the above results for the integrals (\ref{Jbar_integrals_pi0}), (\ref{Jbar_integrals_mixed}) and (\ref{Jbar_integrals_x}), 
it is now a straightforward task to calculate the corresponding $S$-wave projections of the amplitudes ${\cal M}^L_{00}$, 
${\cal M}^L_{x}$ at NLO and, with the help of (\ref{W00_2loop}),  to construct  the absorptive part of the function ${\cal W}_{00}$.
The construction of the full two-loop amplitude ${\cal M}^L_{00}$ then proceeds along the same lines as in the case of equal-mass 
pions described in detail in the previous section. Despite the higher algebraic complexity, from a numerical perspective, 
the dispersive integrals similar to the ones that define the functions ${\widetilde K}_i(s)$, but now featuring
the functions ${\widetilde k}_{\nabla ;i}(s)$ and ${\widetilde k}_{+ ;i}(s)$ in the role of absorptive parts,
present no particular problem.

\section{Summary and conclusions}
\setcounter{equation}{0}

The primary purpose of this study was to present a detailed account of the dispersive 
approach to the construction of $P\pi\to\pi\pi$, $P=K^\pm, K_L, K_S, \eta$, scattering 
amplitudes that possess all the correct analytic properties at the order two loops
in the low-energy expansion. It generalizes the representation of the two-loop
$\pi\pi$ amplitudes first constructed in Ref.~\cite{Stern:1993rg} to the case
where the masses are distinct, and with one of the mesons ($P$) unstable 
through decay into three pions. As compared to $\pi\pi$ scattering, this 
last aspect makes the discussion of the analytic properties 
significantly more involved. We have, therefore, tried to provide the 
necessary information on these aspects. Most notably, we have 
extended the existing discussions \cite{BartonKacser,Bronzan} to the case 
where the charged and the neutral pions have different masses. The most 
remarkable feature is the apparition of an anomalous threshold as soon as 
the final state contains charged pions. This requires a modification 
of both the dispersion relations that provide the starting point of our 
construction and of the manner in which the projection on the partial waves
of the one-loop amplitudes is performed. We plan to come back to these
delicate issues in a forthcoming paper.

We would like to point out that our approach applies as soon as
an expansion under which the counting rules (\ref{chibeh}) are valid
is available. This is, in particular, the case of the combined
chiral and $1/N_C$ expansion \cite{GL1,HerreraSiklody:1996pm,Kaiser:2000gs}. Within this framework,
our construction would apply to further processes, like $\eta'\to\eta\pi\pi$,
which was recently studied in high-precision experiments \cite{Ablikim:2010kp,Ablikim:2017irx,Adlarson:2017wlz}.

From a practical point of view, the two-loop amplitudes constructed this 
way depend on a certain number of subtraction constants, which can be put 
in one-to-one correspondence with the Dalitz-plot parameters (slopes and
curvatures). These representations could, therefore, be used in order to  
analyze experimental high-statistics data for the decay distribution of the $P\to\pi\pi\pi$
processes. The number of parameters to be fitted, for instance, is the same
as in the usual Dalitz-plot expansions,
but the inclusion of the correct analytic properties might allow for 
better fits. Alternatively, these representations can also be 
useful in order to extract information on fundamental quantities, like the 
quark-mass ratio $R$, or the $\pi\pi$ scattering lengths, from the data.
In the former case, we have already illustrated this in Ref.~\cite{Kampf:2011wr},
and we plan to redo a similar analysis using more recent high-statistics 
data on the Dalitz-plot distribution of $\eta\to\pi\pi\pi$ 
\cite{Adlarson:2014aks,Anastasi:2016cdz,Ablikim:2015cmz,Prakhov:2018tou}.

\begin{acknowledgments}
This work is supported in part by grant No.~40652ZE, financed by the French Ministry of Foreign Affairs and the French Ministry 
for Higher Education, Research and Innovation, and by grant No.~8J18FR039, financed by the Czech Ministry of Education, 
Youth and Sports, both within the framework of the bilateral PHC Barrande Mobility Program 2018, and by the Czech Science Foundation Project No.~GACR 18-17224S. 
We also acknowledge useful discussions with J. Gasser, B. Kubis, H. Leutwyler and E. Passemar.
\end{acknowledgments}

\appendix

\renewcommand{\theequation}{\Alph{section}.\arabic{equation}}

\section{Anomalous thresholds}\label{app:anomalous}

\begin{figure}[b]
\begin{center}
  \includegraphics[width=0.40\textwidth]{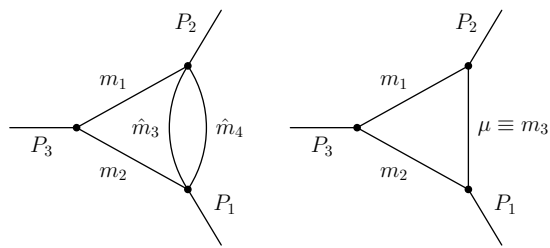}\\
  \caption{The general diagram with the fish topology (left) and its
  reduction to the corresponding triangle diagram (right).}\label{triangle}
\end{center}
\end{figure}

In this appendix, we summarize the analysis of the physical-sheet
singularities of the diagrams with the ``fish'' topology, see Fig.~\ref{triangle}.
Whereas the analysis of Refs.~\cite{BartonKacser,Bronzan} addresses the situation 
where all the pions have the same mass, the more general analysis presented here holds 
for any combinations, allowed by conservation of the electric charge, of charged 
and neutral pions as external and internal states. 
This analysis rests on the study of the Landau singularities \cite{Landau:1959fi,Nakanishi:1959}, 
which is summarized in the monography \cite{Eden:1966dnq}. For more details of this analysis, we also refer to Ref.~\cite{Zdrahal}.

We first recall that, as far as the structure of the singularities is concerned,
one can consider the following dispersive representation of the fish diagram
in terms of the standard triangle diagram \cite{BartonKacser,Gasser:1998qt},
\begin{equation}
\int\limits_{\mu_0^2}^\infty 
d\mu^2 \rho(\mu^2)\int\frac{\mathrm{d}^4q_1}{(2\pi)^{4}}\,\frac{1}{(k_1^2-m_1^2)(k_2^2-m_2^2)(k_3^2-\mu^2)}\,
,
\end{equation}
where the possible values of $\mu_0$ are $2 M_{\pi^0}$, $M_\pi + M_{\pi^0}$, or $2 M_\pi$. 
The precise form of the spectral density $\rho(\mu^2)$ is not important here.
It suffices to know that it provides an adequate renormalization of the ultra-violet
divergence in the sub-graph, but brings in no further singularity, the only additional
singularity being a possible endpoint singularity at the lower end of the 
$\mu^2$ integration. After the introduction of Feynman parameters and integration
over $q_1$, one obtains an integrand whose denominator ${\cal D}$ reads \cite{Petersson:1965zz,tHooft:1978jhc}
\begin{equation}
- {\cal D} = \beta^T Y \beta - i \epsilon,
\end{equation}
where $\beta^T = ( \beta_1 , \beta_2 , \beta_3)$, $0\le \beta_i \le m_i$, and $Y$ is the 
symmetric $3 \times 3$ matrix with entries
\begin{equation}
y_{ii} = 1,\quad y_{ij} = \frac{m_i^2 + m_j^2 - P_k^2}{2 m_i m_j}\,,\ i\neq j,\ k\neq i,j
.
\end{equation}
In what follows, the masses $m_i$ are real and strictly positive numbers.
The virtualities $P_k^2$ of the external lines can be arbitrary complex numbers.
Then, the path of integration over the $\beta_i$ variables need not be distorted if one
of the $y_{ij}$, $i\neq j$, is kept real, while the remaining two are given
negative imaginary parts. In this setting, the $i\epsilon$ contribution to ${\cal D}$ is not
necessary, provided ${\cal D}$ is defined as the boundary value when these imaginary parts
tend to zero \cite{Eden:1966dnq}.

The Landau conditions \cite{Landau:1959fi,Nakanishi:1959} read
\begin{align}
\beta_i\,\frac{\partial {\cal D}}{\partial\beta_i} &= 0 \ \text{ for each internal line } i=1,2,3,~~
\\
{\cal D} &= 0.
\end{align}
The first three equations can be rewritten as
\begin{equation}
\left(
\begin{tabular}{ccc}
$\beta_1$  &  0  &  0
\\
0  &  $\beta_2$  &  0
\\
0  &  0  &  $\beta_3$
\end{tabular}
\right)
Y \beta = 0
.
\end{equation}
\noindent
The leading Landau singularity (LLS), corresponding to $\beta_1\beta_2\beta_3 \neq 0$, 
is then given by the condition 
\begin{equation}
\det Y = 1 + 2 y_{12}y_{23}y_{31} - y_{12}^2 - y _{23}^2 - y_{31}^2 = 0
.
\label{leading_sing}
\end{equation}
The non-leading Landau singularities (NLLSs), corresponding to the vanishing of exactly one $\beta_i$,
require $y_{jk} = \pm 1$, $j\neq k$, $i \neq j,k$. They represent both normal and
anomalous thresholds. Finally, the second-type (or non-Landau) singularity (NLS) curve is
given by
\begin{equation}
\lambda(P_1^2,P_2^2,P_3^2) = 0
\label{non-Landau}
.
\end{equation}
Not all the singularities derived from the Landau conditions do occur on the physical sheet.
Before starting a more detailed analysis for the identification of the physical-sheet 
Landau singularities,
it is useful to identify the domains where ${\cal D}$ never vanishes in the undistorted region of parametric integration.
Apart from the case already mentioned, when two $y_{ij}$ variables are given negative imaginary parts,
such singularity-free domains are, for instance,
\begin{itemize}
\item[SF1:] domains where all the $y_{ij}$ are real and positive; this means (for all $m_i>0$) that all $P_k^2\le m_i^2+m_j^2$;
\item[SF2:] domains where, simultaneously, one of the variables $y_{ij}$, $i\neq j$, is greater than unity, a second
one is greater than zero, and the third one strictly greater than $-1$.
\end{itemize}

\begin{figure}[t]
\begin{center}
  \includegraphics[width=0.48\textwidth]{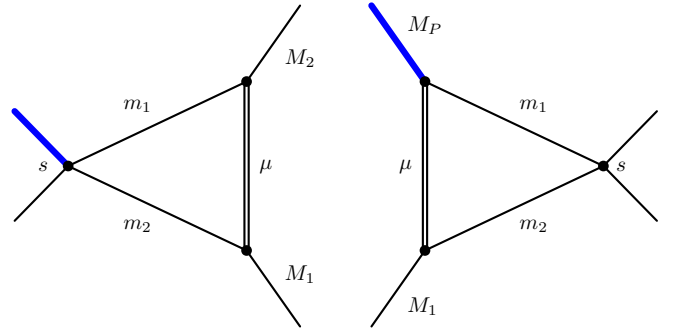}\\
  \caption{The two types of fish/triangle diagrams contributing to the $P\pi\to\pi\pi$
  amplitudes. On the left diagram, the heavy external meson line is combined with 
  an external pion line into the kinematic variable $s$ ($\pi$-diagram).
  On the right diagram, the heavy meson line is an isolated external line ($P$-diagram).
  All external and internal lines are charged or neutral pion lines, except for 
  the thick external line corresponding to the $P$ meson, and the internal 
  double line, which denotes to the dispersive loop corresponding to the exchange of 
  a pion pair.}\label{triangle2}
\end{center}
\end{figure}

We will split the discussion upon considering two types of triangle diagrams, see Fig.~\ref{triangle2}.
In the first one, we will call it a $\pi$-diagram, the on-shell conditions for 
the external lines are, say, $P_i^2=M_i^2$, $i=1,2$, where $M_1$ and $M_2$ are not
necessarily equal, but correspond to a pion mass, $M_\pi$ or $M_{\pi^0}$, whereas 
$P_3^2=s$ is a free variable. The second case corresponds to, say, $P_2^2=M_P^2$,
whereas $P_1^2=M_1^2$ and $P_3^2=s$ are as before. We refer to this situation 
as a $P$-diagram. In both cases, the internal lines
are restricted to neutral or charged pions. A brief discussion of other intermediate
states is to be found at the end of this appendix. Furthermore, since we are interested
in the singularities as $s$ varies in the complex plane, all other quantities, like
$M_1$ and $M_2$ will be kept at their physical values, unless we are forced to modify
them. In particular, unless unavoidable, we will keep the integration over $\mu^2$ fixed 
to occur along the straight line $\mu^2 \ge \mu_0^2$ on the real axis. Since in any case
we have $\mu^2\ge 4 M_{\pi^0}^2 > \Delta_\pi$, we observe that the condition
\begin{equation}
y_{23} \equiv \frac{m_2^2 + \mu^2 - M_1^2}{2 m_2 \mu} > 0
\label{y23_cond}
\end{equation}
is always satisfied.

\subsection{Physical-sheet singularities of the $\pi$-diagrams}

The analysis for the $\pi$-diagrams, which, since the momentum
of the meson $P$ is hidden inside the variable $s$, will also 
hold for $\pi\pi$ scattering, turns out to be quite straightforward.
Indeed, in this case, we also have
\begin{equation}
y_{13} \equiv \frac{m_1^2 + \mu^2 - M_2^2}{2 m_1 \mu} > 0
.
\label{y13_cond}
\end{equation}
Combined with the conditions (\ref{y23_cond}) and SF1, we at once
conclude that we need to worry only about the singularities 
occurring for $s > m_1^2+m_2^2$.

Next, conservation of the electric charge tells us  
either that $M_1=M_2$ and $m_1=m_2$ hold simultaneously, or that $M_1\neq M_2$ 
and $m_1\neq m_2$ hold simultaneously. It leads only to four
possibilities for the charge assignments of the corresponding states: 
\begin{itemize}
\item 
$M_1=M_2=m_1=m_2\equiv M$: Then, we study the amplitudes with fixed $\mu\ge2\mpn$, 
and the singularity-safe domain for the physical amplitude is for $s\le 2 M^2$. The LLS 
occurs for $s=0$ and $s=4M^2-\mu^2$. The NLLSs are $s=0$ and $s=4 M^2$, from $y_{12}=\pm 1$,
and $\mu^2=4 M^2$ from $y_{13, 23}=+1$, the case $y_{13, 23}=-1$ being excluded
for positive masses, cf. Eqs.~(\ref{y23_cond}) and (\ref{y13_cond}). There appears no 
new singularity from pinching at infinity, Eq.~(\ref{non-Landau}). 
At $\mu^2=4 M^2$, there is no singularity on the physical sheet. As expected, it means that the 
only relevant singularities are in $s$, and since only the normal threshold singularity $s=4 M^2$ 
does not belong to the safe region in $s$, the only relevant singularity for the physical 
amplitudes is this normal threshold.
\item 
$M_1=M_2\equiv M$, $m_1=m_2\equiv m$, $M\neq m$: $\mu$ is fixed with $\mu\ge m+M=\mpn+\mpp$, so that
$y_{13}\ge 1$, $y_{23} \ge 1$ and the condition SF2 is satisfied for $s < 4m^2$.
Therefore, physical-sheet singularities only occur for $s \ge 4m^2$. On the other hand,
the possible singular points are: the LLS at
\begin{equation}
s=0,\text{ or } s=-\frac{\lambda(\mu^2,m^2,M^2)}{\mu^2}<0
;
\end{equation}
the NLLSs at $s=0$, $s=4m^2$ (from $y_{12} = \pm 1$), or for $\mu^2=(M \pm m)^2$; and finally, 
there appears, in addition, a NLS at $s=4M^2$. Therefore, in this case, in addition 
to the normal threshold $s=4m^2$, there occurs a NLS at $s=4M^2$, i.e., at the beginning of 
the physical region, provided $M=\mpp$ and $m=\mpn$.
\item 
$M_1=m_1=\mpp$, $M_2=m_2=\mpn$: again $\mu\ge\mpn+\mpp$, and no singularity appears on the 
physical sheet for $s < (\mpp+\mpn)^2$. All the possible singularities are
\begin{equation}
s=\frac{\Delta_\pi^2}{\mu^2} 
, \quad
s=2\Sigma_\pi-\mu^2
,
\end{equation}
both bounded from above by $(M_\pi - M_{\pi^0})^2$, from the LLS, and
\begin{equation}
\mbox{}\quad s=(M_\pi \pm M_{\pi^0})^2, \quad \mu^2=(M_\pi \pm M_{\pi^0})^2
\end{equation}
from the NLLSs and NLS.
Thus, on the physical sheet, we find only the normal threshold singularity at $s=(M_\pi + M_{\pi^0})^2$.
\item 
$M_1=m_2=\mpp$, $M_2=m_1=\mpn$: $\mu\ge2\mpn$, $y_{13} \ge 1$, $y_{23} > 0$, and thus
no singularity occurs on the physical 
sheet for $s <(\mpp+\mpn)^2$. The singularities in this case read
\begin{equation}
\mbox{}\qquad
s=\Sigma_\pi - \frac{\mu^2}{2} \pm \frac{\sqrt{\left(\mu^2-4\mpp^2\right)\left(\mu^2-4\mpn^2\right)}}{2}
\end{equation}
from the LLS, with both solutions smaller than $\Sigma_\pi < (\mpp+\mpn)^2$ whenever they are real,
and
\begin{equation}
\mbox{}\qquad\ s=(M_\pi \pm M_{\pi^0})^2, 
\quad \mu^2 \in \{ 0 , 4M_\pi^2 , 4 M_{\pi^0}^2 \}
\end{equation}
from the NLLSs, with no new constraint from the NLS.
The only singularity lying on the physical sheet is the normal threshold $s=(M_\pi + M_{\pi^0})^2$.
\end{itemize}

In conclusion, the only $\pi$-diagram where there appears an anomalous threshold on the physical 
sheet is the diagram with $M_1=M_2=\mpp$, $m_1=m_2=\mpn$ and $\mu_0=\mpp+\mpn$, which has 
a non-Landau singularity at $s=4\mpp^2$, close to the beginning of the physical region $s\ge 4 M_{\pi^0}^2$.

\subsection{Physical-sheet singularities of the $P$-diagrams}

We next consider the second type of diagrams from Fig.~\ref{triangle2}, which we call $P$-diagrams 
and which needs more careful analysis. Since $\mu_0 \ge 2 M_{\pi^0} > \Delta_\pi$, the condition
(\ref{y23_cond}) is always satisfied. The singularities are most conveniently discussed
through their localization on curves lying in the $(M_P^2,s)$ plane.

In the following, we denote $M_1 \equiv m_5$.
For the LLS, we obtain from Eq.~(\ref{leading_sing})
[$\Delta_{25} = m_2^2 - m_5^2$, $\Delta_{P1} = M_P^2 - m_1^2$]
\begin{multline}\label{Landau P triangle}
\Sigma:\ 2\mu^2 s = -\mu^4+\mu^2(M_P^2+m_1^2+m_2^2+m_5^2)+\Delta_{25}\Delta_{P1}\hspace{1em}
\\
\pm\lambda^{1/2}(\mu^2,M_P^2,m_1^2)\lambda^{1/2}(\mu^2,m_2^2,m_5^2).
\end{multline}
The subleading singularities read
\begin{alignat}{2}
&
\sigma_{s\pm}:&\ s &= (m_1\pm m_2)^2, 
\nonumber\\
&
\sigma_{P\pm}:&\ M_P^2&=(\mu\pm m_1)^2, 
\nonumber\\
&
\sigma_{\mu\pm}:&\ \mu^2&=(m_2\pm m_5)^2.
\end{alignat}
The second-type singularity may occur on the curve
\begin{equation}
\Gamma:\ s=(M_P\pm m_5)^2.
\end{equation}

As before, in the end, we are interested in the analytic properties in $s$ with $M_P$ and all the 
other masses fixed at their physical value. However, in the case $M_P>3M_\pi$, we need to perform 
an analytic continuation in some other variable from the values where the diagram is analytic. 
Inspired by the analysis of Kacser and Bronzan \cite{Bronzan,Kacser}, we start by considering
an analytic continuation in the external variables $P_i^2$, and we deform the 
integration contour in $\mu$ from the original line $\mu\ge\mu_0$  
only when forced to do so. The singularity curve $\sigma_{\mu-}$
is, therefore, again irrelevant, and the curve $\sigma_{\mu+}$, corresponding to the anomalous subleading 
threshold, can be avoided, in the cases of interest here, by the addition of a small imaginary part to $\mu^2$ 
without any change of the analytic structure in $s$.

For the diagrams under consideration, the domains where the $\beta$ integrations do not need to be 
deformed from the original physical integration contour are the following:
\begin{enumerate}
\item[(a)] $\im s>0$ together with $\im M_P^2>0$;
\item[(b)] SF1 holds; the condition (\ref{y23_cond}) holds automatically, 
and the domain $y_{12}, y_{13} \ge0$ corresponds to 
$s\le m_1^2+m_2^2$ together with $M_P^2\le\mu^2+m_1^2$;
\item[(c)] SF2 holds; this happens, in particular, when
\begin{itemize}
\item[-~(c1)] $\mu\ge(m_5+m_2)$ and $M_P^2\le\mu^2+m_1^2$; we can go with $s$ up to $s <(m_1+m_2)^2$;
\item[-~(c2)] $\mu\ge(m_5+m_2)$, $M_P^2 < (\mu+m_1)^2$, and $s\le m_1^2+m_2^2$;
\item[-~(c3)] $M_P^2\le(\mu-m_1)^2$, and $s < (m_1+m_2)^2$;
\item[-~(c4)] $M_P^2\le(\mu+m_1)^2$, and $s\le (m_1-m_2)^2$.
\end{itemize}
\end{enumerate}
Therefore, we see that the anomalous thresholds $\sigma_{s-}$ and $\sigma_{P-}$ bring no singularity to the physical sheet, but for $M_P>3M_\pi$, 
there are always some parts of $\Sigma$ and $\Gamma$ which do not belong to the regions (b) or (c), and thus we need to perform
the analysis very carefully, in order to see whether it is possible to continue the amplitude there without the appearance of singularities.
The answer depends on the relative positions of the individual singularity curves.
An important observation is that the remaining curves of potential singularities meet only at the following points
\begin{align} 
\intertext{$A_{1,2} = \Sigma\cap\Gamma:$}
M_P^2 &=\mu^2+m_1^2 + \frac{\lambda(\mu^2,m_2^2,m_5^2)}{2m_5^2}
\nonumber\\
&\pm\frac{(\mu^2+m_5^2-m_2^2)\sqrt{\lambda(\mu^2,m_2^2,m_5^2)+4m_1^2m_5^2}}{2m_5^2},
\nonumber\\ 
s&=s(A_{1,2}),
\end{align}
\begin{align}
\intertext{$B=\Sigma\cap\sigma_{P+}:$}
M_P^2 &= (\mu+m_1)^2, 
\ s=m_1^2+m_2^2+\frac{m_1}{\mu}(\mu^2+m_2^2-m_5^2),
\end{align}
\begin{align}
\intertext{$C=\Sigma\cap\sigma_{s+}:$}
M_P^2 &= \mu^2+m_1^2+\frac{m_1}{m_2}(\mu^2+m_2^2-m_5^2), 
\nonumber\\
s&=m_1^2+m_2^2+\frac{m_1}{\mu}(\mu^2+m_2^2-m_5^2),
\end{align}
\begin{align}
\intertext{$D_{1,2}=\Gamma\cap\sigma_{P+}:$}
M_P^2 &= (\mu+m_1)^2, 
\quad s=(\mu+m_1\pm m_5)^2,
\end{align}
\begin{align}
\intertext{$E_{1,2}=\Gamma\cap\sigma_{s+}:$}
M_P^2 &= (m_1+m_2\pm m_5)^2, 
\quad s=(m_1+ m_2)^2.
\end{align}

\begin{figure}[t]
\begin{center}
  \includegraphics[width=0.48\textwidth]{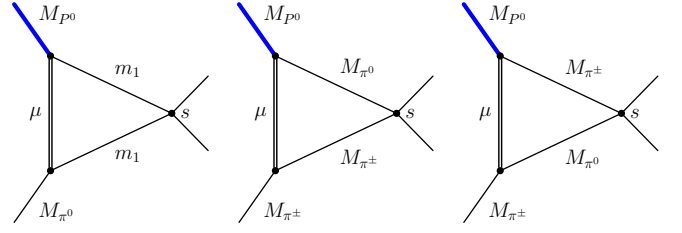}\\
  \caption{The $P$-diagrams contributing to the $P^0\pi\to\pi\pi$
  amplitudes. The assignments of the lines are as in Fig.~\ref{triangle2}.}\label{triangle3}
\end{center}
\end{figure}

We proceed with the analysis of the individual diagrams. 
The distinct types of diagrams with the neutral 
$P^0$ are displayed in Fig.~\ref{triangle3}.

\subsubsection{Analytic properties of the first diagram from Figure~\ref{triangle3}}

\begin{figure}[t]
\begin{center}
  \includegraphics[width=0.48\textwidth]{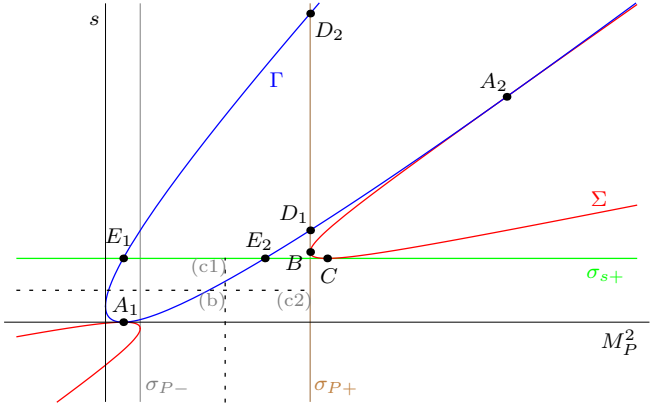}\\
  \caption{The real sections of the singularity curves for the diagrams without
  the occurrence of the anomalous threshold on the physical Riemann sheet.
  For the labels of the curves, as well as for those of the points of intersections, 
  see the main text}\label{Land-triangle1}
\end{center}
\end{figure}

For the first of these diagrams ($m_2=m_1$, $m_5=\mpn$), the singularity curves are
[$\Delta_{10} = m_1^2 - M_{\pi^0}^2$]:
\begin{enumerate}
\item LLS curve $\Sigma$:
\begin{align}
2\mu^2 s &= -\mu^4+\mu^2(M_P^2+\mpn^2+2m_1^2)+\Delta_{10}\Delta_{P1}
\nonumber\\
&
\pm\lambda^{1/2}(\mu^2,M_P^2,m_1^2)\lambda^{1/2}(\mu^2,\mpn^2,m_1^2).
\end{align}
\item NLLS curves:
\begin{gather}
\sigma_{s-}:\ s=0, \qquad \sigma_{s+}:\ s=4m_1^2, \nonumber\\
\sigma_{P\pm}:\ M_P^2=(\mu\pm m_1)^2.
\end{gather}
\item NLS curves:
\begin{equation}
\Gamma:\ s=(M_P\pm \mpn)^2.
\end{equation}
\end{enumerate}
The integration contour for the dispersive loop is the line $\mu\ge(\mpn+m_1)$.
For $m_1=\mpn$, the situation simplifies into the one studied by Kacser and Bronzan in \cite{Kacser,Bronzan}.
However, the relative position of the curves is the same also for $m_1=\mpp$ as is depicted in 
Fig.~\ref{Land-triangle1}, and one therefore
expects that also the singularity structure will remain the same.

Since $\mu\ge(M_{\pi^0}+m_1)$, the denominator ${\cal D}$ of the parametric integrand does not vanish for $\beta_i\ge0$ 
also in the regions (c1) and (c2), and the contribution
of this diagram is without singularities on the physical sheet for all $s$ and $M_P^2$ on the left of or below  
the dashed lines in Fig.~\ref{Land-triangle1}.
Since the normal-threshold lines $\sigma_{s+}$ and $\sigma_{P+}$ correspond to the singularity curves 
with one of the $\beta_i$ equal to zero and
the other two positive, the only part of the real section of the singularity curve $\Sigma$ where ${\cal D}$ 
vanishes for all $\beta_i>0$ is the arc between the points $B$ and $C$, and
this remains true for the appropriate complex surface connected to this real arc. All the 
other parts of $\Sigma$ are non-singular since we can continue
the integral analytically in the following way. We start in the domain (c1), where the 
integration contour in $\beta_i$ is the original one, and we add a small positive imaginary
part to both $s$ and $M_P^2$ [coming so to the domain (a)], which is non-singular for any 
values of $s$ and $M_P^2$ without the need of deformation of the contour. This way,
we can reach any point where the singularity does not occur for positive $\beta_i$, 
upon letting the added imaginary parts tend to zero (without deformation of the integration curve).

However, the arc $BC$ is connected to the non-singular lower-left part of $\Sigma$ by a 
continuous complex part of $\Sigma$ \cite{Eden:1966dnq}. Since all intersections of $\Sigma$ with the other 
curves are just the real points from above, by performing the analytic continuation from 
the lower-left part by the path along $\Sigma$, we do not pass through any branch cut 
until we come to the real arc $BC$, where either $\sigma_{P+}$ or $\sigma_{s+}$ have 
to be crossed. Similarly, in all the points of $\Gamma$, the loop integral can be defined 
by analytic continuation along this complex curve. Therefore, there are no complex singularities 
in this case and the only singularities occurring for the diagram considered on the physical 
sheet are the normal thresholds $\sigma_{s+}$ and $\sigma_{P+}$, together with the anomalous threshold 
on the real arc $BC$, cf. Ref.~\cite{Eden:1966dnq}.
[Note that on the section $BC$, the loop integral can be made analytic without the appearance of singularities by 
deforming the integration contour of the parametric integration to $\beta_1\beta_2<0$.]

The singularity at this anomalous threshold occurs only for 
$M_P^2\in\left((\mu+m_1)^2,2\mu^2+2m_1^2-\mpn^2\right)$. Upon adding to $\mu^2$ a small 
negative imaginary part, we can avoid this singularity. The only point where this is not 
possible is the endpoint of the integration, $\mu=\mpn+m_1$. 

In conclusion, for the fish 
diagram connected with the first diagram of Fig.~\ref{triangle3}, the 
anomalous threshold singularity in the $s$-plane appears only for $M_P^2=(\mpn+2m_1)^2$. 
Since $M_P > 3.5 M_\pi$ for the physical values of the masses of the $K^\pm$, $K_L$, $K_S$ 
and $\eta$ masses, this condition is never fulfilled.\footnote{Note that for the physical masses and 
for $\mu$ corresponding to the endpoint, even the extremal value of the position of $C$, 
$M_P^2=5\mpp^2+\frac{\mpp}{\mpn}\left(5\mpp^2-\mpn^2\right)$ corresponds to $M_P\approx423\,\mathrm{MeV}$. 
For the kaons and the eta, we can, therefore, altogether ignore this complication with the $BC$ section.} 
For the physical masses, the 
only singularities appearing for this diagram are, therefore, just the regular normal thresholds. 
The important observation is that we obtain the correct physical analytic continuation of the 
amplitude also on the leading singularity curve by taking $M_P^2\to M_P^2+i\delta$, where $\delta$
is a small positive number.

\subsubsection{Analytic properties of the second diagram from Figure \ref{triangle3}}

\begin{figure}[t]
\begin{center}
  \includegraphics[width=0.48\textwidth]{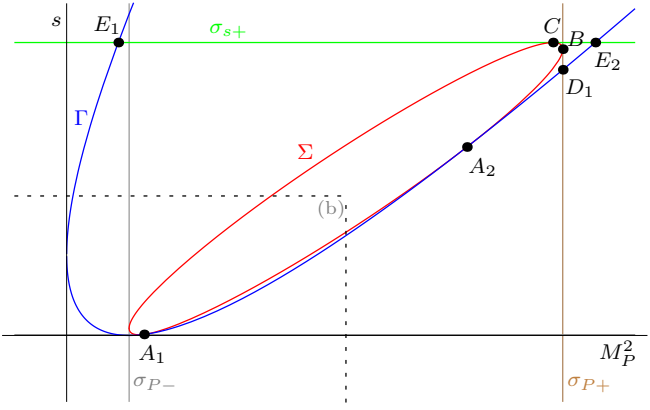}\\
  \caption{The real sections of the singularity curves for the diagrams where
  the anomalous threshold occurs on the physical Riemann sheet.
  For the labels of the curves, as well as for those of the points of intersections, 
  see the main text}\label{Land-triangle2}
\end{center}
\end{figure}

The singularity curves for the second of these diagrams ($m_1=\mpn$, $m_2=m_5=\mpp$) read
\begin{enumerate}
\item LLS curve $\Sigma$:
\begin{equation}\label{anomalni prah}
2s=M_P^2+\mpn^2+2\mpp^2-\mu^2\pm\lambda^{1/2}(\mu^2,M_P^2,\mpp^2)\sigma(\mu^2).
\end{equation}
\item NLLS curves:
\begin{equation}
\sigma_{s\pm}:\ s=(M_\pi \pm M_{\pi^0})^2, \quad \sigma_{P\pm}:\ M_P^2=(\mu\pm\mpn)^2.
\end{equation}
\item NLS curves:
\begin{equation}
\Gamma:\ s=(M_P\pm \mpp)^2.
\end{equation}
\end{enumerate}
For $\mu\ge2\mpp$, the relative position of these curves is again the one depicted in 
Fig.~\ref{Land-triangle1}, and thus, by the same procedure as in the previous 
case, we obtain the contributions whose singularities on the physical sheet are just the 
normal thresholds. However, the integration in the dispersive loop starts at $\mu=2\mpn<2\mpp$.
For these values of $\mu$, the real section of the curves moves into the situation depicted 
in Figure~\ref{Land-triangle2} and the analytic continuation proceeds as follows.
The original integration contour in parametric space is free of singularities in the 
domain (b), i.e., on the left of and below the dashed lines of Figure~\ref{Land-triangle2}.
We can continue the contribution of this diagram along the ellipsis $\Sigma$ further up 
to $B$ and $C$ without the appearance of the singularities on the physical sheet, even without 
deforming the original integration contour, similarly to the previous case, since the only 
part of the real section of $\Sigma$ which corresponds to $\beta_i>0$ is the arc $BC$. 
In order to avoid singularities also on this arc,
we would need to deform the integration contour there. However, all paths from the parts we 
have identified to be non-singular to the arc $BC$ along $\Sigma$ pass through a singularity 
curve, either $\sigma_{P+}$ or $\sigma_{s+}$. Since the curve $\Sigma$ is here real for real 
$M_P^2$, for any complex singularity curve in 
$\beta$-space, there exists a complex conjugated one, and therefore if we want to evade the 
complex singularities on one side, we encounter the complex conjugated one, cf. Ref.~\cite{Eden:1966dnq}. 
We have, therefore, no way how to avoid singularities on the arc $BC$. This diagram thus possesses 
on the physical sheet, in addition to the normal threshold, also the anomalous
threshold (\ref{anomalni prah}) on the arc $BC$ and on all the corresponding complex surfaces.

We can try to avoid these singularities by the addition of a small imaginary part also to $\mu^2$. 
However, this does again not work for the endpoint of the $\mu^2$ integration, in this case for $\mu=2\mpn$. 
Even though the real arc $BC$ is again to the left of the physical value of $M_P^2$ for both the kaons and the eta, 
the singular complex surface connected with this arc extends to the region where $M_P>3\mpn$. 
We are, therefore, left with two complex conjugated anomalous thresholds in $s$ for the physical $M_P>3\mpn$,
\begin{multline}
2s = M_P^2+2\mpp^2-3\mpn^2
\\
\pm\lambda^{1/2}(M_P^2,\mpp^2,4\mpn^2)\sigma(4\mpn^2).
\end{multline}

Thus, for the amplitudes to which this diagram contributes, the starting 
point of our construction, the dispersive representation of Eq.~(\ref{disprel}), 
requires appropriate modifications in order to include the contributions from 
these anomalous thresholds.

\subsubsection{Analytic properties of further diagrams}

We have seen that the appearance of the anomalous thresholds on the physical sheet is connected with the 
position of the real section of the curve $\Sigma$ between the subleading curves
$\sigma_{s\pm}$ and the $\sigma_{P\pm}$, in which case we cannot evade the corresponding normal threshold branch cuts when trying 
to avoid the singularities $\Sigma$ on the complex surfaces connected with the arc $BC$. We can, therefore, 
observe a simple condition for this appearance of the anomalous threshold.
It occurs on the physical sheet only in the case (\ref{Landau P triangle}) is real in the interval 
$M_P^2\in\left((\mu-m_1)^2,(\mu+m_1)^2\right)$. Since
the first triangle function appearing there is imaginary on this interval, the condition means that 
the second triangle function $\lambda(\mu^2,m_2^2,m_5^2)$
has to be imaginary as well. This happens in the interval $\mu^2\in \left((m_2-m_5)^2,(m_2+m_5)^2\right)$.

From this condition, we can formulate the following simple rule of thumb, stating that the $\pi\pi$ fish 
diagram has the anomalous threshold singularity on the physical sheet in the variable $s$ only in the 
case when in the corresponding triangle diagram one of the other vertices (than the one adjacent to $s$) 
is stable and the second one is unstable, when we take for $\mu$ its endpoint value $\mu_0$. 
The vertex is called unstable if the masses on the adjoining lines are such that at least one of them 
is greater than the sum of the other two. Note that this rule does not take 
into account the singularity on the real arc $BC$ of Fig.~\ref{Land-triangle1} (we have found 
that for the pion lines, this singularity never occurs) and the non-Landau singularities, as is obvious 
from its application to the $\pi$-diagrams (the only unstable mass there can be $\mu$, which appears in 
both vertices, i.e.,\ this rule tells there is no anomalous threshold for all $\pi$-diagrams). However, 
in our previous analysis, we have also taken its existence into account, and it does not change the 
above conclusion.

Since the vertex with $M_P$ in $P$-diagrams is always unstable for $\mu=\mu_0$, the anomalous 
threshold appears only in the case $\mu_0<(m_2+m_5)$. Furthermore, since 
$m_2$ and $m_5$ are pion masses and conservation of the electric 
charge has to be respected, the only possibility is $\mu_0=2\mpn$ and $m_2=m_5=\mpp$.

\begin{figure}[t]
\begin{center}
  \includegraphics[width=0.48\textwidth]{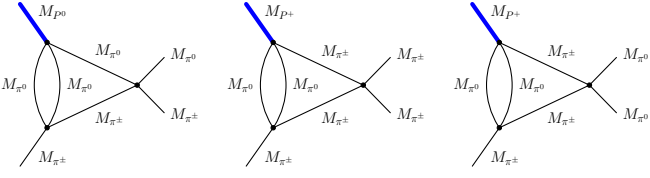}\\
  \caption{The complete set of $P$-diagrams possessing an
  anomalous threshold singularity on the physical Riemann sheet
  for the physical values of the masses $M_P$, $M_\pi$ and $M_{\pi^0}$.}\label{anomfish}
\end{center}
\end{figure}

In conclusion, the only $P$-diagrams possessing the anomalous threshold singularity on the physical sheet 
for the physical values of $M_P$ and of the pion masses are those depicted in Fig.~\ref{anomfish}. 
These diagrams contribute~to the processes $P^0\to\pi^0\pi^+\pi^-$, $P^+\to\pi^-\pi^+\pi^+$ and 
$P^+\to\pi^+\pi^0\pi^0$. Therefore, besides the case $M_\pi = M_{\pi^0}$,
the only $P\to\pi\pi\pi$ decay processes with $M_\pi \neq M_{\pi^0}$ where no such singularity 
occurs are the processes $P^0\to\pi^0\pi^0\pi^0$, with $P^0=K_L$ or $P^0=\eta$.
Kacser's prescription can be extended to these cases, i.e., the required analytic continuation
in $M_P^2$, ${\overline M}_P^2 \to M_P^2 + {\rm i} \delta$, ${\overline M}_P < 3 M_{\pi^0}$, $\delta > 0$, can be performed without 
encountering any singularity.

\subsection{Fish diagrams with other-than-pion internal lines}

The same analysis can be performed for fish diagrams containing
other mesons, kaons and $\eta$'s, in internal lines. Naturally, then it can happen that there will occur anomalous 
threshold singularity for some value of $M_P$. However, thanks to the hierarchy of the masses prohibiting decays 
of the type $P\to P'\pi$, where $P$ and $P'$ are kaons or $\eta$'s, the physical mass $M_P$ will each time be smaller 
than the mass where such anomalous threshold singularity can occur. In other words, in all the cases, the physical 
mass $M_P$ is to the left of the points $B$ and $C$ in one of the situations from Figs.~\ref{Land-triangle1} and 
\ref{Land-triangle2}, i.e., it lies in the region where the anomalous threshold singularity does not appear
on the physical sheet.
The inclusion of these other internal lines does, therefore, not change the conclusions of our analysis from the previous sections.

\section{Integrals of the ${\bar J}$ functions}\label{app:integrals}

In order to compute the partial-wave projections of the one-loop amplitudes,
we need to perform the integrals occurring in Eq.~(\ref{t0_t1_proj}). 
Actually, in order to avoid the cuts on the positive real axis,
the integration has to be performed along a path in the complex plane \cite{Bronzan},
\begin{equation}
\int_{t_-(s)}^{t_+(s)} dt \to \int_{\mathcal{C}\left(t_-(s),t_+(s)\right)} dt
\end{equation}
starting at $t_-(s)$ and ending at $t_+(s)$. It turns out that the complex 
path ${\mathcal{C}\left(t_-(s),t_+(s)\right)}$
can always be chosen such that there exists an open neighborhood of it that also
avoids the cut. This follows from the analysis of Ref.~\cite{Bronzan}
in the equal-mass case, and also holds for $M_\pi \neq M_{\pi^0}$ when the 
singularities of the integrands consist only of the normal branch cut,
starting at $s=4M_\pi^2$ or $s=4M_{\pi^0}^2$,
i.e., when there is no anomalous threshold \cite{Zdrahal}. 
Then, a result of complex analysis \cite{SaksZygmund} tells us that 
the integral is correctly evaluated in the usual way, i.e., upon taking the difference 
of the endpoint values of the  primitive function, provided, of course, that the latter exists.
Given this result, our task in this appendix will be to construct the required 
primitive functions, and then, as a second step, to evaluate them at the endpoints 
$t_\pm (s)$. It will be enough to give the results for the function ${\bar J}_{\pm 0}(t)$
defined in Eq.~(\ref{Jbar_functions}), the expressions corresponding to the two other cases, ${\bar J}(t)$
and ${\bar J}_0(t)$, can easily be obtained upon taking the appropriate limits in the pion masses.
More specifically, we denote the ratio of the pion masses by $q$, 
\begin{equation}
q=\frac{M_{\pi^0}}{M_\pi}
,
\end{equation}
and the two other cases will be obtained simply by taking the limit $q\to 1$ in the end.
We thus first need to know the following primitive functions
\begin{align}\label{integral I1}
I^{(n)}_{\pm 0}(t) &= 16\pi^2\int dt\,t^n \bar{J}_{\pm 0}(t), \quad\text{for\ } n=-1,\dotsc,3;
\nonumber\\
I^{(-2)}_{\pm 0}(t) &= 16\pi^2\int dt\,\frac{\bar{\bar{J}}_{\pm 0}(t)}{t^2}\,
.
\end{align}
These then need to be evaluated at the corresponding endpoints.
We address these two separate issues in turn in the remainder 
of this appendix.

\subsection{Primitive functions}\label{sec:primitive functions}

\begin{figure}[t]
\begin{center}
  \includegraphics[width=0.40\textwidth]{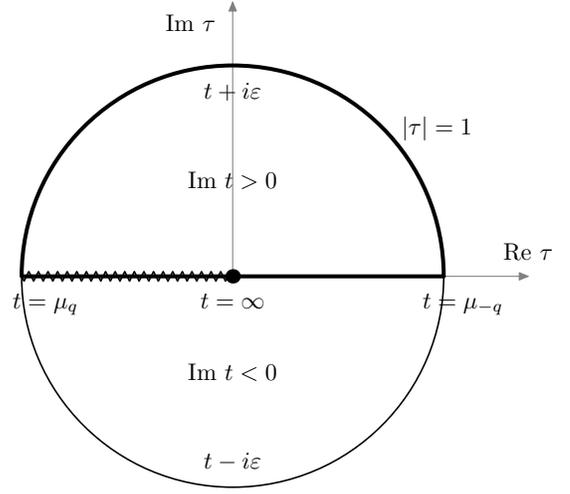}\\
  \caption[Conformal transformation (\ref{tau yz}) mapping the complex $t$-plane onto the unit disk 
in the $\tau$-plane]{Conformal transformation (\ref{tau yz}) mapping the complex $t$-plane onto the unit disk in $\tau$-plane. 
The points on the upper and on the lower semi-circles are identified. In the transformed plane, the one-loop functions to 
which the transformation corresponds have their branch cut located on the line segment 
$\tau\in(-1+{\rm i}\epsilon,0+{\rm i}\epsilon)$.}\label{tau zobrazeni}
\end{center}
\end{figure}

Writing the function $\bar{J}_{\pm 0}(t)$ in the form
\begin{multline}
\bar{J}_{\pm 0}(t) = \frac{1}{16 \pi^2} \Big[
1 + \left( \frac{\Delta_q}{t} - \frac{\Sigma_q}{\Delta_q} \right) \ln q
\\
+ \frac{t-\mu_{-q}}t\, \sigma_q (t) \, \ln \frac{\sigma_q (t) - 1}{\sigma_q(t) + 1}\Big],
\end{multline}
with
\begin{equation}
\mu_q= M_\pi^2 (1+q)^2,\quad \Sigma_q = M_\pi^2 (1+q^2),\quad \Delta_q = M_\pi^2 (1-q^2)
\end{equation}
and
\begin{equation}
\sigma_q (t) = \sqrt{\frac{t-\mu_q}{t-\mu_{-q}}}\,
,
\end{equation}
suggests performing the following transformation of the variable
\begin{equation}\label{tau yz}
\tau = \frac{\sigma_q (t) - 1}{\sigma_q (t) + 1}\,
.
\end{equation}
The function $\sigma_q (t)$ is defined in the complex $t$-plane, with a
cut on the real axis, from $\mu_{-q}$ to $\mu_{q}$, and with 
$\sigma_q (t\pm {\rm i} \epsilon) = \pm {\rm i} \sqrt{(\mu_q - t)/(t-\mu_{-q})}$
when $t$ lies on this cut. 
This transformation then maps the complex plane with the cut $(\mu_{-q},\mu_{q})$ onto the unit disk. 
As illustrated in Figure~\ref{tau zobrazeni}, the points slightly above
the cut ($t+{\rm i}\epsilon$) are mapped slightly below the upper semi-circle while the points slightly 
below this cut ($t-{\rm i}\epsilon$) are mapped slightly above the lower semi-circle 
[the points lying exactly on the upper and on the lower semi-circle have to be identified]. 
The ray $(\mu_q +{\rm i}\epsilon,\infty+{\rm i}\epsilon)$, where the branch cut of the one-loop function 
is located, is mapped onto line segment $(-1+{\rm i}\epsilon,0+{\rm i}\epsilon)$. The inverse transformation
\begin{equation}
t(\tau) = M_\pi^2 (\tau - q) \left( \frac{1}{\tau} - q \right)\!
,
\end{equation}
with
\begin{equation}
\sigma_q (t) = \frac{1 + \tau}{1 - \tau}\,
,\quad
dt = \frac{q M_\pi^2}{\tau^2} (1 - \tau^2) d \tau
,
\end{equation}
can be continued to the whole complex $\tau$-plane and satisfies
\begin{equation}
t(\tau)=t(1/\tau).
\end{equation}
This means that the points $\tau$ and $1/\tau$ should be identified, which also implies the identification of $\tau$ and $\tau^\star=1/\tau$ on the unit-disk boundary 
of Figure~\ref{tau zobrazeni}.

Expressed in terms of the variable $\tau$, the one-loop function ${\bar J}_{\pm 0}$ reads
\begin{multline}
16\pi^2\bar{J}_{\pm 0}(\tau) =
1-\frac{\Sigma_{q}}{\Delta_{q}}\log q-\frac{\tau (1-q^2)}{(\tau -q)(\tau q-1)}\log q
\\
+ \frac{q (1-\tau^2)}{(\tau -q)(\tau q-1)}\log\tau
.
\end{multline}
Its analyticity in the unit disk with the segment $\langle-1,0\rangle$ removed
and the symmetry property $\bar{J}_{\pm 0}(\tau)=\bar{J}_{\pm 0}(1/\tau)$ become manifest in this expression.
The expression that mixes the two variables, 
\begin{equation}
16\pi^2\bar{J}_{\pm 0}(t)=1+\left(\frac{\Delta_q}{t}-\frac{\Sigma_q}{\Delta_q}\right)\log q
+ q \frac{M_\pi^2}{t}\frac{\tau^2-1}{\tau}\log\tau,
\end{equation}
allows obtaining its derivative in a simple way,
\begin{equation}
16\pi^2 \frac{d}{d t}\left(t\bar{J}_{\pm 0}(t)\right) =
\frac{1+\tau^2}{1-\tau^2}\log{\tau}-\frac{1+q^2}{1-q^2}\log{q},
\end{equation}
which in turn makes it easy to check the primitive functions given here.
Finally, we introduce the following function, 
\begin{equation}
\Poly(\tau)=\log{q}\log{\tau}+\Li(1-q\tau)-\Li\left(1-\frac{\tau}{q}\right)
,
\end{equation}
which appears in the results of the integration.

Having prepared all the necessary ingredients, we present here a list of all primitive functions 
that are needed for the computation of the S and P partial waves of the one-loop amplitudes:
\begin{align}
I_{\pm 0}^{(1)}(t) &=
8\pi^2\bar{J}_{\pm 0} (t)\, t(t-\Sigma_{q}) + \frac{t^2}{4}-2 q^2 M_\pi^4 t\,\frac{\log{q}}{\Delta_{q}}
\nonumber\\
&+ q^2 M_\pi^4 \log^2{\tau},\nonumber\\
I_{\pm 0}^{(2)}(t) &=
\frac{8\pi^2}{3}\bar{J}_{\pm 0}(t)\,t\left(2t^2-\Sigma_{q} t-(\Sigma_{q}^2+8 q^2 M_\pi^4)\right)
\nonumber\\
&
+ \frac{t^3}{9}+\frac{\Sigma_q t^2}{12}
-2 q^2 M_\pi^4 t\left(\frac{t}{3}+\Sigma_{q}\right)\frac{\log{q}}{\Delta_{q}}
\nonumber\\
&
+ q^2 M_\pi^4 \Sigma_{q}\log^2{\tau},
\nonumber\\
I_{\pm 0}^{(3)}(t) &=
\frac{t^4}{16}+\Sigma_{q}\frac{t^3}{18}+\frac{t^2}{24} (\Sigma_{q}^2+6q^2 M_\pi^4)
\nonumber\\
&
+ q^2 M_\pi^4 (\Sigma_{q}^2 + q^2 M_\pi^4)\log^2{\tau}
\nonumber\\
&
- q^2 M_\pi^4 t\left[\frac{t^2}{3}+\frac{5\Sigma_{q} t}{6}+2(\Sigma_{q}^2
+ q^2 M_\pi^4)\right]\frac{\log{q}}{\Delta_{q}}
\nonumber\\
&
+ \frac{4\pi^2}{3}\bar{J}_{\pm 0}(t)\,t\Big(3t^3-\Sigma_{q} t^2-(\Sigma_{q}^2+6 q^2M_\pi^4)t
\nonumber\\
&\qquad
- \Sigma_{q}(\Sigma_{q}^2+26 q^2 M_\pi^4)\Big),\nonumber
\\
I_{\pm 0}^{(0)}(t) &=
16\pi^2\bar{J}_{\pm 0} (t)\,t+t+\frac{\Sigma_{q}}{2}\log^2{\tau}-\Delta_{q}\Poly(\tau),
\nonumber\\
I_{\pm 0}^{(-1)}(t) &=
-16\pi^2\bar{J}_{\pm 0}(t)-\frac{1}{2}\log^2{\tau} - 2 + \frac{\Sigma_{q}}{\Delta_{q}}\Poly(\tau),
\nonumber\\
I_{\pm 0}^{(-2)}(t) &=
8\pi^2\bar{J}_{\pm 0}(t)\left(\frac{\Sigma_{q}}{\Delta_{q}^2}-\frac{1}{t}\right)
- \frac{\Sigma_q}{\Delta_q^2} + \frac{13}{36 q M_\pi^2}
\nonumber\\
&
+ \frac{2 q^2 M_\pi^4}{\Delta_{q}^3}\Poly(\tau).
\end{align}
We have adjusted the free integration constants in these primitive functions,
such as to ensure that all of them exhibit a smooth limit for the ratio 
of pion masses $q$ going to unity, and to make them vanish at $t=0$ (in this limit).

\subsection{Endpoint evaluation in the case $M_{\pi^0}=M_\pi$}

In order to perform the explicit calculation of the partial-wave projections,
we should now evaluate the preceding primitives at the endpoints 
given in Eq.~(\ref{limits_int}). In general, this procedure produces complicated expressions. 
However, in particular cases, like for equal-mass pions, the situation simplifies somewhat.
We will, therefore, treat this case in some detail in what follows. For definiteness,
we consider the case of the charged pion. The corresponding expressions for the neutral
pion are obtained from those given below by substituting $M_\pi$ by $M_{\pi^0}$ in all
formulas.

Introducing
\begin{equation}
T_\pm (s) = \frac{4M_\pi^2-t_\pm(s)}{2M_\pi^2}\,
,
\end{equation}
and
\begin{equation}
\sigma_{T_\pm}(s) = \frac{1}{\sigma(t_\pm(s))}  
=\sqrt{\frac{T_\pm (s) -2}{T_\pm (s)}}\, ,
\end{equation}
we obtain 
\begin{align}
\tau(t_\pm(s)) &= T_\pm (s) -1 - T_\pm (s)\, \sigma_{T_\pm} (s),
\nonumber\\
\frac{1}{\tau(t_\pm(s))} &= {T_\pm} (s) - 1 + {T_\pm} (s)\, \sigma_{T_\pm} (s)
,
\end{align}
and 
\begin{equation}
L_{T_\pm} (s) = \log \tau(t_\pm(s))
\equiv \log\left(\frac{1-\sigma_{T_\pm} (s)}{1+\sigma_{T_\pm} (s)}\right).
\end{equation}

Inserting these relations into the primitive functions in the limit $q=1$,
and using the simplification ${T_\pm}\sigma_{T_\pm}^2={T_\pm}-2$, we arrive at
rather simple expressions 
\begin{align}
I^{(1)}(t_\pm) &= M_\pi^4\big[({T_\pm}-2)(5{T_\pm}-8)
\nonumber\\
&
+ 2{T_\pm}({T_\pm}-1) \sigma_{T_\pm} L_{T_\pm}+L_{T_\pm}^2\big],
\nonumber\\
I^{(2)} (t_\pm) &= \frac{M_\pi^6}{9}\big[2({T_\pm}-2)(7{T_\pm}-8)(11-4{T_\pm})
\nonumber\\
&
- 12{T_\pm}({T_\pm}-3)(2{T_\pm}-1)\sigma_{T_\pm} L_{T_\pm} + 18L_{T_\pm}^2\big],
\nonumber\\
I^{(3)}(t_\pm) &= \frac{M_\pi^8}{9}\big[({T_\pm}-2)(81{T_\pm}^3-482{T_\pm}^2+941{T_\pm}
\nonumber\\
&
- 512)+6{T_\pm}(6{T_\pm}^3-34{T_\pm}^2+59{T_\pm}
\nonumber\\
&
-15)\sigma_{T_\pm} L_{T_\pm}+45L_{T_\pm}^2\big],
\nonumber\\
I^{(0}(t_\pm) &= M_\pi^2 \big[6(2-{T_\pm})-2{T_\pm}\sigma_{T_\pm} L_{T_\pm}+L_{T_\pm}^2\big],
\nonumber\\
I^{(-1)}(t_\pm) &=-4-\frac{2{T_\pm}}{{T_\pm}-2}\,\sigma_{T_\pm} L_{T_\pm}-\frac12 L_{T_\pm}^2,\\
I^{(-2)}(t_\pm) &= \frac{1}{18 M_\pi^2}\left[\frac{4(2{T_\pm}-1)}{{T_\pm}-2}+\frac{3{T_\pm}^2}{({T_\pm}-2)^2}\sigma_{T_\pm} L_{T_\pm}\right]\!.
\nonumber
\end{align}
Finally, it is also useful to notice the relations
\begin{align}
16\pi^2\bar{J} (t_\pm) &= 2+\frac{L_{T_\pm}}{\sigma_{T_\pm}}\, ,
\nonumber\\
16\pi^2\bar{\bar{J}} (t_\pm) &= 2+\frac{L_{T_\pm}}{\sigma_{T_\pm}}-\frac{t_\pm}{6 M_\pi^2}\,
.
\end{align}

The link with the expressions given in Eqs.~(\ref{Jbar_integrals}), (\ref{tilde_k_functions}) and (\ref{kappa})
is then provided by the relations
\begin{equation}
\sigma_{T_\pm}(s) = \frac{\Delta_{P\pi} \sigma (s)}{s - 4 M_\pi^2 \pm 2 K(s)}
,~
L_{T_\pm} (s) = \frac{1}{2} \left[ L(s) \mp M(s) \right]
.
\label{sigma_T}
\end{equation}

\begin{figure*}[t]
\begin{tabular}{ccc}
	\includegraphics[width=0.40\textwidth]{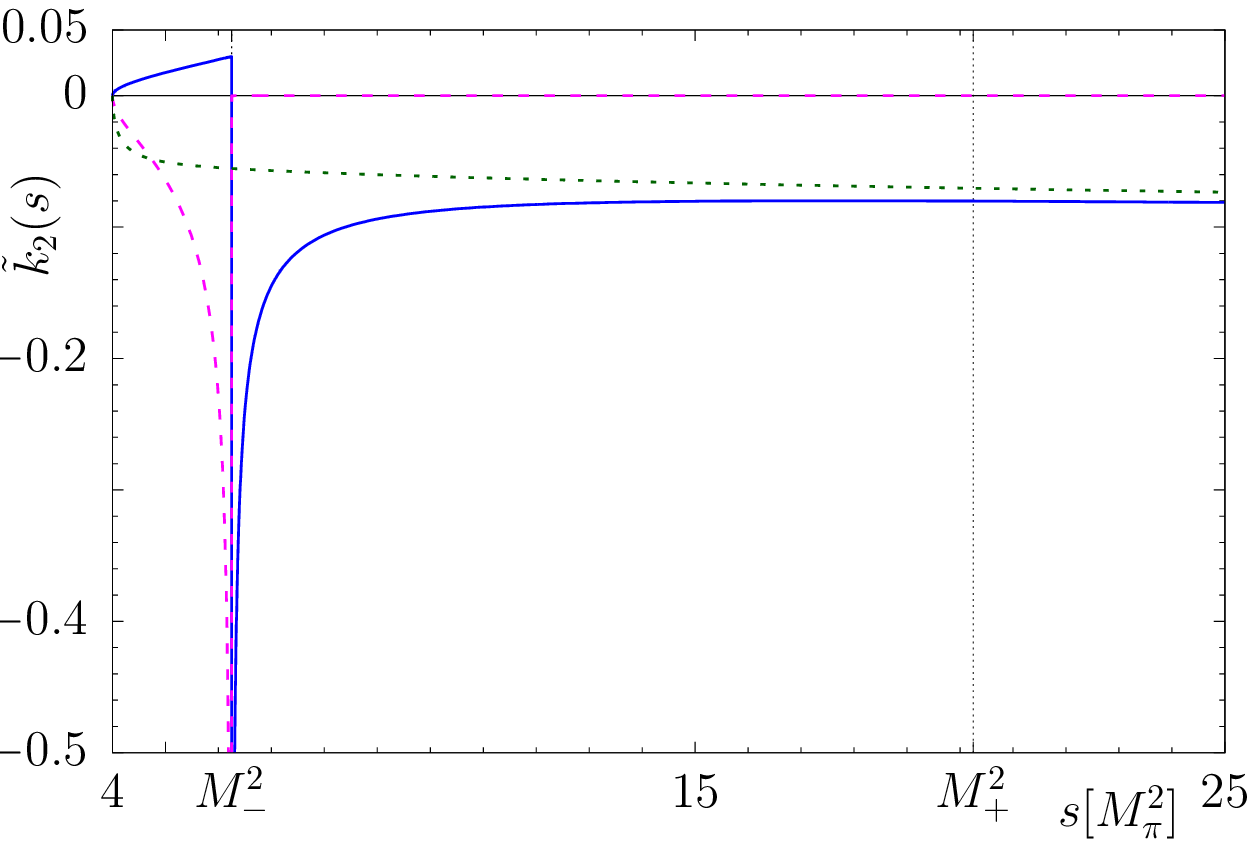} & \hspace{0.02\textwidth} &   \includegraphics[width=0.40\textwidth]{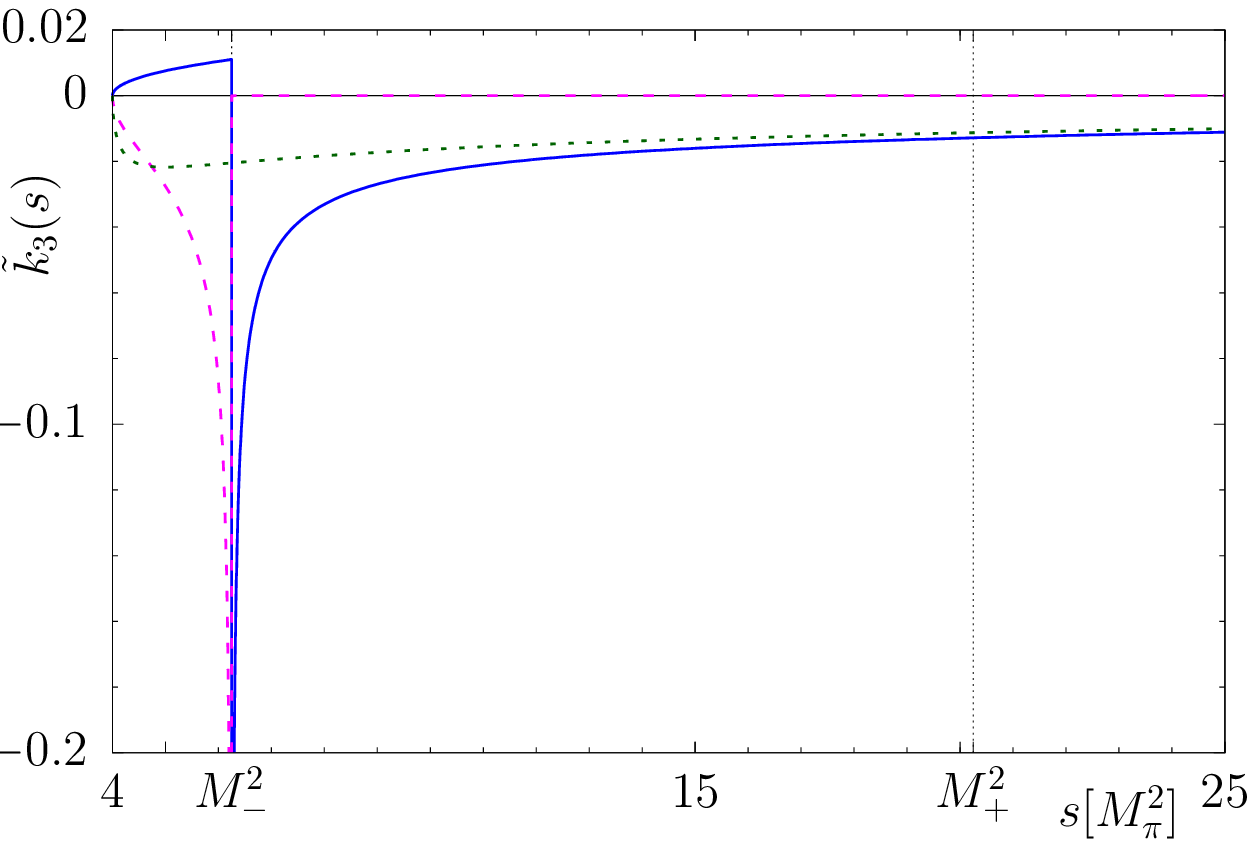} \\
	(a) ${\widetilde k}_2(s)$ & & (b) ${\widetilde k}_3(s)$ \\[6pt]
\end{tabular}
\caption{The real (solid curve in blue) and the imaginary (dashed curve in magenta) parts of functions ${\widetilde k}_{2,3}(s)$ for $M_P= 3.5 M_\pi + {\rm i} 0$. For comparison, also the real part of these functions for $M_P= 2.9 M_\pi$ (dotted curve in green) is plotted (in this case, their imaginary part is zero). The abscissas show $s$ in units of $M_\pi^2$. The vertical lines indicate the positions of $s=M_\pm^2$.
\label{plot_tilde_k}}
\end{figure*}

\begin{figure*}[t]
	\begin{tabular}{ccc}
		\includegraphics[width=0.40\textwidth]{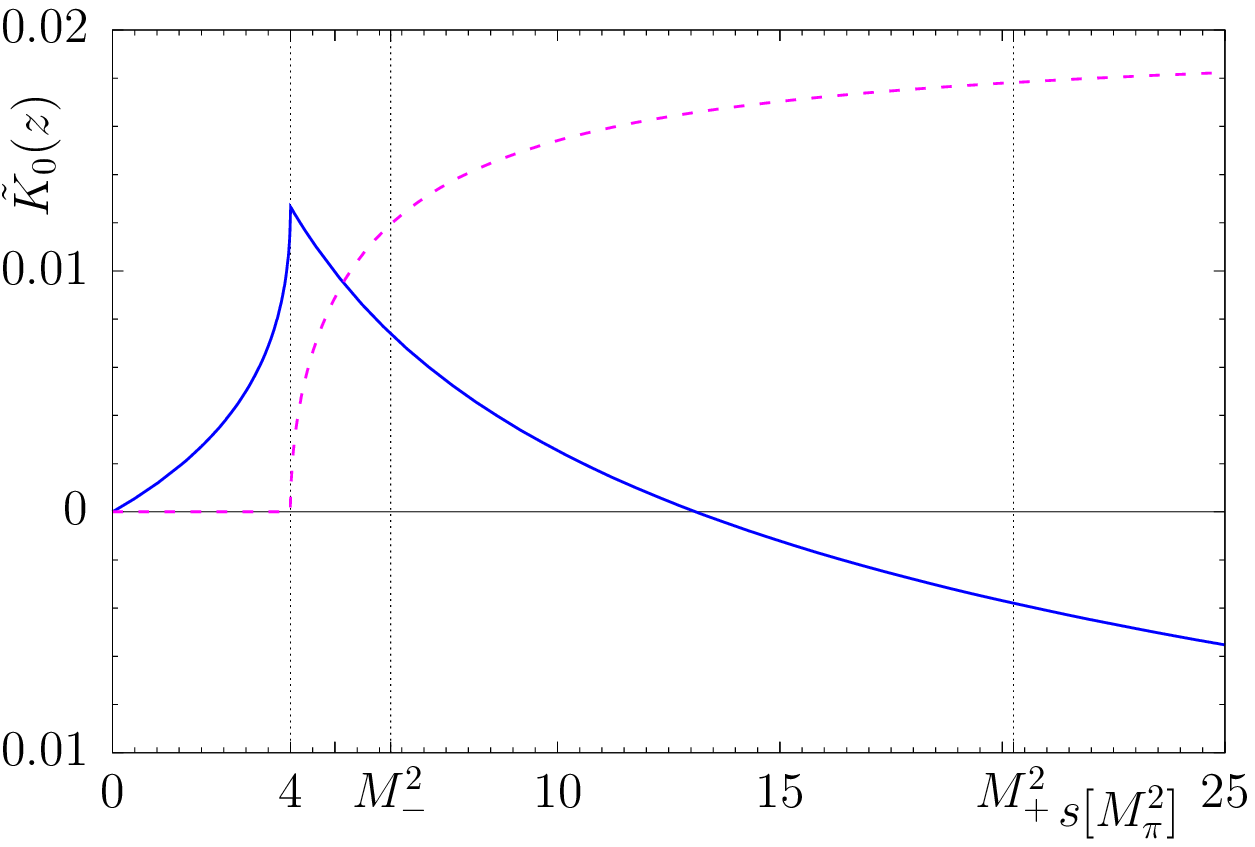} & \hspace{0.02\textwidth} & \includegraphics[width=0.40\textwidth]{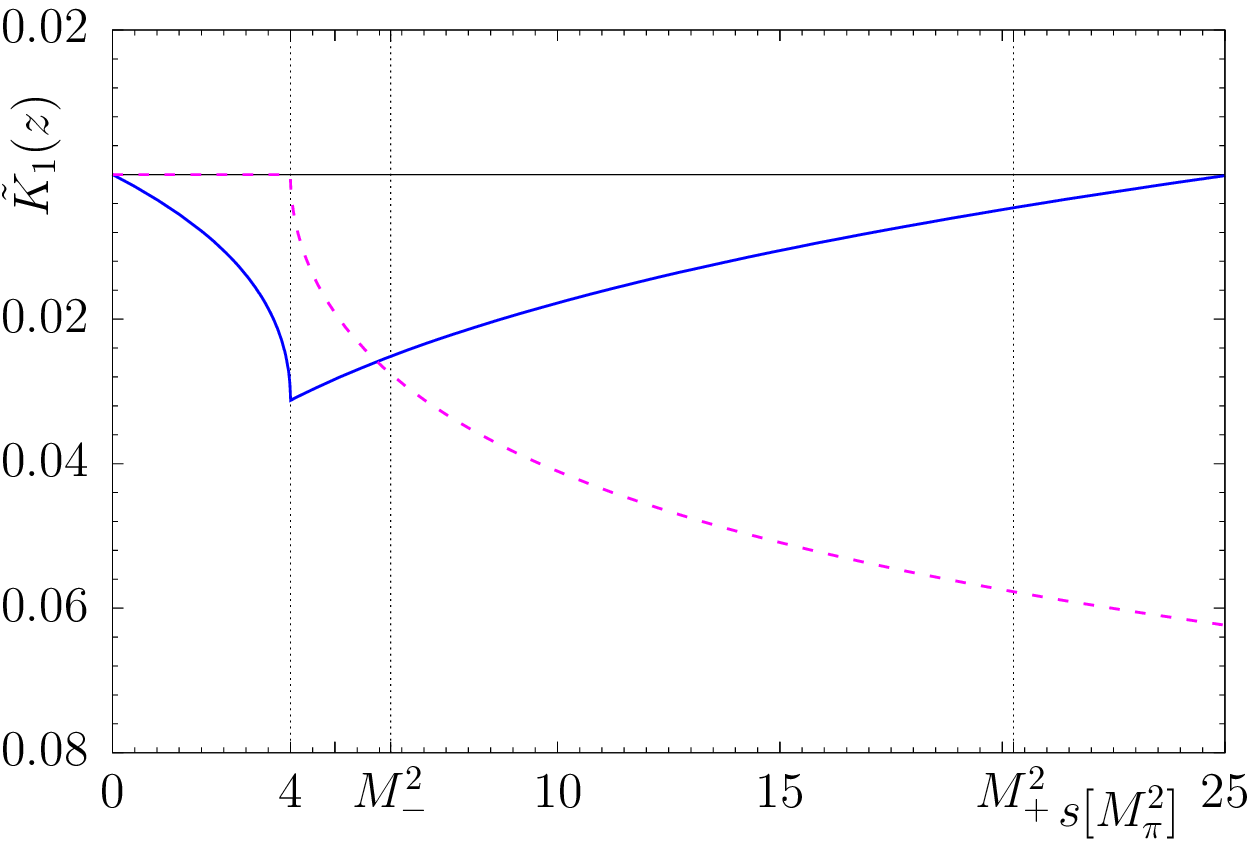} \\
		(a) ${\widetilde K}_0(z)$ & & (b) ${\widetilde K}_1(z)$ \\[6pt]
		\includegraphics[width=0.40\textwidth]{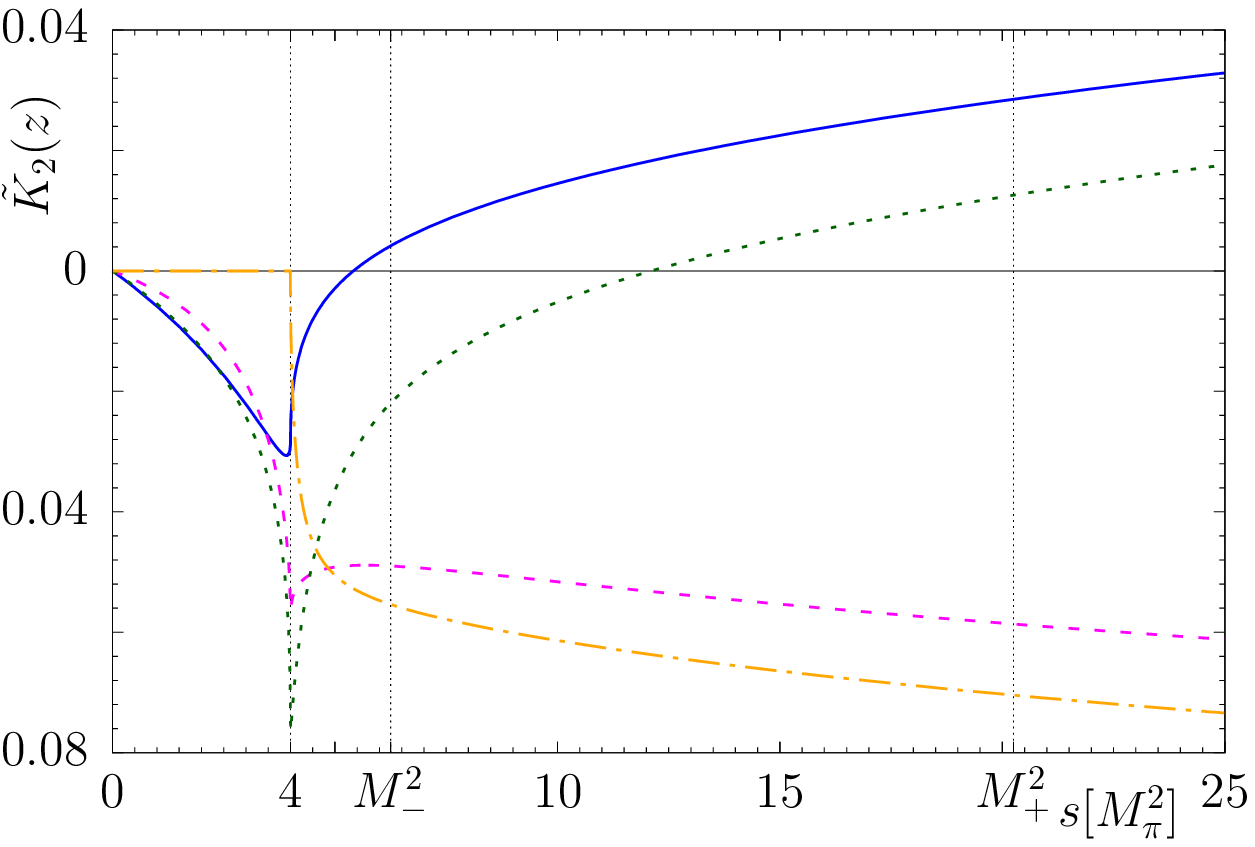} &  & \includegraphics[width=0.40\textwidth]{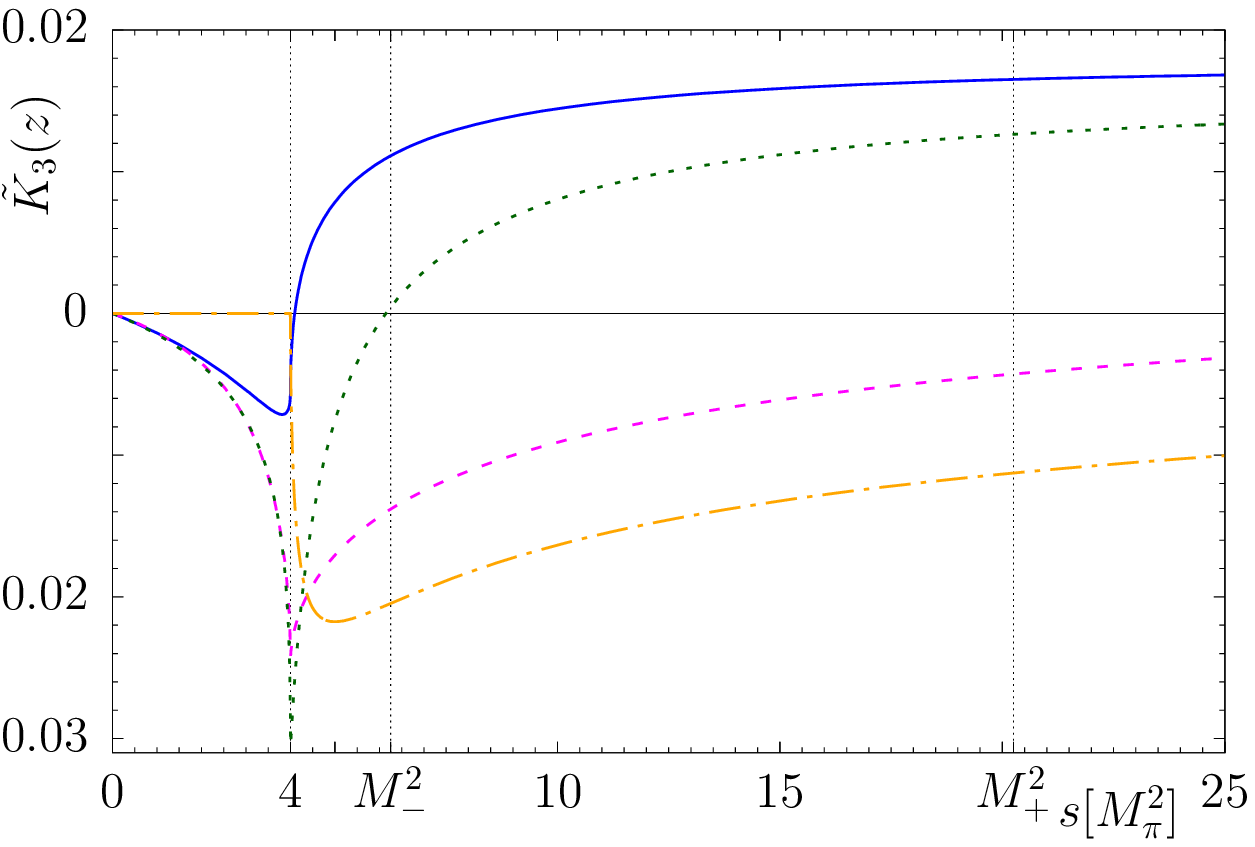} \\
		(c) ${\widetilde K}_2(z)$ & & (d) ${\widetilde K}_3(z)$ \\[6pt]
	\end{tabular}
	\caption{The real (solid curve in blue) and the imaginary (dashed curve in magenta) parts of functions ${\widetilde K}_i(z)$, $z=s+{\rm i} 0$, for $M_P= 3.5 M_\pi + {\rm i} \delta$. For comparison, the real (dotted curve in green) and the imaginary (dot-dashed curve in orange) parts of functions ${\widetilde K}_{2,3}(s)$ for $M_P= 2.9 M_\pi$ are also plotted. The abscissas show $s$ in units of $M_\pi^2$. The positions of $s=\{4M_\pi^2,M_-^2,M_+^2\}$ are indicated by the vertical lines.
		\label{plot_tilde_K}}
\end{figure*}
\section{Properties of the functions ${\widetilde k}_i(s)$ and ${\widetilde K}_i(s)$}\label{App:K-tilde}

In this appendix, we discuss some properties of the functions ${\widetilde k}_i(s)$,
${\widetilde K}_i(s)$ and ${\widetilde K}_i^{(\lambda)}(s)$, introduced
in Eqs.\ (\ref{tilde_k_functions}), (\ref{widetilde_K}) and (\ref{widetilde_K_lambda0}),
respectively. We briefly address their analytic properties and also provide graphical representations. 
We display the functions ${\widetilde k}_i (s)$, with $i=2,3$, on the real
axis in Fig.~\ref{plot_tilde_k} (the real parts of those with $i=0,1$ coincide with the imaginary parts of ${\widetilde K}_i(z)$). 

The functions  ${\widetilde K}_i(z)$ are defined as dispersive integrals in the complex plane,
\begin{equation}
 {\widetilde K}_i(z)=\frac{z^n}{\pi} \int_{4M_\pi^2}^{\infty}\frac{dx}{x^n}\frac{{\widetilde k}_i(x)}{x-z}\,
 ,
 \label{remark}
\end{equation}
and the physical value corresponds to $z=s+{\rm i}0$, which is shown in Fig.~\ref{plot_tilde_K}.
For $z$ which is not located on the integration 
contour, the only potential singularities of the integrand are those of the function ${\widetilde k}_i(x)$ 
where $4M_\pi^2 <x<\infty$. The functions ${\widetilde k}_0 (s)$ and ${\widetilde k}_1 (s)$ do not present
any particular problem. They already occur in the simpler situation provided by $\pi\pi$ scattering,
and the two functions ${\widetilde K}_0(s)$ and ${\widetilde K}_1(s)$ can be expressed in terms of the 
function ${\bar J}(s)$ [see Eq.~(\ref{K0_and_K1})], which has no other singularity than a cut along the
positive-$s$ real axis, starting at $s=4 M_\pi^2$. In the case of ${\widetilde k}_2 (s)$ and ${\widetilde k}_3 (s)$,
there are two possibly problematic points, namely, $x=(M_P\pm M_\pi)^2=M_{\pm}^2$. Closer inspection
reveals that only $x=M_-^2$ corresponds to a singularity, as can actually also be seen directly in Fig.~\ref{plot_tilde_k}.
This singularity is, however, integrable since the integrand in Eq.~(\ref{remark}) behaves as $(x-M_{\pm}^2)^{-1/2}$ 
in the vicinity of this point. 
It means the ${\widetilde K}_i(z)$, defined by the original contour, is at least analytic 
in the complex plane with the exception of the part of the real axis for which $z>4M_{\pi}^2$. 
For $z\to4M_{\pi}^2$, we have a non-integrable endpoint singularity. Therefore, this point constitutes
a branch point of ${\widetilde K}_i(z)$. For $z$ real, $z\ne M_{\pm}^2$ 
we can deform the integration contour in (\ref{remark}), such as to avoid the singularity $(x-z)^{-1}$, 
in an appropriate way, depending on whether we approach the contour from below or from above. The only 
dangerous point is the true singularity of  ${\widetilde k}_i(x)$, i.e., $z=M_{-}^2$. Because we made the 
analytic continuation $M_P^2+{\rm i} \delta$, this singularity is, in fact, avoided by contour deformation when we 
approach $z=M_{-}^2$ from the upper complex half-plane because then there is no pinch of the contour, and 
thus even this point is, in fact, regular. However, when we approach this point from below, the contour is pinched, and the 
singularity cannot be avoided by means of the contour deformation. Nevertheless, this singularity lies on the 
second sheet. Of course, these arguments rest on the fact that the discontinuities ${\widetilde k}_i(z)$ have good 
analytic properties, which allows deforming the contour without encountering other singularities of these functions.

\section{Expressions of the functions ${\cal W}^{P\pi} (s)$}\label{App:w_tilde}
In this appendix, we give the expressions of the functions ${\cal W}^{P\pi} (s)$,
and of the corresponding coefficients ${\widetilde w}_\pm^{(n)}$ and ${\cal B}_{\pm}$
for each function listed in Table \ref{W_for_P-pi_Table}, in the case where the masses
of charged and neutral pions are equal.

For the decay of a charged kaon into three charged pions, 
we have
\begin{align}
&{\cal W}_{++}^{P\pi} (s) =  \frac{\varphi_0^{++} (s)}{2} \bigg\{
\frac{(16\pi)^2}{4}\, \varphi_0^{++} (s) {\widetilde\varphi}_0^{++} (s) {\bar J}^2 (s)
\nonumber\\
&\quad 
+
\left[ \left(C_{++} + \frac{D_{++}}{2} \right) \frac{(s-s_0)^2}{F_\pi^4}  + \frac{2}{3} D_{++} \frac{K^2(s)}{F_\pi^4} \right]{\bar J} (s) \bigg\}
\nonumber\\
&\ 
+
\frac{1}{2} \, [\varphi_0^{++} \sodot {\widetilde\xi}_0^{++}](s)
,
\nonumber\\
&{\cal W}_{+-}^{(0)P\pi} (s) = {\varphi}^{+-}_0 (s) \bigg\{
\frac{(16\pi)^2}{4}  \big[
2 {\widetilde\varphi}^{+-}_0(s)\varphi^{+-}_0(s)
\nonumber\\
&\quad 
+ {\widetilde\varphi}^{x}_0(s) \varphi^{x}_0(s)
\big]     {\bar J}^2 (s) 
+ \left[\frac{1}{4} (C_{++} + 5 D_{++}) \frac{(s-s_0)^2}{F_\pi^4}
\right.
\nonumber\\
&\quad \left.
+ \frac{1}{3} (C_{++} + D_{++}) \frac{K^2(s)}{F_\pi^4} \right]  {\bar J} (s) 
\bigg\} + [\varphi_0^{+-} \sodot {\widetilde\xi}_0^{+-}](s)
\nonumber\\
&\ 
+ \frac{{\varphi}^{x}_0 (s)}{2} \bigg\{
\frac{(16\pi)^2}{4}  \big[
2 {\widetilde\varphi}^{+-}_0(s)\varphi^{x}_0(s)
+ {\widetilde\varphi}^{x}_0(s) \varphi^{00}_0(s)
\big]     {\bar J}^2 (s) 
\nonumber\\
&\quad 
+
\left[ \left(C_{x} + \frac{D_{x}}{2} \right) \frac{(s-s_0)^2}{F_\pi^4}
+ \frac{2}{3} D_{x} \frac{K^2(s)}{F_\pi^4} \right]  {\bar J} (s) \bigg\}
\nonumber\\
&\ 
+
\frac{1}{2} \, [\varphi_0^x \sodot {\widetilde\xi}_0^x](s)
,\\
&{\cal W}_{+-}^{(1)P\pi} (s) = 
\frac{{\rm c}_{+-}}{3 F_\pi^2} \left[
- \frac{B_{++} {\rm c}_{+-}}{18} \frac{16\pi}{2} \frac{(s-4M_\pi^2)^2}{F_\pi^4} \, {\bar J}^2 (s)
\right.
\nonumber\\
&\quad \left.
+ \frac{C_{++} - D_{++}}{6}\,\frac{(s-4M_\pi^2)(s-s_0)}{F_\pi^4} \, {\bar J} (s)
+ {\widetilde\xi}_1^{+-} (s) \right]
\nonumber
\end{align}
with
\begin{equation}
\frac{{\cal B}_{+;++}}{2} = -{\cal B}_{\pm;+-} = -{\cal B}_{-;+-} = B_{++} {\rm c}_{+-}
, \quad {\cal B}_{-;++} = 0
,
\end{equation}
and
\begin{align}
&{\widetilde w}_{+;++}^{(0)} = 
\left[ A_x - \frac{B_x}{3} \frac{M_K^2 + 3 M_\pi^2}{F_\pi^2}  \right] \!
\left( {\rm a}_x - {\rm b}_x \frac{4M_\pi^2}{F_\pi^2} \right)
\nonumber\\
&\quad
+
2 \left[ A_{++} + \frac{B_{++}}{6} \frac{M_K^2 + 3 M_\pi^2}{F_\pi^2}  \right]\!
\left({\rm a}_{+-} - {\rm b}_{+-} \frac{4M_\pi^2}{F_\pi^2} \right)
\nonumber\\
&\quad
+ \frac{4}{3} B_{++} {\rm c}_{+-} \frac{M_\pi^2 (M_K^2 + 3 M_\pi^2)}{F_\pi^4}\,
,
\nonumber\\
&{\widetilde w}_{+;++}^{(1)} =
{\rm b}_x \left[ A_x - \frac{B_x}{3} \frac{M_K^2 + 3 M_\pi^2}{F_\pi^2}  \right]
\nonumber\\
&\quad
+ B_x \left({\rm a}_x - {\rm b}_x \frac{4M_\pi^2}{F_\pi^2} \right)
\nonumber\\
&\quad
+
2 {\rm b}_{+-} \left[ A_{++} + \frac{B_{++}}{6} \frac{M_K^2 + 3 M_\pi^2}{F_\pi^2}  \right]
\nonumber\\
&\quad
- B_{++} \left( {\rm a}_{+-} - {\rm b}_{+-} \frac{4M_\pi^2}{F_\pi^2} \right)
- \frac{B_{++} {\rm c}_{+-}}{3}  \frac{M_K^2 + 7 M_\pi^2}{F_\pi^2}\,
,
\nonumber\\
&{\widetilde w}_{+;++}^{(2)} = 
 B_x {\rm b}_x - {\rm b}_{+-} B_{++} + \frac{B_{++} {\rm c}_{+-}}{3}\,
,
\end{align}
\begin{align}
&{\widetilde w}_{\pm ; +-}^{(0)} = 
  \left[ \frac{A_x}{2} - \frac{B_x}{6} \frac{M_K^2 + 3 M_\pi^2}{F_\pi^2}  \right]
\!\left( {\rm a}_x - {\rm b}_x \frac{4M_\pi^2}{F_\pi^2} \right)
\nonumber\\
&\quad
+ \left[ A_{++} + \frac{B_{++}}{6} \frac{M_K^2 + 3 M_\pi^2}{F_\pi^2}  \right]
\!
\left({\rm a}_{+-} - {\rm b}_{+-} \frac{4M_\pi^2}{F_\pi^2} \right)
\nonumber\\
&\quad
- \frac{2}{3} B_{++} {\rm c}_{+-} \frac{M_\pi^2 (M_K^2 + 3 M_\pi^2)}{F_\pi^4}
\nonumber\\
&\quad
\pm \frac{1}{2}
\bigg[ A_{++}
 - \frac{B_{++}}{3} \frac{M_K^2 + 3 M_\pi^2}{F_\pi^2}  \bigg]\!
\left({\rm a}_{++} - {\rm b}_{++} \frac{4M_\pi^2}{F_\pi^2} \right)\!
,
\nonumber\\
&{\widetilde w}_{\pm ; +-}^{(1)} =
\frac{{\rm b}_x}{2}\left[ A_x - \frac{B_x}{3} \frac{M_K^2 + M_\pi^2 + 2 M_{\pi^0}^2}{F_\pi^2}  \right]
\nonumber\\
&\quad
+
\frac{B_x}{2} \left({\rm a}_x - {\rm b}_x \frac{4M_\pi^2}{F_\pi^2} \right)
- \frac{B_{++}}{2} \left({\rm a}_{+-} - {\rm b}_{+-} \frac{4M_\pi^2}{F_\pi^2} \right)
\nonumber\\
&\quad
+ \frac{{\rm b}_{+-}}{F_\pi^2}\bigg[ A_{++}  +  \frac{B_{++}}{6} \frac{M_K^2 + 3 M_\pi^2}{F_\pi^2}  \bigg]
\nonumber\\
&\quad
+ \frac{B_{++} {\rm c}_{+-}}{6}  \frac{M_K^2 + 7 M_\pi^2}{F_\pi^2}
\pm
\frac{B_{++}}{2} \left({\rm a}_{++} - 4 {\rm b}_{++} \frac{M_\pi^2}{F_\pi^2} \right)
\nonumber\\
&\quad
\pm 
\frac{{\rm b}_{++}}{2} \left[ A_{++} - \frac{B_{++}}{3} \frac{M_K^2 + 3 M_\pi^2}{F_\pi^2}  \right]
,
\\
&{\widetilde w}_{\pm ; +-}^{(2)} = \frac{1}{2}
\Big(\! B_x {\rm b}_x - B_{++} {\rm b}_{+-} - \frac{1}{3} B_{++} {\rm c}_{+-} 
\pm B_{++} {\rm b}_{++} \Big)
.
\nonumber
\end{align}

Next, for the decay of a charged kaon into one charged and two neutral pions, 
we obtain
\begin{align}
&{\cal W}_{x}^{P\pi} (s) = {\varphi}^{x}_0 (s) \bigg\{
\frac{(16\pi)^2}{4}  \big[
2 {\widetilde\varphi}^{+-}_0(s)\varphi^{+-}_0(s) 
\nonumber\\
&\quad
+ {\widetilde\varphi}^{x}_0(s) \varphi^{x}_0(s) \big] {\bar J}^2 (s) 
+
\left[ \frac{1}{4} (C_{++} + 5 D_{++}) \frac{(s-s_0)^2}{F_\pi^4} 
\right.
\nonumber\\
&\quad \left.
+ \frac{1}{3} (C_{++} + D_{++}) \frac{K^2(s)}{F_\pi^4} \right] {\bar J} (s) 
\bigg\} + [\varphi_0^x \sodot {\widetilde\xi}_0^{+-}](s) 
\nonumber\\
&\ 
+ \frac{{\varphi}^{00}_0 (s)}{2} \bigg\{
\frac{(16\pi)^2}{4}  \left[ 2 {\widetilde\varphi}^{+-}_0(s) \varphi^{x}_0(s) 
+ {\widetilde\varphi}^{x}_0(s) \varphi^{00}_0(s)\right]\! {\bar J}^2 (s)
\nonumber\\
&\quad
+
\left[ \left( C_{x} + \frac{D_{x}}{2} \right) \frac{(s-s_0)^2}{F_\pi^4}
+ \frac{2}{3} D_{x} \frac{K_{x}^2(s)}{F_\pi^4} \right]
{\bar J} (s)    
\bigg\}\nonumber\\ 
&\ 
+ \frac{1}{2} \, [\varphi_0^{00} \sodot {\widetilde\xi}_0^x](s),
\nonumber\\
&{\cal W}_{0+}^{(0)P\pi} (s) =  \varphi_0^{+0} (s) \bigg\{
\frac{(16\pi)^2}{2} \varphi_0^{+0} (s)\, {\widetilde\varphi}_0^{0+} (s)\, {\bar J}^2 (s)
\nonumber\\
&\quad \left.
-\left[
\frac{1}{4} \left( C_{x} + 5 D_{x} \right) \frac{(s-s_0)^2}{F_\pi^4}
+ \frac{1}{3} ( C_{x} + D_{x} ) \frac{K^2(s)}{F_\pi^4}
\right]
\right.
\nonumber\\
&\quad
\times {\bar J} (s) \bigg\} + [\varphi_0^{+0} \sodot {\widetilde\xi}_0^{0+}](s),
\nonumber\\
&{\cal W}_{0+}^{(1)P\pi} (s) =
\frac{{\rm c}_{+0}}{3 F_\pi^2} \left[
\frac{B_{x} {\rm c}_{+0}}{18} \, \frac{16\pi}{2} \, \frac{(s-4M_\pi^2)^2}{F_\pi^4} \, {\bar J}^2 (s)
\right.
\nonumber\\
&\quad \left.
- \frac{C_{x} - D_{x}}{6}\,\frac{(s-4M_\pi^2)(s-s_0)}{F_\pi^4} \, {\bar J} (s)
+ {\widetilde\xi}_1^{0+} (s) \right]\!
,
\end{align}
with
\begin{equation}
 \frac{{\cal B}_{+;x}}{2} = {\cal B}_{+;0+} = B_x {\rm c}_{+0},
 \qquad {\cal B}_{-;x} = {\cal B}_{-;0+} = 0
 ,
\end{equation}
and
\begin{align}
&{\widetilde w}_{x}^{(0)} = 
2 \left[ A_x + \frac{B_x}{6} \frac{M_K^2 + 3 M_\pi^2}{F_\pi^2}  \right]\!
\left({\rm a}_{+0} - {\rm b}_{+0} \frac{4M_\pi^2}{F_\pi^2} \right)
\nonumber\\
&\quad+  \frac{4}{3} B_{x} {\rm c}_{+0} \frac{M_\pi^2 (M_K^2 + 3 M_\pi^2)}{F_\pi^4}\,,
\nonumber\\
&{\widetilde w}_{x}^{(1)} = 
2 {\rm b}_{+0} \left[ A_{x} + \frac{B_{x}}{6} \frac{M_K^2 + 3 M_\pi^2}{F_\pi^2}  \right]
\nonumber\\
&\quad 
- B_{x} \left({\rm a}_{+0} - {\rm b}_{+0} \frac{4M_\pi^2}{F_\pi^2} \right)
- \frac{1}{3} B_{x} {\rm c}_{+0} \frac{M_K^2 + 7 M_\pi^2}{F_\pi^2}\,,
\nonumber\\
&{\widetilde w}_{x}^{(2)} = B_x \left(   - {\rm b}_{+0}  + \frac{1}{3} {\rm c}_{+0} \right)
,
\end{align}
\begin{align}
&{\widetilde w}_{\pm ; 0+}^{(0)} = 
- \left[ A_x + \frac{B_x}{6} \frac{M_K^2 + 3 M_\pi^2}{F_\pi^2}  \right]
\!\left({\rm a}_{+0} - {\rm b}_{+0} \frac{4M_\pi^2}{F_\pi^2} \right)
\nonumber\\
&\quad
+\frac{2}{3} B_{x} {\rm c}_{+0} \frac{M_\pi^2 (M_K^2 + 3 M_\pi^2)}{F_\pi^4}
\nonumber\\
&\quad
\mp \left[ A_{++} + \frac{B_{++}}{6} \frac{M_K^2 + 3 M_\pi^2}{F_\pi^2}  \right]
\!
\left({\rm a}_{x} - {\rm b}_{x} \frac{4M_\pi^2}{F_\pi^2} \right)
\nonumber\\
&\quad
\mp 
 \left[ A_{x} - \frac{B_x}{3} \, \frac{M_K^2 + 3 M_\pi^2}{F_\pi^2}  \right] \frac{{\rm a}_{00}}{2}\,
,
\nonumber\\
&{\widetilde w}_{\pm ; 0+}^{(1)} =
- {\rm b}_{+0} \left[ A_x + \frac{B_x}{6} \frac{M_K^2 + 3 M_\pi^2}{F_\pi^2}  \right]
\nonumber\\
&\quad
+ \frac{B_x}{2}\left({\rm a}_{+0} - {\rm b}_{+0} \frac{4M_\pi^2}{F_\pi^2} \right)
-
\frac{1}{6} B_{x} {\rm c}_{+0} \frac{M_K^2 + 7 M_\pi^2}{F_\pi^2}
\nonumber\\
&\quad 
\mp {\rm b}_{x} \left[ A_{++} + \frac{B_{++}}{6} \, \frac{M_K^2 + 3 M_\pi^2}{F_\pi^2}  \right]
\nonumber\\
&\quad 
\pm 
\frac{B_{++}}{2} \left({\rm a}_{x} - {\rm b}_{x} \frac{4M_\pi^2}{F_\pi^2} \right)
\mp 
\frac{B_x {\rm a}_{00}}{2}\,
,
\nonumber\\
&{\widetilde w}_{\pm ; 0+}^{(2)} = \frac{1}{2} 
\left[ B_x \left( {\rm b}_{+0} + \frac{{\rm c}_{+0}}{3} \right) \pm B_{++} {\rm b}_{x} \right]\!
.
\end{align}

For the amplitude of $K_L$ decaying into one charged and two neutral pions,
we have
\begin{align}
&{\cal W}_{L;x}^{P\pi} (s) = {\varphi}^{+-}_0 (s) \bigg\{
\frac{(16\pi)^2}{4}  \big[
2 {\widetilde\varphi}^{L;x}_0(s)\varphi^{+-}_0(s) 
\nonumber\\
&\quad
+ {\widetilde\varphi}^{L;00}_0(s) \varphi^{x}_0(s) \big] {\bar J}^2 (s) 
+
\left[ \left(C_{x}^L + \frac{D_{x}^L}{2} \right) \frac{(s-s_0)^2}{F_\pi^4} 
\right.
\nonumber\\
&\quad \left.
+ \frac{2}{3}\, D_x^L \frac{K^2(s)}{F_\pi^4} \right] {\bar J} (s) 
\bigg\} + [\varphi_0^{+-} \sodot {\widetilde\xi}_0^{L;x}](s)
\nonumber\\
&\ 
+ \frac{{\varphi}^{x}_0 (s)}{2} \bigg\{C_{00}^L \left[ \frac{3}{2} \frac{(s-s_0)^2}{F_\pi^4} 
+ \frac{2}{3} \frac{K_{x}^2(s)}{F_\pi^4} \right]
{\bar J} (s)    
\nonumber\\
&\quad
+\frac{(16\pi)^2}{4}  \left[ 2 {\widetilde\varphi}^{L;x}_0(s) \varphi^{x}_0(s) 
+ {\widetilde\varphi}^{L;00}_0(s) \varphi^{00}_0(s)\right]  {\bar J}^2 (s)
\bigg\}
\nonumber\\
&\  
+   \frac{1}{2} \, [\varphi_0^x \sodot {\widetilde\xi}_0^{L;00}](s),
\nonumber\\
&{\cal W}_{L;+0}^{(0)P\pi} = \varphi_0^{+0} (s) \bigg\{
\frac{(16\pi)^2}{2}\, \varphi_0^{+0} (s)\, {\widetilde\varphi}_0^{L;+0} (s) {\bar J}^2 (s)
\nonumber\\
&\quad
-
\left[
\frac{1}{4} \left( C_{x}^L + 5 D_{x}^L \right) \frac{(s-s_0)^2}{F_\pi^4} + \frac{1}{3} ( C_{x}^L + D_{x}^L ) \frac{K^2(s)}{F_\pi^4}
\right]
\nonumber\\
&\quad
\times 
 {\bar J} (s)
 \bigg\} + [\varphi_0^{+0} \sodot {\widetilde\xi}_0^{L;0+}](s),
\\
&{\cal W}_{L;+0}^{(1)P\pi} =
\frac{{\rm c}_{+0}}{3 F_\pi^2} \left[
\frac{B_{x}^L {\rm c}_{+0}}{18} \, \frac{16\pi}{2} \, \frac{(s-4M_\pi^2)^2}{F_\pi^4} \, {\bar J}^2 (s)
\right.
\nonumber\\
&\quad  \left.
-\, \frac{C_{x}^L - D_{x}^L}{6}\,\frac{(s-4M_\pi^2)(s-s_0)}{F_\pi^4} \, {\bar J} (s)
+ {\widetilde\xi}_1^{L;+0} (s) \right]
\nonumber
\end{align}
with
\begin{equation}
 \frac{{\cal B}_{+;L;x}}{2} = {\cal B}_{\pm;L;+0} = B_x^L {\rm c}_{+0},
 \qquad {\cal B}_{-;L;x} = 0
 ,
\end{equation}
and
\begin{align}
{\widetilde w}_{L;x}^{(0)} &= 
2 \left[ A_x^L + \frac{B_x^L}{6} \frac{M_{K_L}^2 + 3 M_\pi^2}{F_\pi^2}  \right]
\! \left({\rm a}_{+0} - {\rm b}_{+0} \frac{4M_\pi^2}{F_\pi^2} \right)
\nonumber\\
&
+ \frac{4}{3} B_{x}^L {\rm c}_{+0} \frac{M_\pi^2 (M_{K_L}^2 + 3 M_\pi^2)}{F_\pi^4}\,,
\nonumber\\
{\widetilde w}_{L;x}^{(1)} &= 
2 {\rm b}_{+0} \left[ A_{x}^L + \frac{B_{x}^L}{6} \frac{M_{K_L}^2 + 3 M_\pi^2}{F_\pi^2}  \right]
\nonumber\\
&
- B_{x}^L \left({\rm a}_{+0} - {\rm b}_{+0} \frac{4M_\pi^2}{F_\pi^2} \right)
- \frac{1}{3} B_{x}^L {\rm c}_{+0} \frac{M_{K_L}^2 + 7 M_\pi^2}{F_\pi^2}\,,
\nonumber\\
{\widetilde w}_{L;x}^{(2)} &= B_x^L \left(   - {\rm b}_{+0}  + \frac{1}{3} {\rm c}_{+0} \right)
,
\end{align}
\begin{align}
&{\widetilde w}_{\pm ; L ; +0}^{(0)} = 
- \left[ A_x^L + \frac{B_x^L}{6} \frac{M_{K_L}^2 + 3 M_\pi^2}{F_\pi^2}  \right]
\! \left(\!{\rm a}_{+0} - {\rm b}_{+0} \frac{4M_\pi^2}{F_\pi^2} \right)
\nonumber\\
&\quad 
+
\frac{2}{3} B_{x}^L {\rm c}_{+0} \frac{M_\pi^2 (M_{K_L}^2 + 3 M_\pi^2)}{F_\pi^4}
\mp \frac{A_{00}^L}{2}
\left({\rm a}_{x} - {\rm b}_{x} \frac{4M_\pi^2}{F_\pi^2} \right)
\nonumber\\
&\quad 
\mp 
 \left[ A_{x}^L - \frac{B_x^L}{3} \, \frac{M_{K_L}^2 + 3 M_\pi^2}{F_\pi^2}  \right] \!
 \left({\rm a}_{+-} - {\rm b}_{+-} \frac{4M_\pi^2}{F_\pi^2} \right)\!
,
\nonumber\\
&{\widetilde w}_{\pm ; L ; +0}^{(1)} =
- {\rm b}_{+0} \left[ A_x^L + \frac{B_x^L}{6} \frac{M_{K_L}^2 + 3 M_\pi^2}{F_\pi^2}  \right]
\nonumber\\
&\quad 
+ \frac{B_x^L}{2}\left({\rm a}_{+0} - {\rm b}_{+0} \frac{4M_\pi^2}{F_\pi^2} \right)
-
\frac{1}{6} B_{x}^L {\rm c}_{+0} \frac{M_{K_L}^2 + 7 M_\pi^2}{F_\pi^2}
\nonumber\\
&\quad 
\mp {\rm b}_{+-} \left[ A_{x}^L - \frac{B_{x}^L}{3} \, \frac{M_{K_L}^2 + 3 M_\pi^2}{F_\pi^2}  \right]
\nonumber\\
&\quad 
\mp
B_{x}^L \left({\rm a}_{+-} - {\rm b}_{+-} \frac{4M_\pi^2}{F_\pi^2} \right)
\mp 
\frac{A_{00}^L{\rm b}_x}{2}\,
,
\nonumber\\
&{\widetilde w}_{\pm ; L;+0}^{(2)} =
 \frac{1}{6}\, B_x^L \left(3 {\rm b}_{+0} + {\rm c}_{+0}\mp 6 {\rm b}_{+-}\right) 
.
\end{align}

For the decay of $K_L$ into three neutral pions, we obtain
\begin{equation}
\begin{split}
&{\cal W}_{L;00}^{P\pi} (s) = {\varphi}^{x}_0 (s) \bigg\{
\frac{(16\pi)^2}{4}  \big[
2 {\widetilde\varphi}^{L;x}_0(s)\varphi^{+-}_0(s) 
\\
&\qquad
+ {\widetilde\varphi}^{L;00}_0(s) \varphi^{x}_0(s) \big] {\bar J}^2 (s) +
\left[ \left(C_{x}^L + \frac{D_{x}^L}{2} \right) \frac{(s-s_0)^2}{F_\pi^4} 
\right.
\\
&\qquad \left.
+ \frac{2}{3}\, D_x^L \frac{K^2(s)}{F_\pi^4} \right] {\bar J} (s) 
\bigg\} + [\varphi_0^x \sodot {\widetilde\xi}_0^{L;x}](s)
\\
&\quad
+ \frac{{\varphi}^{00}_0 (s)}{2} \bigg\{
\frac{(16\pi)^2}{4}  \left[ 2 {\widetilde\varphi}^{L;x}_0(s) \varphi^{x}_0(s) 
\right.
\\
&\qquad \left.
+ {\widetilde\varphi}^{L;00}_0(s) \varphi^{00}_0(s)\right]  {\bar J}^2 (s)
+
C_{00}^L \left[ \frac{3}{2} \frac{(s-s_0)^2}{F_\pi^4} 
\right.
\\
&\qquad \left.
+ \frac{2}{3} \frac{K_{x}^2(s)}{F_\pi^4} \right]
{\bar J} (s)    
\bigg\} + \frac{1}{2} \, [\varphi_0^{00} \sodot {\widetilde\xi}_0^{L;00}](s)
,
\end{split}
\end{equation}
with
\begin{equation}
 {\cal B}_{\pm;L;00} = 0
 ,
\end{equation}
and
\begin{align}
{\widetilde w}_{L;00}^{(0)} &=
A_{00}^L {\rm a}_{00}
+
2 \left[ A_x^L - \frac{B_x^L}{3} \, \frac{M_{K_L}^2 + 3 M_\pi^2}{F_\pi^2}  \right]
\nonumber\\
& \times
\left({\rm a}_x - {\rm b}_x \frac{4M_\pi^2}{F_\pi^2} \right)\!,
\nonumber\\
{\widetilde w}_{L;00}^{(1)} &= 
2 {\rm b}_x \left[ A_x^L - \frac{B_x^L}{3} \, \frac{M_{K_L}^2 + 3 M_\pi^2}{F_\pi^2}  \right]
\nonumber\\
&
+ 2 B_x^L \left({\rm a}_x - {\rm b}_x \frac{4M_\pi^2}{F_\pi^2} \right)\!,
\nonumber\\
{\widetilde w}_{L;00}^{(2)} &= 2 B_x^L {\rm b}_x 
.
\end{align}

Finally, for the three-pion decay of $K_S$, we find
\begin{align}
&{\cal W}_{S;x}^{(1)P\pi} = 
\frac{{\rm c}_{+-}}{3 F_\pi^2} \left[
\frac{B_{x}^S {\rm c}_{+-}}{9} \, \frac{16\pi}{2} \, \frac{(s-4M_\pi^2)^2}{F_\pi^4} \, {\bar J}^2 (s)
\right.
\nonumber\\
& \left.
- \frac{D_{x}^S}{3}\frac{(s-4M_\pi^2)(s-s_0)}{F_\pi^4} \, {\bar J} (s)
+ {\widetilde\xi}_1^{S;x} (s) \right]\!,
\nonumber\\
&{\cal W}_{S;+0}^{(0)P\pi} = \varphi_0^{+0} (s) \left\{
\frac{(16\pi)^2}{2}\, \varphi_0^{+0} (s)\, {\widetilde\varphi}_0^{S;+0} (s) \,{\bar J}^2 (s)
\right.
\nonumber\\
&\left.
+ \frac{D_x^S}{12} \left[
 9 \frac{(s-s_0)^2}{F_\pi^4} - 4 \frac{K^2(s)}{F_\pi^4} 
\right]
 {\bar J} (s)
 \right\}
 + [\varphi_0^{+0} \sodot {\widetilde\xi}_0^{S;0+}](s),
\nonumber\\
&{\cal W}_{S;+0}^{(1)P\pi} = 
\frac{{\rm c}_{+0}}{3 F_\pi^2} \left[
-\frac{B_{x}^S {\rm c}_{+0}}{18} \, \frac{16\pi}{2} \, \frac{(s-4M_\pi^2)^2}{F_\pi^4} \, {\bar J}^2 (s)
\right.
\nonumber\\
& \left.
+ \frac{D_{x}^S}{6}\,\frac{(s-4M_\pi^2)(s-s_0)}{F_\pi^4} \, {\bar J} (s) + {\widetilde\xi}_1^{S;+0} (s)
\right]
,
\end{align}
with
\begin{gather}
{\cal B}_{+;S;x} = 2 B_x^S {\rm c}_{+0},\qquad {\cal B}_{+;S;x} = 0,\\
\ {\cal B}_{\pm;S;+0} =(1 \pm 2) B_{x}^S {\rm c}_{+0}
,
\end{gather}
and
\begin{align}
{\widetilde w}_{S;x}^{(0)} &= 
- B_x^S \frac{M_{K_S}^2+3M_\pi^2}{F_\pi^2}\left({\rm a}_{+0} - {\rm b}_{+0} \frac{4M_\pi^2}{F_\pi^2} \right)
\nonumber\\
&
+ \frac{4}{3} B_x^S {\rm c}_{+0} \frac{M_\pi^2 (M_{K_S}^2 + 3 M_\pi^2)}{F_\pi^4}\,
,
\nonumber\\
{\widetilde w}_{S;x}^{(1)} &= 
3 B_x^S \left({\rm a}_{+0} - {\rm b}_{+0} \frac{4M_\pi^2}{F_\pi^2} \right)
- B_x^S {\rm b}_{+0} \frac{M_{K_S}^2 + 3 M_\pi^2}{F_\pi^2}
\nonumber\\
&- \frac{1}{3} B_x^S {\rm c}_{+0} \frac{M_{K_S}^2 + 7 M_\pi^2}{F_\pi^2}\,,
\nonumber\\
{\widetilde w}_{S;x}^{(2)} &= B_x^S \left( 3  {\rm b}_{+0} + \frac{1}{3} {\rm c}_{+0} \right)\! 
,
\end{align}
\begin{align}
{\widetilde w}_{\pm ; S ; +0}^{(0)} &= 
\frac{B_{x}^S}{2} \, \frac{M_{K_S}^2 + 3 M_\pi^2}{F_\pi^2} 
\left({\rm a}_{+0} - {\rm b}_{+0} \frac{4M_\pi^2}{F_\pi^2} \right)
\nonumber\\
&
- \frac{2}{3} B_{x}^S {\rm c}_{+0} \frac{M_\pi^2 (M_{K_S}^2 + 3 M_\pi^2)}{F_\pi^4} (1 \pm 2)
,
\nonumber\\
{\widetilde w}_{\pm ; S ; +0}^{(1)} &=
- \frac{3 B_{x}^S}{2} \left({\rm a}_{+0} - {\rm b}_{+0} \frac{4M_\pi^2}{F_\pi^2} \right)
\nonumber\\
&
+ \frac{B_{x}^S {\rm b}_{+0}}{2} \, \frac{M_{K_S}^2 + 3 M_\pi^2}{F_\pi^2} 
\nonumber\\
&
- \frac{1}{6} B_{x}^S {\rm c}_{+0} \frac{M_\pi^2 (M_{K_S}^2 + 7 M_\pi^2)}{F_\pi^4} (1 \pm 2)
,
\nonumber\\
{\widetilde w}_{\pm ;S; +0}^{(2)} &= -\frac{3}{2} \, B_{x}^S
\left[ {\rm b}_{+0} - \frac{{\rm c}_{+0}}{9} (1\pm 2)\right]
.
\end{align}


\begin{thebibliography}{99}    



\bibitem{Batley:2005ax} 
  J.~R.~Batley {\it et al.} [NA48/2 Collaboration],
  Phys.\ Lett.\ B {\bf 633}, 173 (2006)
  [hep-ex/0511056].                 
  
  
\bibitem{Batley:2006mu} 
  J.~R.~Batley {\it et al.} [NA48/2 Collaboration],
  Phys.\ Lett.\ B {\bf 634}, 474 (2006)
  [hep-ex/0602014].
  
  
\bibitem{Batley:2006tt} 
  J.~R.~Batley {\it et al.} [NA48/2 Collaboration],
  Phys.\ Lett.\ B {\bf 638}, 22 (2006)
  [Phys.\ Lett.\ B {\bf 640}, 297 (2006)]   
  [hep-ex/0606007].
  
  
\bibitem{Batley:2007md} 
  J.~R.~Batley {\it et al.} [NA48/2 Collaboration],
  Phys.\ Lett.\ B {\bf 649}, 349 (2007)
  [hep-ex/0702045].
  
  
\bibitem{Batley:2007aa} 
  J.~R.~Batley {\it et al.} [NA48/2 Collaboration],
  Eur.\ Phys.\ J.\ C {\bf 52}, 875 (2007)
  [arXiv:0707.0697 [hep-ex]].
  
  
\bibitem{Batley:2000zz} 
  J.~R.~Batley {\it et al.},
  Eur.\ Phys.\ J.\ C {\bf 64}, 589 (2009)
  [arXiv:0912.2165 [hep-ex]].
  

\bibitem{Batley:2010fj} 
  J.~R.~Batley {\it et al.} [NA48/2 Collaboration],
  Phys.\ Lett.\ B {\bf 686}, 101 (2010)
  [arXiv:1004.1005 [hep-ex]].
  
  
\bibitem{Abouzaid:2008aa} 
  E.~Abouzaid {\it et al.} [KTeV Collaboration],
  Phys.\ Rev.\ D {\bf 78}, 032009 (2008)
  [arXiv:0806.3535 [hep-ex]].
  
  
\bibitem{Bashkanov:2007aa} 
  M.~Bashkanov {\it et al.},
  Phys.\ Rev.\ C {\bf 76}, 048201 (2007)
  [arXiv:0708.2014 [nucl-ex]].
  
  
\bibitem{Ambrosino:2008ht} 
  F.~Ambrosino {\it et al.} [KLOE Collaboration],
  JHEP {\bf 0805}, 006 (2008)
  [arXiv:0801.2642 [hep-ex]].
  
  
\bibitem{Adolph:2008vn} 
  C.~Adolph {\it et al.} [WASA-at-COSY Collaboration],
  Phys.\ Lett.\ B {\bf 677}, 24 (2009)
  [arXiv:0811.2763 [nucl-ex]].
  
  
\bibitem{Prakhov:2008ff} 
  S.~Prakhov {\it et al.} [Crystal Ball at MAMI and A2 Collaborations],
  Phys.\ Rev.\ C {\bf 79}, 035204 (2009)
  [arXiv:0812.1999 [hep-ex]].
  
\bibitem{Unverzagt:2008ny} 
M.~Unverzagt {\it et al.} [Crystal Ball at MAMI and TAPS and A2 Collaborations],
Eur.\ Phys.\ J.\ A {\bf 39}, 169 (2009)
[arXiv:0812.3324 [hep-ex]].
  
  
\bibitem{Ambrosinod:2010mj} 
  F.~Ambrosino {\it et al.} [KLOE Collaboration],
  Phys.\ Lett.\ B {\bf 694}, 16 (2010)
  [arXiv:1004.1319 [hep-ex]].
  
  
\bibitem{Adlarson:2014aks}                                                 
  P.~Adlarson {\it et al.} [WASA-at-COSY Collaboration],
  Phys.\ Rev.\ C {\bf 90}, 045207 (2014)
  [arXiv:1406.2505 [hep-ex]].
  
  
\bibitem{Anastasi:2016cdz} 
  A.~Anastasi {\it et al.} [KLOE-2 Collaboration],
  JHEP {\bf 1605}, 019 (2016)
  [arXiv:1601.06985 [hep-ex]].

\bibitem{Ablikim:2015cmz} 
M.~Ablikim {\it et al.} [BESIII Collaboration],
Phys.\ Rev.\ D {\bf 92}, 012014 (2015)
[arXiv:1506.05360 [hep-ex]].
  
  
\bibitem{Prakhov:2018tou}  
  S.~Prakhov {\it et al.} [A2 Collaboration],
  Phys.\ Rev.\ C {\bf 97}, 065203 (2018)
  [arXiv:1803.02502 [hep-ex]].
  
  
\bibitem{Li:2009jd} 
  H.~B.~Li,
  J.\ Phys.\ G {\bf 36}, 085009 (2009)
  [arXiv:0902.3032 [hep-ex]].
  
  
\bibitem{Gan:2015nyc} 
  L.~Gan,
  PoS CD {\bf 15}, 017 (2015).
  
\bibitem{Bijnens:2002vr} 
  J.~Bijnens, P.~Dhonte and F.~Borg,
  Nucl.\ Phys.\ B {\bf 648}, 317 (2003)
  [hep-ph/0205341].
  
  
\bibitem{Bijnens:2004ku} 
  J.~Bijnens and F.~Borg,
  Nucl.\ Phys.\ B {\bf 697}, 319 (2004)
  [hep-ph/0405025].                            
  
  
\bibitem{Bijnens:2004vz} 
  J.~Bijnens and F.~Borg,
  Eur.\ Phys.\ J.\ C {\bf 39}, 347 (2005)
  [hep-ph/0410333].
  
  
\bibitem{Bijnens:2004ai} 
  J.~Bijnens and F.~Borg,
  Eur.\ Phys.\ J.\ C {\bf 40}, 383 (2005)
  [hep-ph/0501163].
  
  
\bibitem{Gamiz:2003pi} 
  E.~Gamiz, J.~Prades and I.~Scimemi,
  JHEP {\bf 0310}, 042 (2003)
  [hep-ph/0309172].


\bibitem{Cabibbo:2004gq} 
  N.~Cabibbo,
  Phys.\ Rev.\ Lett.\  {\bf 93}, 121801 (2004)
  [hep-ph/0405001].
  

\bibitem{Cabibbo:2005ez}    
  N.~Cabibbo and G.~Isidori,
  JHEP {\bf 0503}, 021 (2005)
  [hep-ph/0502130].


\bibitem{Gamiz:2006km} 
  E.~Gamiz, J.~Prades and I.~Scimemi,
  Eur.\ Phys.\ J.\ C {\bf 50}, 405 (2007)
  [hep-ph/0602023].

\bibitem{Colangelo:2006va} 
G.~Colangelo, J.~Gasser, B.~Kubis and A.~Rusetsky,
Phys.\ Lett.\ B {\bf 638}, 187 (2006)
[hep-ph/0604084].

  
\bibitem{Bijnens:2007pr} 
  J.~Bijnens and K.~Ghorbani,
  JHEP {\bf 0711}, 030 (2007)
  [arXiv:0709.0230 [hep-ph]].
  
  
\bibitem{Bissegger:2007yq} 
  M.~Bissegger, A.~Fuhrer, J.~Gasser, B.~Kubis and A.~Rusetsky,
  Phys.\ Lett.\ B {\bf 659}, 576 (2008)
  [arXiv:0710.4456 [hep-ph]].          
  
  
\bibitem{Bissegger:2008zz} 
  M.~Bissegger, A.~Fuhrer, J.~Gasser, B.~Kubis and A.~Rusetsky,
  Prog.\ Part.\ Nucl.\ Phys.\  {\bf 61}, 178 (2008).
  
  
\bibitem{Bissegger:2008ff} 
  M.~Bissegger, A.~Fuhrer, J.~Gasser, B.~Kubis and A.~Rusetsky,
  Nucl.\ Phys.\ B {\bf 806}, 178 (2009)
  [arXiv:0807.0515 [hep-ph]].
  
  
\bibitem{Gullstrom:2008sy} 
  C.-O.~Gullstrom, A.~Kupsc and A.~Rusetsky,
  Phys.\ Rev.\ C {\bf 79}, 028201 (2009)
  [arXiv:0812.2371 [hep-ph]].
  
  
\bibitem{Schneider:2010hs} 
  S.~P.~Schneider, B.~Kubis and C.~Ditsche,
  JHEP {\bf 1102}, 028 (2011)
  [arXiv:1010.3946 [hep-ph]].
  
  
\bibitem{Kampf:2011wr} 
  K.~Kampf, M.~Knecht, J.~Novotn\'y and M.~Zdr\'ahal,
  Phys.\ Rev.\ D {\bf 84}, 114015 (2011)
  [arXiv:1103.0982 [hep-ph]].
  
\bibitem{Gasser:2011ju}
J.~Gasser, B.~Kubis and A.~Rusetsky,
Nucl.\ Phys.\ B {\bf 850}, 96 (2011)
[arXiv:1103.4273 [hep-ph]].
  
\bibitem{Descotes-Genon:2014tla} 
S.~Descotes-Genon and B.~Moussallam,
Eur.\ Phys.\ J.\ C {\bf 74}, 2946 (2014)
[arXiv:1404.0251 [hep-ph]].  

\bibitem{Kolesar:2016jwe} 
M.~Koles\'{a}r and J.~Novotn\'{y},
Eur.\ Phys.\ J.\ C {\bf 77}, 41 (2017)
[arXiv:1607.00338 [hep-ph]].

 
\bibitem{Colangelo:2016jmc} 
  G.~Colangelo, S.~Lanz, H.~Leutwyler and E.~Passemar,
  Phys.\ Rev.\ Lett.\  {\bf 118}, 022001 (2017)
  [arXiv:1610.03494 [hep-ph]].

\bibitem{Albaladejo:2017hhj} 
M.~Albaladejo and B.~Moussallam,
Eur.\ Phys.\ J.\ C {\bf 77}, 508 (2017)
[arXiv:1702.04931 [hep-ph]].
  
\bibitem{Kolesar:2017xrl}
M.~Koles\'{a}r and J.~Novotn\'{y},
Eur.\ Phys.\ J.\ C {\bf 78}, 264 (2018)
[arXiv:1709.08543 [hep-ph]].
  
\bibitem{Colangelo:2018jxw} 
G.~Colangelo, S.~Lanz, H.~Leutwyler and E.~Passemar,
Eur.\ Phys.\ J.\ C {\bf 78}, 947 (2018)
[arXiv:1807.11937 [hep-ph]].  


\bibitem{DescotesGenon:2012gv} 
S.~Descotes-Genon and M.~Knecht,
Eur.\ Phys.\ J.\ C {\bf 72}, 1962 (2012)
[arXiv:1202.5886 [hep-ph]].

  
\bibitem{Stern:1993rg} 
  J.~Stern, H.~Sazdjian and N.~H.~Fuchs,
  Phys.\ Rev.\ D {\bf 47}, 3814 (1993)
  [hep-ph/9301244].
  
  
\bibitem{Knecht:1995tr} 
  M.~Knecht, B.~Moussallam, J.~Stern and N.~H.~Fuchs,
  Nucl.\ Phys.\ B {\bf 457}, 513 (1995)
  [hep-ph/9507319].
  
  
\bibitem{Zdrahal:2008bd} 
  M.~Zdr\'ahal and J.~Novotn\'y,
  Phys.\ Rev.\ D {\bf 78}, 116016 (2008)
  [arXiv:0806.4529 [hep-ph]].  
  
  
\bibitem{Bernard:2013faa} 
  V.~Bernard, S.~Descotes-Genon and M.~Knecht,
  Eur.\ Phys.\ J.\ C {\bf 73}, 2478 (2013)
  [arXiv:1305.3843 [hep-ph]].
  
  
\bibitem{Colangelo:2015kha} 
  G.~Colangelo, E.~Passemar and P.~Stoffer,
  Eur.\ Phys.\ J.\ C {\bf 75}, 172 (2015)
  [arXiv:1501.05627 [hep-ph]].


\bibitem{plan1}  
K.~Kampf, M.~Knecht, J.~Novotn\'y and M.~Zdr\'ahal, in preparation.


\bibitem{plan2}  
K.~Kampf, M.~Knecht, J.~Novotn\'y and M.~Zdr\'ahal, in preparation.


\bibitem{Kampf:2008ts} 
  K.~Kampf, M.~Knecht, J.~Novotn\'y and M.~Zdr\'ahal,
  Nucl.\ Phys.\ Proc.\ Suppl.\  {\bf 186}, 334 (2009)
  [arXiv:0810.1906 [hep-ph]].
  
  
\bibitem{Zdrahal:2009ns} 
  M.~Zdr\'ahal, K.~Kampf, M.~Knecht and J.~Novotn\'y,
  PoS EFT {\bf 09}, 063 (2009)
  [arXiv:0905.4868 [hep-ph]].        
  
  
\bibitem{Zdrahal:2009cp} 
  M.~Zdr\'ahal, K.~Kampf, M.~Knecht and J.~Novotn\'y,
  PoS CD {\bf 09}, 122 (2009)
  [arXiv:0910.1721 [hep-ph]].


\bibitem{Zdrahal}
M. Zdr\'ahal, {\it Construction of pseudoscalar meson amplitudes in chiral perturbation theory using a dispersive approach},
PhD thesis, Charles University, Prague (2011); available at \url{https://is.cuni.cz/webapps/zzp/download/140012164}.
  

\bibitem{GL1}
  J.~Gasser and H.~Leutwyler,
  Nucl.\ Phys.\  B {\bf 250}, 465 (1985).
  

\bibitem{KMW1}
  J.~Kambor, J.~H.~Missimer and D.~Wyler,
  Nucl.\ Phys.\  B {\bf 346}, 17 (1990).
  

\bibitem{KMW2}
  J.~Kambor, J.~H.~Missimer and D.~Wyler,
  Phys.\ Lett.\  B {\bf 261}, 496 (1991).
  

\bibitem{Sutherland:1966zz}
  D.~G.~Sutherland,
  Phys.\ Lett.\  {\bf 23}, 384 (1966).
  
  
\bibitem{Bell:1996mi}
  J.~S.~Bell and D.~G.~Sutherland,
  Nucl.\ Phys.\ B {\bf 4}, 315 (1968).
  
  
\bibitem{Baur:1995gc}
  R.~Baur, J. Kambor and D. Wyler, 
  Nucl.\ Phys.\ B {\bf 460}, 127 (1996)
  [hep-ph/9510396].
  
  
\bibitem{Ditsche:2008cq}
  C.~Ditsche, B. Kubis and U.-G. Mei{\ss}ner, 
  Eur.\ Phys.\ J.\ C {\bf 60}, 83 (2009)
  [arXiv:0812.0344 [hep-ph]].
  
  
\bibitem{Cronin:1967jq}
  J.~A.~Cronin,
  Phys.\ Rev.\  {\bf 161}, 1483 (1967).
  
  
\bibitem{Gasser:1984pr}
  J.~Gasser and H.~Leutwyler,
  Nucl.\ Phys.\ B {\bf 250}, 539 (1985).
  
  
\bibitem{Kambor:1995yc}
  J.~Kambor, C. Wiesendanger and D. Wyler, 
  Nucl.\ Phys.\ B {\bf 465}, 215 (1996)
  [hep-ph/9509374].
  
  
\bibitem{Anisovich:1996tx} 
A.~V.~Anisovich and H.~Leutwyler,
Phys.\ Lett.\ B {\bf 375}, 335 (1996)
[hep-ph/9601237].
  
  
\bibitem{Khuri:1960zz} 
  N.~N.~Khuri and S.~B.~Treiman,
  Phys.\ Rev.\  {\bf 119}, 1115 (1960).

\bibitem{Kacser}
C.~Kacser, 
Phys.\ Rev.\ \textbf{132}, 2712 (1963).

\bibitem{Bonnevay:1963}
G.~Bonnevay, Nuovo Cim. {\bf 30}, 1325 (1963).

\bibitem{Bronzan:1964zz} 
J.~B.~Bronzan,
Phys.\ Rev.\  {\bf 134}, B687 (1964).

\bibitem{Aitchison:1965}
I.~J.~R.~Aitchison, 
Nuovo Cim. {\bf 35}, 434 (1965).

\bibitem{Neveu:1970tn} 
A.~Neveu and J.~Scherk,
Annals Phys.\  {\bf 57}, 39 (1970).

\bibitem{Anisovich:1993kn} 
A.~V.~Anisovich,
Phys.\ Atom.\ Nucl.\  {\bf 58}, 1383 (1995)
[Yad.\ Fiz.\  {\bf 58N8}, 1467 (1995)].


\bibitem{Angelopoulos:1998aw} 
  A.~Angelopoulos {\it et al.} [CPLEAR Collaboration],
  Eur.\ Phys.\ J.\ C {\bf 5}, 389 (1998).

  
\bibitem{Angelopoulos:2003hm} 
  A.~Angelopoulos {\it et al.} [CPLEAR Collaboration],
  Phys.\ Rept.\  {\bf 374}, 165 (2003).
  

\bibitem{Batley:2005zp}
  J.~R.~Batley {\it et al.} [NA48 Collaboration],
  Phys.\ Lett.\ B {\bf 630}, 31 (2005)
  [hep-ex/0510008].       


\bibitem{Eden:1966dnq} 
  R.~J.~Eden, P.~V.~Landshoff, D.~I.~Olive and J.~C.~Polkinghorne,
  The analytic S-matrix, Cambridge Univ. Press (1966).
  
  
\bibitem{Kennedy1962}
J.~Kennedy and T.~D.~Spearman, 
Phys. Rev. {\bf 126}, 1596 (1962).

\bibitem{Bronzan} 
J.~B.~Bronzan and C.~Kacser,
Phys.\ Rev.\ \textbf{132}, 2703 (1963).


\bibitem{Anisovich}
V.~V.~Anisovich and A.~A.~Anselm,
Usp.\ Fyz.\ Nauk \textbf{88}, 287 (1966) [Sov. Phys. Usp. \textbf{9}, 117 (1966)].


\bibitem{Tanabashi2019}
	M.~Tanabashi et al. (Particle Data Group),
	Phys.\ Rev. D {\bf 98}, 030001 (2018) and 2019 update.

\bibitem{Adler:1964um}     
  S.~L.~Adler,
  Phys.\ Rev.\  {\bf 137}, B1022 (1965).
  
  
\bibitem{Adler:1965ga} 
  S.~L.~Adler,
  Phys.\ Rev.\  {\bf 139}, B1638 (1965).
  

\bibitem{KnechtUrech98}
M. Knecht and R. Urech, 
Nucl. Phys. B {\bf 519}, 329 (1998).
[hep-ph/9709348].


\bibitem{BartonKacser}
G. Barton and C. Kacser, 
Nuovo Cim. {\bf 21}, 593 (1961).
  
\bibitem{HerreraSiklody:1996pm} 
P.~Herrera-Siklody, J.~I.~Latorre, P.~Pascual and J.~Taron,
Nucl.\ Phys.\ B {\bf 497}, 345 (1997)
[hep-ph/9610549].

\bibitem{Kaiser:2000gs} 
R.~Kaiser and H.~Leutwyler,
Eur.\ Phys.\ J.\ C {\bf 17}, 623 (2000)
[hep-ph/0007101].

\bibitem{Ablikim:2010kp} 
M.~Ablikim {\it et al.} [BESIII Collaboration],
Phys.\ Rev.\ D {\bf 83}, 012003 (2011)
[arXiv:1012.1117 [hep-ex]].


\bibitem{Adlarson:2017wlz} 
P.~Adlarson {\it et al.},
Phys.\ Rev.\ D {\bf 98}, 012001 (2018)
[arXiv:1709.04230 [hep-ex]].

\bibitem{Ablikim:2017irx} 
M.~Ablikim {\it et al.} [BESIII Collaboration],
Phys.\ Rev.\ D {\bf 97}, 012003 (2018)
[arXiv:1709.04627 [hep-ex]].
  
  
\bibitem{Landau:1959fi}               
  L.~D.~Landau,
  Nucl.\ Phys.\  {\bf 13}, 181 (1959).
  
  
\bibitem{Nakanishi:1959} 
  N.~Nakanishi,
  Prog.\ Theor.\ Phys.\  {\bf 22}, 128 (1959); ibid. {\bf 23}, 284 (1960).


\bibitem{Gasser:1998qt} 
  J.~Gasser and M.~E.~Sainio,
  Eur.\ Phys.\ J.\ C {\bf 6}, 297 (1999)
  [hep-ph/9803251].
  
\bibitem{Petersson:1965zz} 
  B.~Petersson,
  J.\ Math.\ Phys.\  {\bf 6}, 1955 (1965).


\bibitem{tHooft:1978jhc}
  G.~'t Hooft and M.~J.~G.~Veltman,
  Nucl.\ Phys.\ B {\bf 153}, 365 (1979).

\bibitem{SaksZygmund}
S. Saks and A. Zygmund, Analytic functions, volume 28 of Monografie Matematyczne, 
Polskie Towarzystwo Matematyczne, Warsaw (1952); trans. by
E. J. Scott, Elsevier Science Ltd, 3rd revised edition (1971).

\end{thebibliography}
\end{document}